\documentclass[review]{elsarticle}

\usepackage{lineno,hyperref}
\usepackage{amsfonts}
\usepackage{amssymb}
\usepackage{graphicx}
\usepackage[cmex10]{amsmath}
\usepackage[]{algorithm}
\usepackage{algorithmic}
\usepackage{subfigure}
\usepackage{array}
\usepackage{setspace}
\usepackage{chngpage}
\usepackage[english]{babel}
\usepackage{diagbox}
\usepackage{multirow}
\usepackage{bibspacing}
\journal{Journal of \LaTeX\ Templates}

\begin{document}

\begin{frontmatter}

\title{Deep Learning Methods for Solving Linear Inverse Problems: Research Directions and Paradigms}

\tnotetext[mytitlenote]{The paper is published in (Elsevier) Signal Processing, 2020. { https://www.sciencedirect.com/science/article/pii/S0165168420302723?dgcid=author}}

\author{Yanna Bai$^{1}$, Wei Chen$^{1,}$*, Jie Chen$^{2}$ and Weisi Guo$^{3,4}$}
\address{$1.$ State Key Laboratory of Rail Traffic Control and Safety, Beijing Jiaotong University, Beijing, China \\
$2.$ Northwestern Polytechnical University, Xian, China\\
$3.$ Cranfield University, Milton Keynes, UK  \\
$4.$ Alan Turing Institute, London, UK}
\cortext[my corresponding author]{Corresponding author}

\begin{abstract}
The linear inverse problem is fundamental to the development of various scientific areas. Innumerable attempts have been carried out to solve different variants of the linear inverse problem in different applications. Nowadays, the rapid development of deep learning provides a fresh perspective for solving the linear inverse problem, which has various well-designed network architectures results in state-of-the-art performance in many applications. In this paper, we present a comprehensive survey of the recent progress in the development of deep learning for solving various linear inverse problems. We review how deep learning methods are used in solving different linear inverse problems, and explore the structured neural network architectures that incorporate knowledge used in traditional methods. Furthermore, we identify open challenges and potential future directions along this research line.
\end{abstract}

\end{frontmatter}


\section{Introduction}
The study of the inverse problem begins early from the 20th century and is still attractive today. The inverse problem refers to using the results of actual observations to infer the values of the parameters that characterize the system and to estimate data that are not easily directly observed.

\par
The inverse problem exists in many applications. In geophysics, the inverse problem is solved to detect mineral deposits such as underground oil based on the observations of an acoustic wave which is sent from the surface of the earth. In medical imaging, the inverse problem is solved to reconstruct an image of the internal structure of the human body based on the X-ray signal passing through the human body. In mechanical engineering, the inverse problem is solved to perform nondestructive testing by processing the scattered field on the surface, which avoids expensive and destructive evaluation. In imaging, the inverse problem is solved to recover images of high quality from the lossy image, for example, image denoising and image super-resolution (SR).

\par
Mathematically, the inverse problem can be described as the estimation of hidden parameters of the model $\mathbf{m} \in \mathbb{R}^{N} $ from the observed data $\mathbf{d} \in \mathbb{R}^{M}$, where $N$ (possibly infinite) is the number of model parameters and $M$ is the dimension of observed data. A general description of the inverse problem is
\begin{equation}\label{inverse}
\setlength{\abovedisplayskip}{3pt}
\mathbf{d}= \mathcal{A}(\mathbf{m}),
\setlength{\belowdisplayskip}{3pt}
\end{equation}
where $\mathcal{A}$ is the forward operator mapping the model space to the data space.

An inverse problem is well-posed if it satisfies the following three properties \cite{backus1968resolving}.
\begin{itemize}
  \item Existence. For any data $\mathbf{d}$, there exists an $\mathbf{m}$ that satisfies (\ref{inverse}), which means there exists a model that fits the observed data.
  \item Uniqueness. For every $\mathbf{d}$, if there are $\mathbf{m}_{1}$ and $\mathbf{m}_{2}$ that satisfy (\ref{inverse}), then $\mathbf{m}_{1}=\mathbf{m}_{2}$, which means the model that fits the observed data is unique.
  \item Stability. $\mathcal{A}^{-1}$ is a continuous map, which means small changes in the observed data $\mathbf{d}$ will make small changes in the estimated model parameters $\mathbf{m}$.
\end{itemize}
If any of the three properties does not hold, the inverse problem is ill-posed.

\subsection{The Linear Inverse Problem}
In linear inverse problems (LIPs), the forward operator $\mathbf{A}$ in (\ref{inverse}) is linear and can be written as a matrix $\mathbf{A} \in \mathbb{R}^{M\times N}$. When $M = N$ and the matrix $\mathbf{A}$ has a full rank, the solution of the LIP is unique, and the model parameters are given by multiplying the matrix inverse $\mathbf{A}^{-1}$ with the data $\mathbf{d}$. In the situation $M > N$, it becomes an over-determined problem that may have no solution. In situations where $M < N$, the LIP is undetermined, and the solution of the undetermined LIP is not unique, which means this LIP is ill-posed. To solve the ill-posed LIP, extra knowledge of the system model is usually needed, which is also known as prior information.

\par
In the presence of noisy observed data $\mathbf{d}$, the LIP can be expressed as an optimization problem as following
\begin{equation}\label{optimize with noise}
\setlength{\abovedisplayskip}{3pt}
\min_{\mathbf{m}} \|\mathbf{d}-\mathbf{A}\mathbf{m}\|^{2}_{2}+ J(\mathbf{m}),
\setlength{\belowdisplayskip}{3pt}
\end{equation}
where $J(\cdot)$ incorporates the prior information. For example, the Tikhonov regularization is popularly used, where $J(\mathbf{m})= \| \boldsymbol{\Gamma} \mathbf{m}\|^{2}_{2}$ and $\boldsymbol{\Gamma}$ represents the Tikhonov matrix (e.g. $\boldsymbol{\Gamma} =\alpha \mathbf{I}$).

\par
Based on the different prior information and the structure of the operator $\mathbf{A}$, the LIP can be classified into different categories \cite{kabanikhin2008definitions}. In the following two subsections, we review LIPs that attract extensive interests in recent years.

\subsection{LIPs With Various Parameterized Models}

In this subsection, we introduce LIPs with various parameterized models, which correspond to different prior information.

\subsubsection{Sparse LIPs}
In LIPs, one popular prior information is the sparsity of the parameters, which has been applied in communication systems \cite{8770109,8122055}, sensor networks \cite{7390294,7031966} and many other applications \cite{7293676,986949}.

\par
In sparse LIPs, $\mathbf{m}$ is a sparse vector where only several elements of $\mathbf{m}$ are non-zeros, and the prior information $J(\cdot)=\alpha \|\mathbf{m}\|_{0}$, where $\alpha$ is some regularization parameter and $\|\mathbf{m}\|_{0}$ denotes the $\ell_{0}$ norm of the vector $\mathbf{m}$ that counts the number of non-zeros in $\mathbf{m}$.
While the optimization problem in sparse LIPs has non-continuous objective function and is NP-hard, we usually resort to solve an alternative problem with a smoothed objective function \cite{8332502}. The regularizer $J(\cdot)$ is replaced by a sparsity-enforcing function, e.g., the $\ell_{1}$ norm function $J(\cdot)= \|\mathbf{m}\|_{1}$ and the log penalty function $J(\cdot)= \sum_{i=1}^N\log(1+m_i^{2})$ in \cite{lee2007efficient}. Under certain conditions on the matrix $\mathbf{A}$ and the sparsity level of $\mathbf{m}$, the solution of the new optimization problem is equivalent to the original problem \cite{8123942}.

\par
In addition to the sparse structure, real world signals exhibit many other structures, e.g., block-sparsity \cite{7101819}, group-sparsity \cite{6826555}, tree-sparsity \cite{5437428} and others \cite{6967808,7478129}, which can be exploited in solving $\mathbf{m}$ from the observations $\mathbf{d}$. Considering the block-sparsity or the group-sparsity, $\mathbf{m}$ can be written as $\mathbf{m}=[\mathbf{m}_{1};\mathbf{m}_{2};\ldots;\mathbf{m}_{r}]$ with $\mathbf{m}_{i}\in \mathbb{R}^{L} (i=1,\ldots,r)$ for $M=Lr$, where only several of the $\mathbf{m}_{i}$ are non-zeros vectors. For the tree-sparsity, the non-zeros cluster along the branches of the tree. That means, if a node is non-zero, then the other nodes that are on the branch from the root to the node are non-zeros. The tree-sparsity wildly exists in the wavelet coefficients of nature signals and images.

\par
\begin{figure}[!t]%
\centering%
\vspace{-0.8cm}
\setlength{\abovecaptionskip}{-0.05cm}
\setlength{\belowcaptionskip}{-0.5cm}
\includegraphics[width=0.6\textwidth]{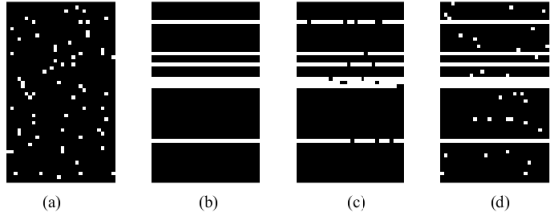}%
\DeclareGraphicsExtensions. \caption{Various structured sparse models \cite{7558157}. (a)sparsity, (b)row-sparsity (c)row-sparsity with embedded element-sparsity, (d)row-sparsity plus element-sparsity.}
\label{fig:sparse}
\end{figure}

Another popular structure exists in the multiple measurement vector (MMV) problem which is the extension of the basic sparse LIP. The hidden parameter is $\mathbf{M}=[\mathbf{m}_{1},\mathbf{m}_{2},\ldots,\mathbf{m}_{L}]\in \mathbb{R}^{N \times L}$, and the measurements $\mathbf{D}=[\mathbf{d}_{1},\mathbf{d}_{1},\ldots,\mathbf{d}_{L}]\in \mathbb{R}^{M \times L}$. In many MMV problems, columns of $\mathbf{M}$ are considered to be jointly sparse \cite{8039502}. The simplest MMV structure is row-sparsity where the non-zeros of each column share the same supports (Fig.~\ref{fig:sparse}(b)). There are various jointly sparse structures in MMV problems, some of which are illustrated in Fig.~\ref{fig:sparse} \cite{7558157}. More structures can be formed by combining the jointly sparse structure in the MMV and the structure in each vector, e.g., the forest sparsity \cite{6800127} which combines the joint sparsity and the tree-sparsity.

\subsubsection{Low-rank LIPs}

Low-rank matrix recovery is another rapid-developed research topic with broad applications, such as saliency detection \cite{zou2013segmentation}, face recognition \cite{1177153} and others \cite{koren2009matrix,7914672}.

\par
The low-rank matrix recovery aims to estimate a low-rank matrix $\mathbf{M} \in \mathbb{R}^{N_{1} \times N_{2}}$ from the observed data $\mathbf{d}$, which is obtained by using a linear operator $\mathcal{A}: \mathbb{R}^{N_{1} \times N_{2}} \rightarrow \mathbb{R}^{M} (M<N_{1}N_{2})$. In low-rank matrix recovery problem, the prior information $J(\cdot)=\alpha\cdot \text{rank}(\mathbf{M})$, where $\text{rank}(\cdot)$ denotes the matrix rank and $\alpha$ denotes the regularization parameter. This optimization problem is also NP-hard. Alternatively, under certain conditions on the linear mapping and the matrix rank, one can replace $J(\cdot)$ with $J(\cdot)=\alpha\|\mathbf{M}\|_{\ast}$, where $\|\cdot\|_{\ast}$ denotes the matrix nuclear norm that sum the singular values of the matrix. As the tightest convex relaxation of rank minimization, the nuclear norm minimization problem can be solved via various convex optimization algorithms \cite{8318613}.

\par
In real-world signals, the low-rank structure can be combined with other structures. In simultaneously sparse and low-rank matrix reconstruction problem, which exists in sub-wavelength optical imaging, hyperspectral image unmixing, graph denoising, the matrix $\mathbf{M}$ is simultaneously sparse and low-rank \cite{8453881}. The corresponding regularizer $J(\cdot)=\alpha \|\mathbf{M}\|_{0}+ \beta \cdot \text{rank}(\mathbf{M})$, where $\alpha$ and $\beta$ are positive parameters that balance the sparsity, the matrix rank, and the data fitting term. A popular convex relaxation of this problem is to replace the $\ell_{0}$ norm and rank function with the $\ell_{1}$ norm and the nuclear norm, respectively. The sparse plus low-rank matrix reconstruction aims to recover a matrix $\mathbf{M}$ which is the superposition of a low-rank matrix $\mathbf{L}$ and a sparse matrix $\mathbf{S}$. This problem arises in applications such as network anomalous detection, magnetic resonance imaging (MRI) and single voice extraction. An alternative optimization problem with convex relaxed terms can be used to facilitate algorithm development, e.g., the robust principal component analysis (RPCA) with an identity matrix as the mapping $\mathcal{A}$ \cite{candes2011robust}.

\par
The low-rank structure also exists in tensor. Tensor is a higher dimensional generalization of the matrix that attracts great attention recent years. Low-rank tensor recovery aims to recover the low-rank tensor $\boldsymbol{\mathcal{M}} \in \mathbb{R}^{N_{1} \times \ldots \times N_{n}} $ from a limited number of observations, where $\mathcal{A}: \mathbb{R}^{N_{1} \times \ldots \times N_{n}} \rightarrow \mathbb{R}^{M} $ (typically $M \ll \prod_{i=1}^n N_i $). The corresponding prior information $J(\cdot)=\text{rank}(\boldsymbol{\mathcal{M}})$, where $\text{rank}(\boldsymbol{\mathcal{M}})$ denotes some form of tensor rank. One popular approach is to use tensor nuclear norm $\|\boldsymbol{\mathcal{M}}\|_{\ast}$, which is a convex combination of the nuclear norms of all matrices unfolded along different modes~\cite{6138863}. There also exists nonconvex method, for example, in \cite{8861380}, Chen et al. propose an empirical Bayes method that has state-of-the-art performance in sparse and low-rank matrix recovery.

\subsection{LIPs With Different Structures of $\mathbf{A}$}
In this subsection, we introduce the LIPs with various linear operators $\mathbf{A}$, which arises in different applications.

The linear operator A is an identity matrix in denoising. Solving the inverse problem in denoising is to remove the noise $\mathbf{n}$ from the observed data $\mathbf{d}$. In LIPs, the observed data may contain noise that comes from the measurement process, the transmission process, the quantization or the compression process for storage. Imperfect instruments and interfering natural phenomena can also introduce noise. There are various types of noise in different applications. For example, images may be corrupted by gaussian noise, salt and pepper noise, speckle noise, Brownian noise and other \cite{saxena2014noises}. Denoising is the process of removing the noise from the observed data, which is an essential and important problem that can be found in astronomy, medical imaging and many other applications. Existing algorithms for denoising include non-local means~\cite{8705542}, curvelet transform~\cite{7740078}, statistical modeling~\cite{4517848}, and nonlocal self-similarity (NSS) models \cite{4271520}. The NSS models are popular in advanced methods such as BM3D \cite{4271520}, NCSR \cite{6392274} and WNNM \cite{6909762}. For blind denoising, the techniques based on dictionary learning and transform learning are popular \cite{7194811,7004012,6566099,8438535}.

Image SR is another typical LIP where the linear operator $\mathbf{A}=\mathbf{D}\mathbf{B}\mathbf{M}$ refers to the image acquisition process which contains a set of degradations that involve warping, blurring, down-sampling and noise \cite{1203207}. Image SR aims to reconstruct a high-resolution (HR) image from a single low-resolution (LR) image or multiple LR images. Since the number of known parameters in LR images exceeds the number of unknown variables in HR images, image SR is an ill-posed LIP. Classic methods for image SR include edge-based methods~\cite{8759425}, image statistical methods~\cite{7816724}, sparse coding~\cite{8532893} and example-based methods~\cite{7351320}.

\par
Compressed sensing (CS) is a LIP whose linear operator $\mathbf{A}$ has more columns than rows. CS is a sampling paradigm that breaks the Nyquist theory and can restore the entire desired signal from fewer measured values by using sparse signal characteristics. In CS, the linear operator $\mathbf{A}$ has fewer rows than columns, i.e., $M<N$, which leads to an underdetermined system. To reconstruct the signal $\mathbf{m}$ from a reduced number of observations, the reconstructed signal $\mathbf{m}$ is required to be sparse, or represented as a sparse vector under certain transformations, e.g., wavelet transform, Fourier transform and discrete cosine transform.

\par
Feature Selection (FS) is a LIP whose linear operator $\mathbf{A}$ has fewer columns than rows. FS is the process that finds features having the most contribution to our prediction or the output we are interested in. It is a useful tool to simplify models for interpretation, reduce overfitting and avoid the curse of dimensionality in machine learning and signal processing. FS has been applied in many applications such as text categorization, bioinformatics and data mining. One approach to conduct FS is to formulate the problem as a LIP. For example, to classify handwritten digits, each row of $\mathbf{A}$ includes the feature coefficients of one data sample \cite{6588338}. Since the number of data samples could be large, the linear operator $\mathbf{A}$ has $M>N$. A key premise of FS is that the data contains redundant or irrelevant features, and thus removing those features does not result in loss of information in the prediction \cite{7530147}.

\par
Dictionary learning denotes a LIP whose linear operator $\mathbf{A}$ and its representation $\mathbf{m}$ are learned from the observed data $\mathbf{d}$, which exists in many applications such as image classification\cite{7924323}, outliers detection \cite{qi2018learning}, and distributed CS \cite{6880772}. With the learned dictionary $\mathbf{A}$, the high-dimensional signal performs dimensionality reduction to remove redundant information generated in the sampling process. Generally, only some of the atoms in the dictionary are used to construct the sparse representation of the high-dimensional signal. Compared with the predefined dictionary, e.g., wavelets, the learned one would be more appropriate for the signals of the same ensemble and could lead to improved performance in various tasks, e.g., denoising and classification. We refer interested readers to \cite{5714407} for more details on various dictionary learning methods including the probabilistic learning methods, the learning methods based on clustering or vector quantization, and the methods for learning dictionaries with a particular construction. While the traditional dictionary learning relies on the one level of the dictionary, the new deep dictionary learning (DDL), which combines the concept of dictionary learning and deep learning (DL), uses multiple layers of dictionaries to represent the signal \cite{7779008}. The dictionary learning can also combine with other techniques, for example, Gong et al. propose a simultaneously sparse and low-rank tensor representation model to enhance the capability of dictionary learning for hyperspectral image denoising \cite{8982090}, and Xin et al. jointly optimize the sensing matrix and sparsifying dictionary for tensor CS \cite{7914777}.

\section{DL and LIPs}

In this section, we first illustrate the motivation and advantages of using DL in solving LIPs. Then, we summarize the earlier efforts of using DL in inverse problems and clarify the novelty of this review. Then, we briefly introduce the categorization of different methods in section 3.

\subsection{Motivation and Advantages}

As a long-standing problem, plenty of algorithms have been proposed in kinds of literature to solve LIPs, for example, in CS, under certain conditions on the sensing matrix $\mathbf{A}$, e.g., the restricted isometry property (RIP) \cite{candes2008restricted}, the LIP has a unique solution and can be solved with algorithms with relatively low computational complexity, e.g., iterative hard thresholding \cite{blumensath2009iterative}, orthogonal matching pursuit \cite{4385788}, message-passing algorithms \cite{donoho2009message} and the sparse Bayesian learning based algorithms \cite{8074806}. However, in applications, these conditions are often unattainable.

\par
In recent years, DL attracts wide attention as a promising approach to solve the LIP. For example, by unfolding an iterative algorithm into a neural network (NN), we can learn the parameters of iterative algorithms from training data, which differs from traditional algorithms that employ predetermined parameters.

\par
\begin{figure}[!t]%
\centering%
\vspace{-0.8cm}
\setlength{\abovecaptionskip}{-0.05cm}
\setlength{\belowcaptionskip}{-0.5cm}
\includegraphics[width=0.6\textwidth]{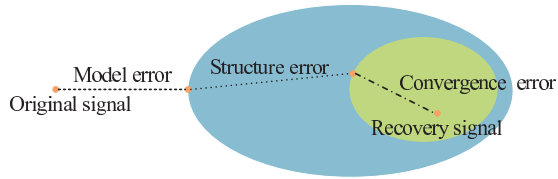}%
\DeclareGraphicsExtensions. \caption{The decomposition of the error in the solution of inverse problems. }
\label{fig:error}
\end{figure}

Using DL to solve inverse problems has several advantages. Firstly, in comparison to traditional iterative algorithms, DL can significantly increase the speed of convergence. For example, Gregor and LeCun validate that the DL based method is 10 times faster than the iterative coordinate descent method with the same approximation error~\cite{gregor2010learning}. In addition, DL based methods are capable to decrease the average recovery error. As shown in Fig.~\ref{fig:error}, the recovery error of all algorithms results comes from several aspects. Imperfect modeling of the problem leads to the model error, the approximation (e.g., using convex relaxation) of the original objective function leads to the structure error, and the sub-optimal solution of algorithms leads to the convergence error. Instead of dealing with the imperfect mathematical models and approximated optimization problems, the DL based method learns the mapping from the input to the output directly and has the potential to overcome or relieve challenges brought by the model error, the structure error and the convergence error in traditional algorithms. The success of DL methods for inverse problems has been observed in a number of works \cite{6247952,6781616,xin2016maximal,gregor2010learning,wang2016learning,7905837,7934066}.

\begin{table}[]
\centering  
\setlength{\abovecaptionskip}{0pt}%
\setlength{\belowcaptionskip}{4pt}%
\caption{The denoising results of real-world images.}\label{result_dataset}
\footnotesize
\begin{tabular}{|c|c|c|c|}  
\hline
Criterion   &Method   &DND  &PolyU   \\ \hline

\multirow{5}*{PSNR}

&BM3D \cite{4378954}  &$34.51$ &$37.84$ \\  \cline{2-4}

&KSVD \cite{1710377} &$36.49$ &$36.3726$ \\  \cline{2-4}

&MCWNNM \cite{8237387} &$37.38$ &$35.2274$ \\  \cline{2-4}

&TWSC \cite{xu2018trilateral} &$37.94$ &$36.4771$  \\  \cline{2-4}

&\textbf{CBDNet~\cite{guo2019toward}} &$\textbf{38.06}$ &$\textbf{37.00}$  \\  \hline

\multirow{5}{*}{SSIM}

&BM3D  &$0.8507$ &$0.9619$ \\  \cline{2-4}

&KSVD &$0.8978$ &$0.9244$ \\  \cline{2-4}

&MCWNNM &$0.9294$ &$0.9453$ \\  \cline{2-4}

&TWSC &$0.9403$ &$0.9281$  \\  \cline{2-4}

&\textbf{CBDNet} &$\textbf{0.9421}$ &$\textbf{0.9457}$  \\  \hline

\end{tabular}
\end{table}

\begin{table}[]
\centering  
\setlength{\abovecaptionskip}{0pt}%
\setlength{\belowcaptionskip}{4pt}%
\caption{Test time for different methods on a single image denoising.}\label{time}
\footnotesize
\begin{tabular}{c|c|c|c|c|c}  
\hline
Methods  &\textbf{CBDNet}   &KSVD  &BM3D &MCWNNM &TWSC      \\ \hline\hline
Time(s)   &$\textbf{0.0165}$ &$0.8991$ &$1.3575$ &$298.21$ &$391.47$\\  \hline
\end{tabular}
\end{table}
\par
To unveil the advantages of the DL based method in solving LIPs, we show the performance of the DL model and the state-of-the-art traditional algorithms in real-word image denoising. The results in DND \cite{8099777} dataset come from the work of Guo et al.~\cite{guo2019toward} and the results in PolyU \cite{xu2018real} dataset is from our experiments. As shown in Table \ref{result_dataset} and Table \ref{time}, the CBDNet outperforms most of traditional algorithms in both PSNR/SSIM and computing time. These simulations are conducted on a computer with a quad-core 4.2GHz CPU, 16 GB RAM, a GTX1080Ti GPU, and the Microsoft Windows 10 operating system.

\subsection{Related Surveys and Categorization Methodology}

Several remarkable works have compiled comprehensive reviews on using DL in inverse problems. However, existing reviews mainly focus on the application of imaging \cite{8103129,8253590,8723565,8962949,9084378}. In \cite{8103129}, McCann et al. summarize the use of the convolutional NN (CNN) to solve imaging problems such as denoising, SR, and reconstruction. They focus on the design of the CNNs including the training data, the architecture, and the problem formulation. Lucas et al. also focus on imaging problems, but they summarize a wild range of NNs, including the multilayer perceptron (MLP), CNNs, autoencoders (AEs), and generative adversarial networks (GANs) \cite{8253590}. In the recent work \cite{9084378}, Ongie et al. propose a taxonomy for DL in imaging according to the forward model and the learning process. Other reviews include the review of using DL for MRI image reconstruction \cite{8962949} and image SR \cite{8723565}, which are also focus on a special application of inverse problems. A survey for data-driven methods in inverse problems is given in \cite{arridge2019solving}, which aims to promote more theoretical research.

\par
In this paper, we categorize the LIPs according to various parameterized models according to different prior information in the linear operator A and the data d, then we focus on the innovation of constructing a specified NN for various parameterized models, instead of considering the NN as a black box. We aim to provide a comprehensive review of state-of-the-art DL techniques in solving LIPs, not limited to imaging problems. Our hope is that this article can provide guidance for designing NNs for various LIPs. At last, we discuss the existing challenges and promising directions for further research, which are not all covered in literature.

\par
\begin{figure}[!t]%
\centering%
\vspace{-0.8cm}
\setlength{\abovecaptionskip}{-0.05cm}
\setlength{\belowcaptionskip}{-0.5cm}
\includegraphics[width=1\textwidth]{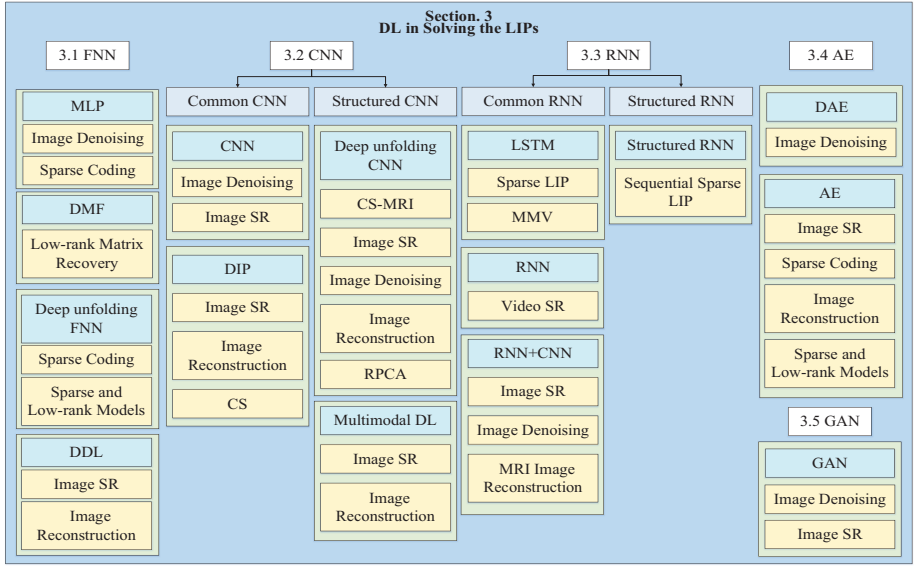}%
\DeclareGraphicsExtensions. \caption{Schematic diagram of the structure of section 3. }
\label{fig:structure}
\end{figure}

In Fig.~\ref{fig:structure}, we show the structure of section 3. Our taxonomy in section 3 is according to the type of NNs, as the architecture of the NN is the most pivotal element of DL and determines whether the NN can effectively capture the deep features of the training data. We summarize the use of fully connected NNs (FNNs), CNNs, recurrent NNs (RNNs), AEs, and GANs in dealing with various LIPs, including CS, denoising, image SR, and others. In addition to the generic NN, we summarize various structured NN, which defines the NN that combines the prior information in various forms. Among the structured NN, the most famous one is the deep unfolding methods which unfold the iterations of an iterative inference method into layer-wise structure analogous to a NN \cite{hershey2014deep}. In addition to the deep unfolding networks, we also consider the structured networks that get inspiration from the traditional analyzed-based methods. For example, the DDL combines the concept of DL and dictionary learning.

\section{DL in Solving LIPs}

In this section, we introduce how DL is exploited to handle LIPs in different applications and provide detailed instructions on the construction of the NN and the training process. Different settings in DL based methods are summarized in Table \ref{fnn in LIP}-Table \ref{GAN in LIP}, which include the input/output, loss function, learning rate, initialization and training algorithms. With these settings, we can easily train NNs using popular DL platforms such as Tensorflow~\cite{abadi2016tensorflow} and PyTorch \cite{paszke2019pytorch}.

\subsection{FNNs}
The FNN, also known as MLP is one of the most basic structures in DL and a powerful tool in solving LIPs. In addition to the basic FNN, some modifications can be employed to enhance the performance, such as skip connections between layers \cite{gregor2010learning}, well-designed activation functions \cite{wang2016learning} and weight constrains \cite{daubechies2004iterative}. Here we introduce common FNNs and structured FNNs related to LIPs. Various FNNs for LIPs are summarized in Table \ref{fnn in LIP}.

\begin{table*}[]
\centering  
\setlength{\abovecaptionskip}{0pt}%
\setlength{\belowcaptionskip}{6pt}%
\caption{FNNs for LIPs.}\label{fnn in LIP}
\begin{adjustwidth}{-1cm}{-1cm}
\tiny
\begin{tabular}{|p{0.5cm}|p{1.1cm}|p{1.5cm}|p{1.5cm}|p{1.5cm}|p{1.5cm}|p{2cm}|p{1.5cm}|}
\hline
Ref. &Application &Input &Output &Loss Function &Initialization &Learning Rate &Optimizer   \\ \hline\hline

\cite{6247952}  &Image Denoising &Normalized overlapping patches  &Clean patches &The quadratic error  &Normal distribution  & $0.1/N$($N$ is the number of layer units) &SGD \\ \hline

\cite{6781616} &Image Denoising &Pre-processed grey image and depth image &Denoised image &Proposed edge based weighted loss function &Not given &Not given &Not given  \\ \hline

\cite{xin2016maximal} &Sparse Coding  &The observed signal &Non-zero probability &Softmax loss function &Follows \cite{7410480} &$0.01$(reduce by 90 percent every 50 epoches)  &SGD(momentum 0.9, weight decay $10^{-4}$ ) \\ \hline

\cite{gregor2010learning} &Sparse Coding &The observed signal &The recovered signal  &The quadratic error &The weights given in (\ref{ISTA}) &Not given &SGD  \\ \hline

\cite{8545232} &Sparse Coding  &The observed signal &The recovered signal  &The quadratic error &The weights given in (\ref{ISTA}) &$10^{-4}$  &Not given  \\ \hline

\cite{chen2018theoretical} &Sparse Coding &The observed signal &The recovered signal  &The quadratic error plus the L1 norm &Pre-trained network &Stage-wise &Not given  \\ \hline

\cite{aberdam2020ada} &Sparse Coding &The observed signal &The recovered signal  &The quadratic error &Identity matrix &Not given &Not given  \\ \hline

\cite{liu2018alista} &Sparse Coding &The observed signal &The recovered signal  &The quadratic error &Analytic weights(fixed) &$10^{-4}$(exponentially decaying)  &Not given  \\ \hline

\cite{ablin2019learning} &Sparse Coding &The observed signal &The recovered signal  &The quadratic error &Follow ISTA(fixed) &Not given  &Not given  \\ \hline

\cite{wang2016learning} &Sparse Coding &The observed signal &The recovered signal  &Softmax loss function  &The weights in IHT &$10^{-2}$ &SGD  \\ \hline

\cite{7905837}  &Sparse Coding  &The observed signal &The recovered signal  &The quadratic error &The weights in AMP &Not given &ADAM \\ \hline

\cite{7934066} &Sparse Coding  &The observed signal &The recovered signal  &The quadratic error  &The weights in VAMP &Not given &ADAM  \\ \hline

\cite{scetbon2019deep}  &image denoising  &Noisy patches &Clean patches &The quadratic error  &Kaiming Uniform
 &$10^{-4}$ &SGD  \\ \hline

\cite{8695874} &Sparse Coding  &The observed signal &The recovered signal  &The quadratic error  &Not given &$4\times10^{-2}$ to $8\times10^{-4}$  &ADAM  \\ \hline

\cite{8683182} &Sparse Coding  &The observed signal &The recovered signal  &The quadratic error  &Not given &Not given &ADAM  \\ \hline

\end{tabular}
\end{adjustwidth}
\end{table*}

\par
Perhaps the most straightforward DL based method for LIPs is the use of common FNNs. Especially for image denoising~\cite{6247952,6781616} and sparse LIPs~\cite{xin2016maximal}. Considering that the ordinary MLP can approximate more functions than the CNNs, Burger et al. firstly apply an ordinary MLP for image denoising and obtained competitive results compared to the classical BM3D~\cite{6247952}. To achieve the start-of-the-art performance, they adopt a large network that consists of sufficient parameters, a large patch size and large training set. The network is effective in noisy images which contain the additive white Gaussian noise. However, the accuracy of this method is sensitive to the mismatch of the noise distributions in the training data set and the testing data set. To against varying noise levels, Wang and Morel employ a linear mean shift before the denoising network to improves the robustness of the network \cite{6781616}. To solve the sparse LIP, Xin et al. incorporate some powerful techniques such as batch normalization and residual connection into the FNNs, and uses the support of the vector as labels to train the network, which reduces the burden of the NN in solving the sparse inverse problem~\cite{xin2016maximal}.

In addition to the image denoising and sparse LIPs, the FNN is also used in low-rank matrix recovery. One of the typical low-rank matrix recovery problems is the matrix completion problem where the matrix to be completed is assumed to be low-rank. In \cite{fan2018matrix}, Fan and Cheng propose a deep-structure NN named deep matrix factorization (DMF) for matrix completion, which is more computationally efficient than the nuclear norm and truncated nuclear norm related methods. In DMF, the input is low-dimensional unknown latent variables and is jointly optimized with the parameters. The output of the network is the incomplete low-rank matrix. The DMF aims to recover an incomplete low-rank matrix by learning a nonlinear latent variable model. Exploring the implicit regularization, Arora et al. prove that the deeper DMF can lead to more accurate low-rank solutions \cite{arora2019implicit}.


\par
FNNs can also benefit from the unfolding of traditional iterative algorithms, which leads to deep unfolding FNNs \cite{hershey2014deep}. Generally, the $t$-th iteration of an iterative algorithm can be written as
\begin{equation}\label{iteration}
\setlength{\abovedisplayskip}{3pt}
\mathbf{\hat{m}}_{t+1} = g(\mathbf{W}\mathbf{d}+\mathbf{S}\mathbf{\hat{m}}_{t}),
\setlength{\belowdisplayskip}{3pt}
\end{equation}
where $\mathbf{W}$ and $\mathbf{S}$ are algorithm-dependent parameters, and $g$ is a nonlinear function. In view of the fact that the update rule in (\ref{iteration}) shares great similarities to one layer of a FNN, various iterative algorithms are unfolded and transformed into different deep unfolding FNNs for solving LIPs.

\begin{figure}
\centering
\vspace{-0.8cm}
\setlength{\abovecaptionskip}{-0.05cm}
\setlength{\belowcaptionskip}{-0.5cm}
\subfigure[Left: Block diagram of the ISTA. Right: The structure of the LISTA, uses a time-unfolded version of the ISTA block diagram of three iterations (The network can have arbitrary layers). ]{
\label{fig:subfig:ista}
\includegraphics[width=0.78\textwidth]{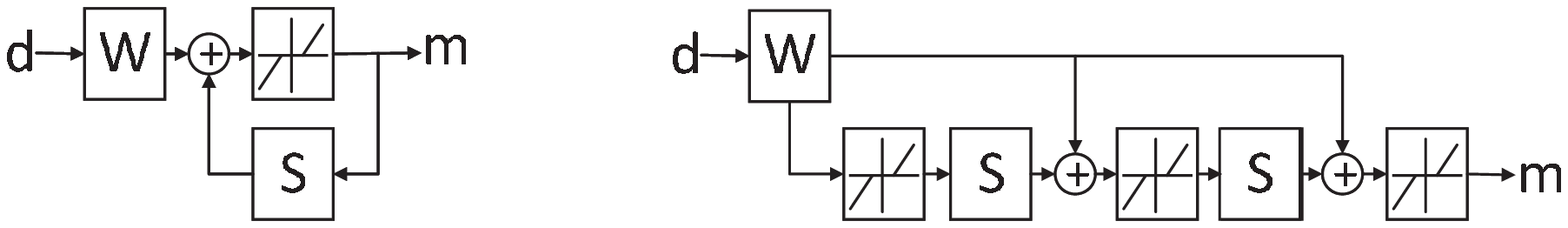}}
\hspace{1in}
\subfigure[Left: Block diagram of the IHT algorithm. Right: The time-unfolded version of the IHT algorithm. In the deep $\ell_{0}$ encoder, the hard thresholding function is decomposed into two linear scaling operators plus a HELU.]{
\label{fig:subfig:iht}
\includegraphics[width=0.78\textwidth]{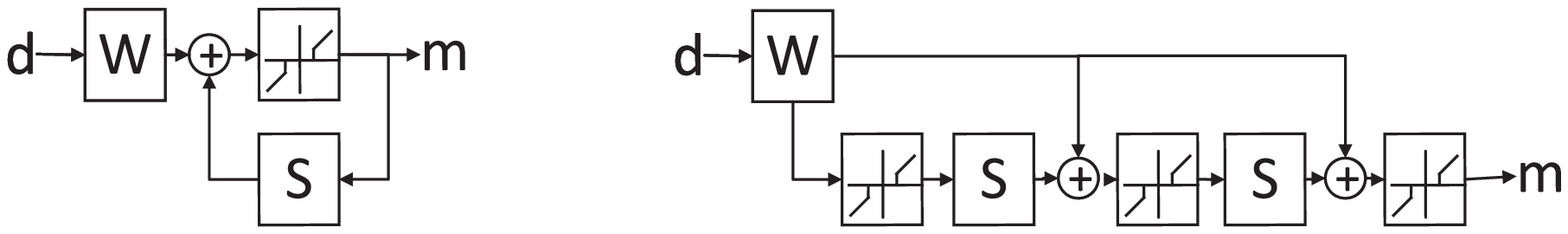}}
\hspace{1in}
\subfigure[A different form of LISTA, with learnable parameters $\mathbf{A}_{t},\mathbf{B}_{t}$ and $\theta_{t}$. ]{
\label{fig:subfig:ista2}
\includegraphics[width=0.78\textwidth]{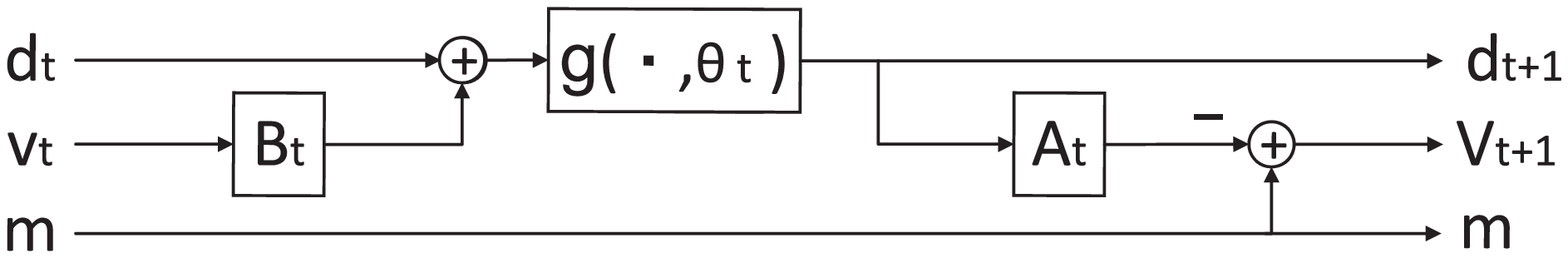}}
\hspace{1in}
\subfigure[The structure of LAMP, with learnable parameters $\mathbf{A}_{t},\mathbf{B}_{t}$ and $\theta_{t}$.]{
\label{fig:subfig:amp}
\includegraphics[width=0.78\textwidth]{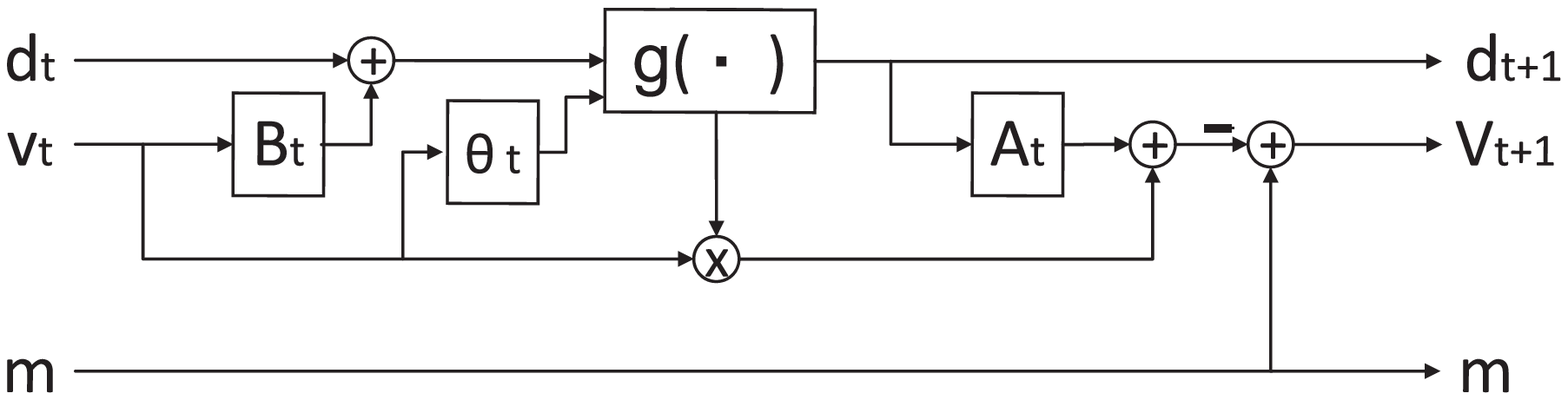}}
\caption{Various deep unfolding FNNs for sparse LIPs.}
\label{fig:subfig0} 
\end{figure}

\par
As the high computation time of traditional sparse coding methods fail to meet the requirement of real-time applications, Gregor and LeCun unfold the iterative shrinkage and thresholding algorithm (ISTA)~\cite{daubechies2004iterative}, and propose a new network for fast sparse coding, namely learned ISTA (LISTA), which is shown in Fig.~\ref{fig:subfig:ista}~\cite{gregor2010learning}. The iterative steps of the ISTA is given by
\begin{equation}\label{ISTA}
\setlength{\abovedisplayskip}{3pt}
\mathbf{v}_{t}=\mathbf{d}-\mathbf{A}\mathbf{\hat{m}}_{t},\
\mathbf{\hat{m}}_{t+1} = g_{\theta}(\mathbf{\hat{m}}_{t}+\mathbf{A}^{T}\mathbf{v}_{t}),
\setlength{\belowdisplayskip}{3pt}
\end{equation}
where $\mathbf{v}_{t}$ is the residual error, $g_{\theta}(x)=\mathrm{sign}(x)\max\{|x|-\theta,0\}$ is the element-wise soft-thresholding function and $\theta$ is the shrinkage parameter. Equation (\ref{ISTA}) can be rewritten as (\ref{iteration}) with the $\mathbf{W}$ and $\mathbf{S}$ given by
\begin{equation}\label{ISTA weights}
\setlength{\abovedisplayskip}{3pt}
\mathbf{W}=\mathbf{A}^{T},\
\mathbf{S}=\mathbf{I}-\mathbf{A}^{T}\mathbf{A}.
\setlength{\belowdisplayskip}{3pt}
\end{equation}

The LISTA adopts the element-wise soft-thresholding function in the ISTA as the activation function and limits the parameters of all layers to share the same weight as the unfolded ISTA. Different from the hand-designed parameters in the ISTA, the parameters $\mathbf{W}$, $\mathbf{S}$, and $\theta$ in the LISTA are learned from the training data. The parameters in the ISTA (\ref{ISTA weights}) can be used as a good initialization for training the LISTA. To generate the label $\mathbf{\tilde{m}_{i}}$ in the training data, Gregor and LeCun use the Coordinate Descent (CoD) algorithm to solve the $\ell_{1}$ norm minimization problem for each $\mathbf{d}_i$, which may not be the most sparse solution owing to the structure error as illustrated in Fig.\ref{fig:error}.

\par
To improve the performance of LISTA, various variants of LISTA are proposed. In \cite{8545232}, Zhang et al. propose cascade LISTA and cascade learned CoD (LCoD), which are used to reconstruct the sparse signal and predict image sparse code. In cascade LISTA and cascade LCoD, several individual LISTA and LCoD are trained in parallel to decrease the accumulated error, and when test, those networks are in series. To obtain a linear convergence, Chen et al. introduce a partial weight coupling structure into the LISTA \cite{chen2018theoretical}. While LISTA is trained for a certain $\mathbf{A}$, it lacks scalability for various models. Even a small deviation in $\mathbf{A}$ can deteriorate its performance. To this end, Aberdam et al. propose Ada-LISTA, which uses both signals and their dictionaries as inputs \cite{aberdam2020ada}. In Ada-LISTA, the input dictionaries are embedded into the network, and two auxiliary learned matrices are used to wrap the dictionary. In addition to the learned weight matrix, the deep unfolding FNNs can also be designed to only learn the step-size and threshold parameters, for example, the Analytic LISTA (ALISTA) in \cite{liu2018alista}, where the weight matrix is obtained from the analysis of corresponding optimization problem. In \cite{ablin2019learning}, Ablin et al. choose to only learn the step-size of LISTA, which is confirmed to be competitive in sufficiently sparse cases.

\par
To avoid the structure error produced in generating the training data, Wang et al. propose the deep $\ell_{0}$ encoder to solve the $\ell_{0}$ norm minimization problem directly, where the label $\mathbf{\tilde{m}_{i}}$ is the original sparse signal \cite{wang2016learning}. The deep $\ell_{0}$ encoder is obtained based on the unfolding of the iterative hard thresholding (IHT)~\cite{blumensath2009iterative} algorithm (Fig.~\ref{fig:subfig:iht}), which is similar to the ISTA except the nonlinear function. The nonlinear function in the IHT algorithm is the hard thresholding function $g_{\theta}(x)=x\cdot\mathrm{sign}(\max\{|x|-\theta,0\})$ and $\theta$ is the activation threshold. To update $\theta$, the authors decompose the original hard thresholding function $g_{\theta}(x)$ into two linear scaling operators plus a hard thresholding linear unit (HELU)
\begin{equation}\label{HELU}
\setlength{\abovedisplayskip}{3pt}
\text{HELU}_{\theta}(x)=\left\{
\begin{array}{rcl}
0 & & {|x/\theta| < 1}\\
x & & {|x/\theta| \geq 1}
\end{array} \right..
\setlength{\belowdisplayskip}{3pt}
\end{equation}
However, the HELU is a discontinuous function that destroys the universal approximation capability of the network and is hard to train. To this end, a novel continuous function HELU$_{\sigma}$ is proposed, which is given in
\begin{equation}\label{HELU_sigma}
\setlength{\abovedisplayskip}{3pt}
\text{HELU}_{\sigma}(x)=\left\{
\begin{array}{rcl}
0 & & {|x|\leq 1-\sigma}\\
\frac{x-1+\sigma}{\sigma} & & {1-\sigma < x < 1}\\
\frac{x+1-\sigma}{\sigma} & & {-1 <  x < \sigma -1}\\
x & & {|x| \geq 1}
\end{array} \right..
\setlength{\belowdisplayskip}{3pt}
\end{equation}
Obviously, HELU$_{\sigma}$ is equivalent to the HELU in (\ref{HELU}) when $\sigma \rightarrow 0$. At the beginning of the training, $\sigma$ can be set as a small constant and then gradually decreased during the training phase. Besides, for the case with a known sparse level $k$, the HELU layer can be replaced by a max-$k$ pooling layer and a max-$k$ unpooling layer. Similar to the LISTA, the weights of the deep $\ell_{0}$ encoder are learned and shared among layers.

\par
Based on the approximate message passing (AMP) algorithm~\cite{donoho2009message}, a network that adopts the independent weights among layers is proposed by Borgerding and Schniter~\cite{7905837}. Compared with the ISTA (\ref{ISTA}), the residual error of the AMP algorithm depends on the $t$-th iterative and the $(t-1)$-th iterative. The $t$-th iterative of the AMP algorithm is given by
\begin{equation}\label{AMP}
\setlength{\abovedisplayskip}{3pt}
\mathbf{v}_{t}=\mathbf{d}-\mathbf{A}\mathbf{\hat{m}}_{t}+b_{t}\mathbf{v}_{t-1},\
\mathbf{\hat{m}}_{t+1} = g_{\theta}(\mathbf{\hat{m}}_{t}+\mathbf{A}^{T}\mathbf{v}_{t}),
\setlength{\belowdisplayskip}{3pt}
\end{equation}
where $b_{t}=\frac{1}{M}\|\mathbf{\hat{m}}_{t}\|_{0}$, $\theta_{t}=\frac{\alpha}{M}\|\mathbf{\hat{v}}_{t}\|_{2}$ and $\alpha$ is a tuning parameter.
The difference of the LISTA and the learned AMP (LAMP) can be found in Fig.~\ref{fig:subfig:ista2} and Fig.~\ref{fig:subfig:amp}. In~\cite{7934066}, Borgerding et al. further extend the vector AMP (VAMP) algorithm~\cite{8006797} into the learned VAMP (LVAMP) network. Compared with the LAMP network, the LVAMP network offers increased robustness to deviations of the matrix $\mathbf{A}$ from i.i.d. Gaussian.

\par
The deep unfolding method can also be used in low-rank models. In \cite{9054280}, Pu et al. design a specific deep unfolding network based on the alternating direction method of multipliers (ADMM) for sparse and low-rank matrices. In particular, to make the network differentiable and learnable, they use a special non-linear activation function $f(x)=\text{ReLU}(x-\theta)-\text{ReLU}(-x-\theta)$ to replace the shrinkage operator in ADMM, and use the online RPCA for the low-rank term.

\par
In addition to get inspiration from the unfolding the iterative algorithm which follows (\ref{iteration}), the NN can be combined with traditional algorithms in other forms. By using a NN to perform each step of the traditional K-SVD algorithm, Scetbon et al. unfold the K-SVD into an end-to-end deep architecture and train it in a supervised manner \cite{scetbon2019deep}. The proposed scheme boosts the performance of the famous K-SVD denoising algorithm. By embedding the minimum mean squared error (MMSE) estimator into the NN, Ito et al. propose the trainable iterative soft thresholding algorithm (TISTA) \cite{8695874}, where the MMSE estimator is used as a shrinkage function to improves the speed of convergence. Similar to TISTA, Yao et al. combine the Stein’s unbiased risk estimate into the ISTA (SURE-TISTA) based network \cite{8683182}. Both TISTA and SURE-TISTA use fewer learnable variables while achieving a performance close to LAMP.

\par
DDL is another type of structured FNN that combines the knowledge of traditional algorithms. It can be used in inverse problems such as image SR and image reconstruction \cite{8554065,singhal2020reconstructing,singhal2019a,8461651} and image SR \cite{8461651}. While solving the inverse problems in imaging with DDL, Lewis D. et al. reform the entire inversion process with the variable splitting augmented Lagrangian approach, then segregate it into several subproblems, and solve all the variables jointly \cite{8554065}. To reconstruct the multi-echo MRI with DDL, Singhal and Majumdar propose two variants of DDL, including the joint-sparse dictionary learning based DDL and low-rank based DDL \cite{singhal2020reconstructing}. In \cite{singhal2019a}, they introduce the coupled dictionary learning technique into DDL, and propose a domain adaptation approach for different imaging tasks. For image SR, Huang and Dragotti design an $L$-layer FNN which includes $L-1$ analysis dictionaries and one synthesis dictionary \cite{8461651}. The analysis dictionaries are used for feature extraction, and are learned in an unsupervised manner with the geometric analysis operator learning method. The synthesis dictionary is designed for image SR, and is learned in a supervised manner via an approach which is similar to ADMM.

\subsection{CNNs}

The CNN has effectively reduced the number of parameters by replacing the fully connected layers with the convolutional layers. CNN inspired by the biological visual cortex can capture the local similarity of images and thus is employed as a key technique in most image-related applications. Various CNNs for LIPs are summarized in Table \ref{cnn in LIP}.

\begin{table*}[]
\centering  
\setlength{\abovecaptionskip}{0pt}%
\setlength{\belowcaptionskip}{6pt}%
\caption{CNNs for LIPs.}\label{cnn in LIP}
\begin{adjustwidth}{-1cm}{-1cm}
\tiny
\resizebox{1.2\textwidth}{8cm}{
\begin{tabular}{|p{0.5cm}|p{1cm}|p{1.5cm}|p{1.5cm}|p{2cm}|p{2cm}|p{2.5cm}|p{2cm}|}
\hline
Ref. &Application &Input &Output &Loss Function &Initialization &Learning Rate &Optimizer   \\ \hline\hline


\cite{7341021} &Image Denoising &Cropped noisy image &Denoised image &Mean squared error  &Gaussian distribution & $10^{-3}$ for the final layer and $10^{-1}$ for other layers &SGD  \\ \hline

\cite{8365806} &Image Denoising &Sub-images &Denoised sub-images &Mean squared error &Orthogonal initialization & $10^{-3}$ to $10^{-4}$ (reduced when the training error stops decreasing)  & ADAM  \\ \hline

\cite{7472127} &Depth Image Denoising &Pre-processing grey image and Depth image &Denoised images &Weighted Euclidean-based distance function &Not given &Not given  &Not given \\ \hline

\cite{8435923} &Hyperspectral Image Denoising  &Noisy image &Denoised image &Mean squared error &Follows \cite{7780459} &$10^{-3}$ to $10^{-5}$ &ADAM \\ \hline

\cite{8454887} &Hyperspectral Image Denoising &Noisy image &Denoised image &The mean squared error &Not given &$10^{-2}$  &ADAM  \\ \hline

\cite{8944535} &Medical Image Denoising &Noisy image &Denoised image  &Perceptual losses and squared Euclidean distance  &Follows \cite{7780459}  & $10^{-1}$ (decreased by ten times after every ten epochs)  &ADAM  \\ \hline

\cite{8923780} &Medical Image Denoising &Noisy image &Denoised image &Mean squared error &Gaussian distribution & $10^{-2}$ to $10^{-4}$ (decay exponentially every 50 epochs) &ADAM  \\ \hline

\cite{7839189} &Image Denoising &Noisy image &Residual image &Averaged mean squared error &Follows \cite{7410480} & $10^{-1}$ to $10^{-4}$ (decay exponentially every 50 epochs) &SGD(momentum 0.9, weight decay $10^{-4}$) \\ \hline

\cite{8372095} &Image Denoising &Noisy image &Residual image &The L2 loss function &Follows \cite{7410480} & $10^{-3}$ to $10^{-4}$ (reduced after 30 out of 40 epochs) &SGD(momentum 0.9) \\ \hline

\cite{8741329} &Image Denoising  &Noisy image &Residual image
&Huber loss \cite{huber1992robust} &Not given &$5\times10^{-4}$ to $5\times10^{-5}$ &ADAM  \\ \hline

\cite{7115171} &Image SR &Interpolated LR subimage  &HR subimage &Mean squared error &Gaussian distribution with zero mean and standard deviation $10^{-3}$ & $10^{-4}$ for the first two layers, and $10^{-5}$ for the last layers.  &SGD  \\ \hline

\cite{7780576} &Image and Video SR  &LR subimage &HR subimage &Mean squared error &Not given &$10^{-2}$ to $10^{-4}$ (reduced when the change of training error is smaller than a threshold) &Not given  \\ \hline

\cite{dong2016accelerating} &Image SR &LR subimage  &HR subimage &Mean squared error &Not given & $10^{-3}$ for the convolution layers, and $10^{-4}$ for the deconvolution layer.  &Not given \\ \hline

\cite{8100101} &Image SR &LR subimage  &HR subimage &Charbonnier penalty function  &Follows \cite{7410480} & $10^{-5}$ &ADAM (momentum 0.9, weight decay $10^{-4}$) \\ \hline

\cite{8578277} &Image SR &LR image &HR image &Mean squared error &Follows \cite{7410480} &$10^{-4}$ to $10^{-6}$ (decreased by a factor of 10 for every $5\times10^{5}$ iterations.) &ADAM (momentum 0.9, weight decay $10^{-4}$)   \\ \hline


\cite{7780551}  &Image SR &LR image &HR image &Mean squared error &Not given  &$10^{-1}$ &Not given   \\ \hline

\cite{8014885}  &Image SR &LR image patch &HR image patch &The L1 loss &Pretrained network &$10^{-4}$ (halved at every
$2\times10^{-5}$ minibatch updates.) &ADAM   \\ \hline

\cite{8099781}  &Image SR &LR image patch &HR image patch &Mean squared error &Follows \cite{7410480} &$10^{-4}$ to $10^{-1}$ (decreased to half every 10 epochs.) &The adjustable gradient clipping \cite{7780551}  \\ \hline

\cite{8578442} &Image SR &LR image patch &HR image patch &Mean squared error &Not given &$10^{-3}$ to $10^{-5}$ &ADAM    \\ \hline

\cite{8578427} &Image SR &LR image &HR image &The L1 loss &Not given &$10^{-3}$ to $10^{-6}$ &ADAM   \\ \hline

\cite{sun2016deep} &CS MRI &Sampling data in k-space &Reconstructed MR image &Normalized mean square error  &Based on ADMM &Not given  &L-BFGS  \\ \hline

\cite{8578294} &Image CS &CS measurement &Image block &Self designed &Based on the linear mapping matrix &$10^{-4}$ &ADAM  \\ \hline

\cite{dong2014learning} &Image SR &Interpolated LR subimage &HR image &Mean squared error &Gaussian distribution with zero mean and standard deviation 0.001 & $10^{-4}$ for the first two layers, and $10^{-5}$ for the last layers &SGD  \\ \hline

\cite{7410407} &Image SR &Interpolated LR subimage &HR image &Mean squared error &Harr-like gradient filters and uniform weights &Not given &SGD  \\ \hline

\cite{8100106}  &Image Denoising &Noisy image &Clean image &Corresponds to the negative peak signal-to-noise-ratio (PSNR) &Learned parameters following a greedy-training strategy  &Not given &L-BFGS \\ \hline

%



\end{tabular}}
\end{adjustwidth}
\end{table*}

\subsubsection{Common CNNs}

\begin{figure}
\centering
\vspace{-0.8cm}
\setlength{\abovecaptionskip}{-0.05cm}
\setlength{\belowcaptionskip}{-0.5cm}
\subfigure[The network structure of the DE-CNN.]{
\label{fig:subfig:decnn}
\includegraphics[width=0.9\textwidth]{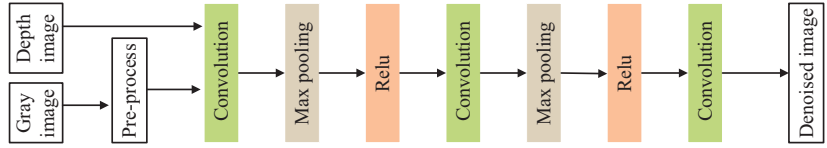}}
\hspace{1in}
\subfigure[The network structure of the FFDNet.]{
\label{fig:subfig:ffdnet}
\includegraphics[width=0.9\textwidth]{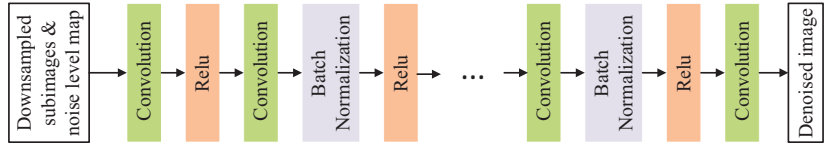}}
\hspace{1in}
\subfigure[The network structure of the DnCNN.]{
\label{fig:subfig:dncnn}
\includegraphics[width=0.9\textwidth]{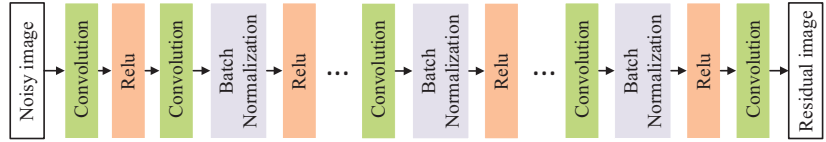}}
\caption{Common CNNs for image denoising.}
\label{fig:subfig1} 
\end{figure}

For image denoising, FNNs introduced in the previous subsection require a predetermined input image size, while CNNs are more flexible for dealing with images with arbitrary sizes. In \cite{7341021}, Wang et al. propose a two-layer CNN, where they use the Relu activation function for the first layer and the sigmoid activation function for the second layer. Besides, under the inspiration of lateral inhibition in real neurons and computational neuroscience models, a novel local response normalization is employed after the output of ReLU, which leads to the local competition and enhances the feature of gray images. For real-world images where the noise is signal-dependent, non-Gaussian and spatially variant, a fast and flexible denoising CNN (FFDNet) is proposed \cite{8365806}. To handle the varying noise level in the FFDNet, the noise level $\sigma$ is first extended to a noise level map. The noise level map is then concatenated with the down sampled sub-images to form a tensor that is used as the input of the network (Fig.~\ref{fig:subfig:ffdnet}).

\par
Various CNNs are designed for different denoising applications. Zhang et al. extend the CNN to depth image denoising and propose a denoising and enhancement CNN (DE-CNN) \cite{7472127}. In the DE-CNN, the input of the network contains both the depth image and pre-processed gray image, as shown in Fig.~\ref{fig:subfig:decnn}. The authors also propose a novel edge based weighted loss function and a data augmentation strategy that expands useful depth images. For hyperspectral image denoising, Chang et al. use the CNN to extract the spectral and the spatial information, where spectral correlation is depicted by the multiple channels \cite{8435923}. In \cite{8454887}, Yuan et al. use the spatial and spectral information as input. They capture and fuse multiscale spatial-spectral feature for the final restoration. For medical image denoising, Panda et al. propose a wide residual CNNs for medical image denoising \cite{8944535}. In order to solve the problem that the use of squared Euclidean distance will lead to over-smoothed image, they combine the perceptual loss and squared Euclidean distance for training, which is confirmed to be helpful in keeping structural or anatomical details. Wang et al. design a local receptive field smoothing network which remains the smoothing properties of the receptive field by weighting their local neighborhoods \cite{8923780}.

\par
Instead of expecting the clean image as network output, Zhang et al. propose a denoising CNN (DnCNN) that outputs the residual between a clean image and a noisy image \cite{7839189}. By using residual learning, the network is able to handle unknown noise levels and can be also transferred to other tasks such as single image SR and image deblocking. Wang et al. further combine the dilated convolution \cite{yu2015multi} with residual learning to improve computational efficiency and enlarge the receptive field \cite{8372095}. In \cite{8741329}, Su et al. propose a deep multi-scale cross-path concatenation residual network (MC$^{2}$RNet) for Poisson denoising, where they use cross-path concatenation and the skip connection to obtain multi-scale context representations of images.

%

\begin{figure}
\centering
\vspace{-0.8cm}
\setlength{\abovecaptionskip}{-0.05cm}
\setlength{\belowcaptionskip}{-0.5cm}
\subfigure[The structure of the SRCNN.]{
\label{fig:subfig:srcnn}
\includegraphics[width=0.6\textwidth]{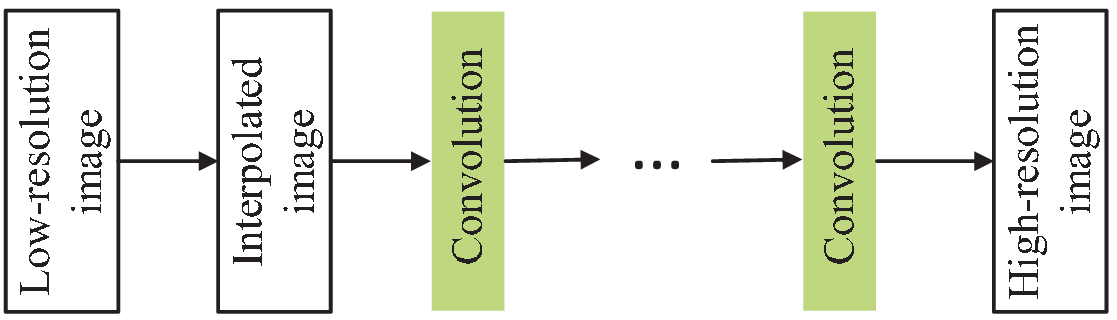}}
\hspace{1in}
\subfigure[The structure of the ESPCN.]{
\label{fig:subfig:espcn}
\includegraphics[width=0.6\textwidth]{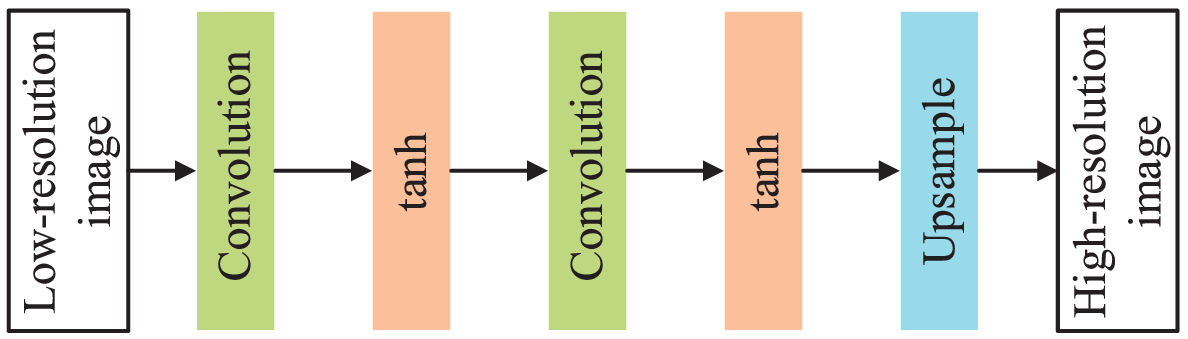}}
\hspace{1in}
\subfigure[The structure of the FSRCNN.]{
\label{fig:subfig:fsrcnn}
\includegraphics[width=0.6\textwidth]{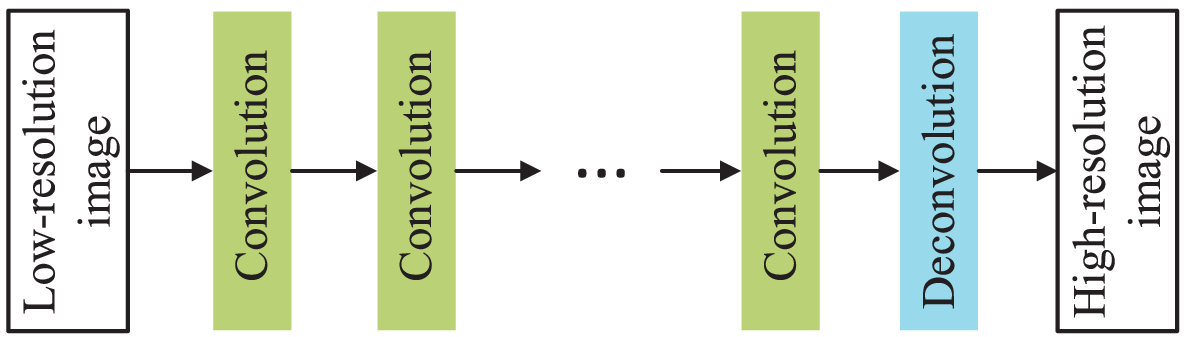}}
\hspace{1in}
\subfigure[The structure of the LapSRN.]{
\label{fig:subfig:lapsan}
\includegraphics[width=0.6\textwidth]{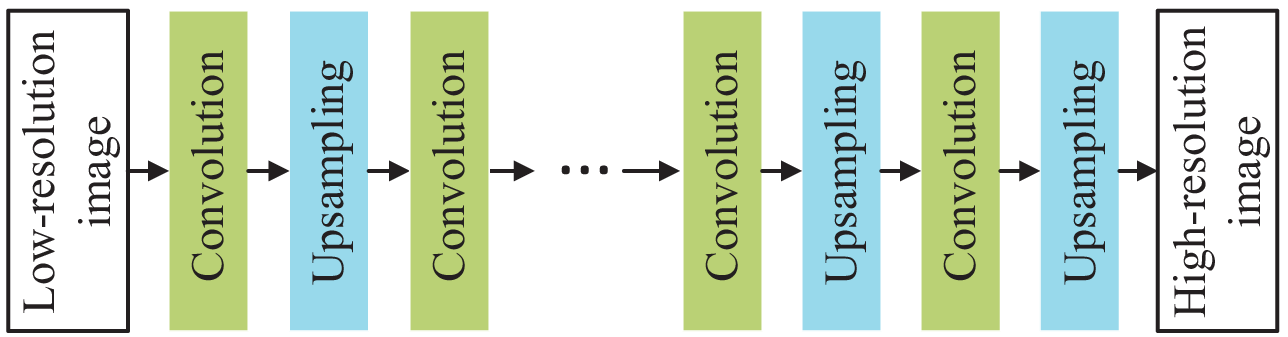}}
\hspace{1in}
\subfigure[The structure of the DBPN.]{
\label{fig:subfig:dbpn}
\includegraphics[width=0.6\textwidth]{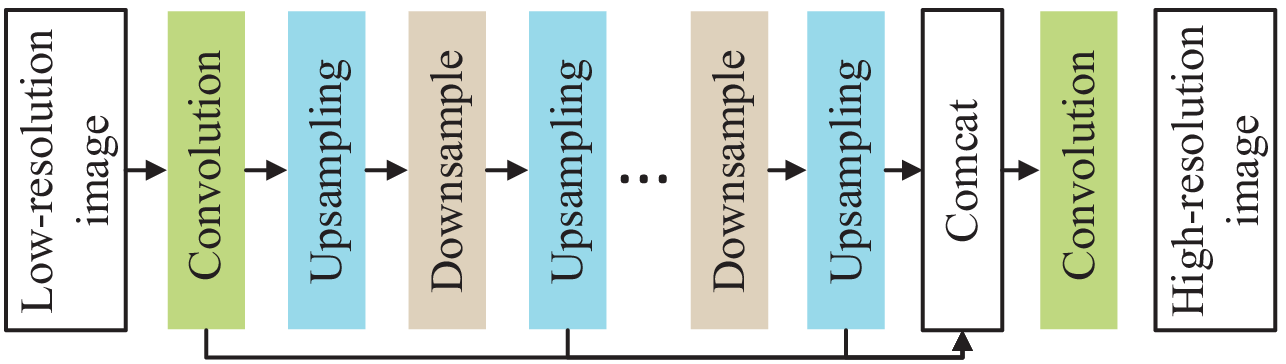}}
\caption{Common CNNs for image SR.}
\label{fig:subfig2} 
\end{figure}

\begin{table}[]
\centering  
\setlength{\abovecaptionskip}{4pt}%
\setlength{\belowcaptionskip}{4pt}%
\caption{The SR results: average PSNR/SSIM for scale factors 2× and 4×.}\label{cnn for SR}
\footnotesize
\begin{tabular}{|c|c|c|c|c|c|}  
\hline
Algorithm  &Scale   &PSNR(Set5)  &SSIM(Set5)  &PSNR(Set14)  &SSIM(Set14)  \\ \hline
Bicubic &$2$ &$33.66$ &$0.93$ &$30.32$ &$0.87$\\
SRCNN &$2$ &$36.65$ &$0.95$ &$32.42$ &$0.91$\\
FSRCNN &$2$ &$37.00$ &$0.96$ &$32.65$ &$0.91$\\
ESPCN &$2$ &$37.26$ &$0.95$ &$32.88$ &$0.91$\\
LapSRN &$2$ &$37.52$ &$0.96$ &$33.08$ &$0.91$\\
D-DBPN  &$2$ &$\textbf{38.09}$ &$\textbf{0.96}$ &$\textbf{33.85}$ &$\textbf{0.92}$\\  \hline
Bicubic &$4$ &$28.42$ &$0.81$ &$26.00$ &$0.70$\\
SRCNN &$4$ &$30.49$ &$0.86$ &$27.50$ &$0.75$ \\
FSRCNN &$4$ &$30.72$ &$0.86$ &$27.62$ &$0.76$\\
ESPCN &$4$ &$30.90$ &$0.86$ &$27.73$ &$0.76$ \\
LapSRN  &$4$ &$31.54$ &$0.88$ &$28.19$ &$0.77$\\
D-DBPN  &$4$ &$\textbf{32.47}$ &$\textbf{0.89}$ &$\textbf{28.82}$ &$\textbf{0.78}$\\
  \hline
\end{tabular}
\end{table}

\par
Different with image denoising, in image SR the dimension of the output is higher than the input. To explore the information in different dimension space, various network architectures are designed. In super-resolution convolutional neural network (SRCNN) \cite{7115171}, the input of the network is an interpolated LR image (Fig.~\ref{fig:subfig:srcnn}). The SRCNN uses a relatively large filter size to utilize the information from more pixels and simultaneously processes multiple channels, which leads to superior performance in comparison to traditional example-based approaches. Considering that the SRCNN is sub-optimal and computationally inefficient owing to the use of the interpolated image as input, more efficient networks such as efficient sub-pixel CNN (ESPCN) \cite{7780576} and fast SRCNN (FSRCNN) are proposed \cite{dong2016accelerating}. Both ESPCN and FSRCNN use LR image as input and perform the upsampling in the last layer. The last layer of ESPCN is a sub-pixel convolution layer, which firstly generates multiple feature maps and then conducts a periodic shuffling to the pixels to produce the final HR image. The last layer of FSRCNN is a deconvolution layer, and FSRCNN uses smaller filter sizes and specially designed shrinking layer to accelerate the network. While the ESPCN and FSRCNN get the HR image in the last layer, Lai et al. propose Laplacian pyramid SR network (LapSRN) which progressively increases the dimension of the output of each layer (Fig.~\ref{fig:subfig:lapsan}) \cite{8100101}. The deep back-projection network (DBPN) which uses the iterative up- and down-sampling layers to explore the mutual dependencies of LR images and HR images, as shown in Fig.~\ref{fig:subfig:dbpn} \cite{8578277}. Each pair of sampling layers represents a type of the degradation and corresponding components. Furthermore, Haris et al. propose the dense DBPN (D-DBPN), which adds skip connections to allow the concatenation of features between layers. It is observed the dense DBPN can further improve the performance of the SR, especially in large scaling factors. In Table \ref{cnn for SR} and Table \ref{num}, we compare the performance of different CNNs for image SR in datasets Set5 \cite{bevilacqua2012low} and Set14 \cite{zeyde2010single}, and compare the different CNNs for image SR. Compared with SRCNN, FSRCNN is deeper, but uses less filters and smaller filter sizes. Thus, the FSRCNN has fewer parameters and is faster ($41.3\times$) without performance degradation. ESPCN uses the same filter sizes as SRCNN, but decreases the number of filters and extracts the features in the LR space to reduces the computational complexity and obtain the real-time speed. Compared with the previous networks, LapSRN is much deeper (27 layers) and uses the residual learning to assist the training. Charbonnier loss function used in LapSRN has a higher gradient magnitude than the $\ell_{2}$ loss and decreases the ringing artifacts. For D-DBPN, the network has a depth up to 48 layers and uses smaller filter sizes than the SRCNN, FSRCNN and LapSRN. Even with a shallow depth (18 layers), the DBPN outperforms the LapSRN (31.54 dB) with 0.05 dB.

\begin{table}[]
\centering  
\setlength{\abovecaptionskip}{0pt}%
\setlength{\belowcaptionskip}{4pt}%
\caption{Comparisons among various CNNs for SR.}\label{num}
\footnotesize
\begin{tabular}{|p{1.2cm}|p{1.5cm}|p{3.5cm}|p{2cm}|p{4cm}|}  
\hline
network  &Parameters  &Training data  &Loss function &Network\\ \hline\hline
SRCNN  &$57k$  &ImageNet subset (over $5$ million sub-images)  &Mean squared error &Conv(9,64,1)-Conv(5,32,64)-Conv(5,1,32)  \\  \hline
FSRCNN  &$12k$  &General-100 dataset and 91-image dataset (19 times more images after data augmentation) &Mean squared error &Conv(5,56,1)-Conv(1,12,56)-4Conv(3,12,12)-Conv(1,56,12)-Conv(9,1,56) \\  \hline
ESPCN  &$20k$  &91-image dataset and ImageNet subset  &Mean squared error &Conv(5,64,1)-Conv(3,32,64)-Conv(3,1,32) \\  \hline
LapSRN  &$812k$  &Berkeley segmentation dataset and 91-image dataset   &Charbonnier penalty function &Conv(3,64,3)-2(10Conv(3,64,64)-Conv(3,256,64)-Conv(3,3,64)-Conv(3,12,3))\\  \hline
DBPN   &$10M$  &DIV2K and Flickr and ImageNet subset  &Mean squared error &Conv(256,3,3)-Conv(32,1,1)-7(Conv(32,2,2)-Conv(32,6,6)-Conv(32,2,2)-Conv(32,6,6)-Conv(32,2,2)-Conv(32,6,6))
-Conv(32,2,2)-Conv(32,6,6)-Conv(32,2,2)-Conv(3,3,3)
 \\  \hline
\end{tabular}
\end{table}

\par
In addition to the different methods for tackling high dimension data sets, we also explore the design of CNNs for SR. The simplest method is to find inspiration from famous networks that have obtained success in other tasks. For example, the very deep convolutional networks (VDSR) \cite{7780551} is inspired by VGG-net. In VDSR, Kim et al. boost the performance by directly increasing the network depth, which usually leads to training difficulties. Thus, they propose to increase the learning rate and learn residuals only to prevent training difficulties. Similar to VDSR, Lim et al. design their enhanced deep super-resolution network (EDSR) according to famous network ResNet \cite{8014885}. Instead of increasing the network depth, they cut down the unnecessary batch normalization layers in residual blocks, which also improves the performance. Other interesting networks contain the deep recursive residual network (DRRN) \cite{8099781}, and residual dense network (RDN) \cite{8578360}. The multi-scale deep super-resolution system (MDSR) consider to solve the SR with different upscaling factors in a single model \cite{8014885}. They use several parallel layers on the front and back of the network, and several shared layers in the middle of the network. In \cite{8578442}, Zhang et al. consider the problem that the LR images may contain multiple degradations. They propose a dimensionality stretching strategy to enable the blur kernel and noise level as input. For the mismatch problem between the training data and real SR situations, Shocher et al. design a zero-shot SR (ZSSR), which relies on the input image itself to train the network \cite{8578427}.

\par
While it is generally assumed that the success of previous networks relies on a large amount of training data, Ulyanov et al. show the contrary conclusion that the structure of the network is natural to capture the image statistics prior with a deep image prior (DIP) \cite{8579082}. They apply the DIP in single image SR by using the structure of a randomly-initialized CNN as image prior to upsample an image without learning. Sidorov and Hardeberg further apply the DIP to hyperspectral imaging and 3D-convolutional networks \cite{9022040}. Besides, the DIP is also popular in medical image reconstruction \cite{8581448,9060001}, where a large amount of training pairs is not always feasible. In \cite{van2018compressed}, Van Veen et al. propose the DIP for CS (CS-DIP). To overcome the overfitting of DIP, they propose to regularize the weights of the network during the optimization process. In \cite{8968714}, Ren et al. further use the DIP for CS problems where the sparse dictionary is uncertain, and the proposed computational intelligent CS algorithm is used for soil PH measurement. DIP can also be combined with other techniques, for example, the combination of DIP and traditional total variation regularization for image denoising \cite{8682856} and high dynamic range imaging \cite{9054218}. There is also some progress on the theoretical aspects of DIP. For the linear CS problem and nonlinear compressive phase retrieval problem, Jagatap and Hegde prove that compared with the hand-designed priors, the DIP can achieve better compression rates under the same image quality \cite{jagatap2019algorithmic}. Different mathematical interpretations of DIP are shown by Dittmer et al. in \cite{dittmer2019regularization}, Dittmer et al. introduce the idea of viewing the DIP as the optimization of Tikhonov functionals. In \cite{heckel2020compressive}, Heckel and Soltanolkotabi show the self-regularizing property of DIP and prove that sufficiently structured signals and images can be approximately reconstructed by the untrained CNN.

\subsubsection{Structured CNNs}

\begin{figure}
\centering
\vspace{-0.8cm}
\setlength{\abovecaptionskip}{-0.05cm}
\setlength{\belowcaptionskip}{-0.5cm}
\subfigure[The network structure of the ADMM-Net.]{
\label{fig:subfig:admm}
\includegraphics[width=0.9\textwidth]{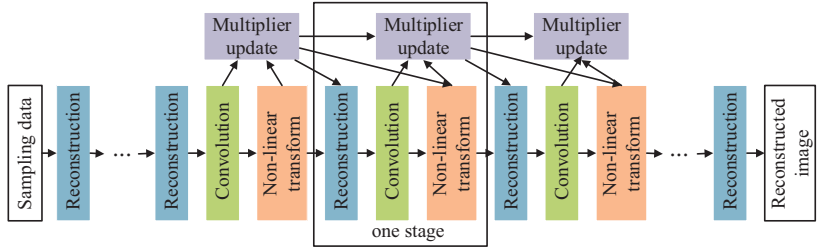}}
\hspace{1in}
\subfigure[The network structure of the ISTA-Net.]{
\label{fig:subfig:istanet}
\includegraphics[width=0.9\textwidth]{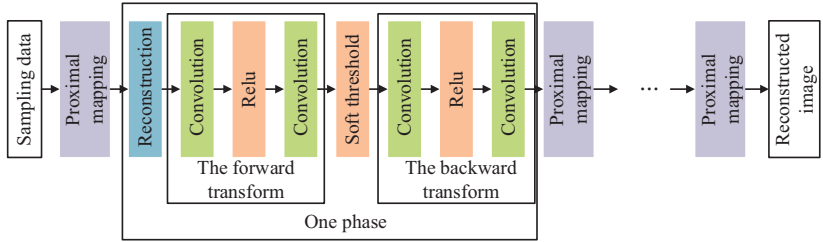}}
\caption{Deep unfolding CNNs for image CS.}
\label{fig:subfig1} 
\end{figure}

Akin to the structured FNNs, structures in traditional algorithms can also be employed in the design of structured CNNs.

\par
One example of the deep unfolding networks is the application of image reconstruction. Following in the iterative procedure of the ADMM algorithm, Yang et al. construct a CNN based ADMM-Net for CS-MRI, where each layer represents a subproblem in the ADMM optimization problem (Fig.~\ref{fig:subfig:admm}) \cite{sun2016deep}. Especially, in the ADMM-Net, all the parameters are learned, including the transforms, penalty parameters and shrinkage functions. Furthermore, in \cite{8550778}, they redesign the ADMM algorithm and unfold it to the more powerful ADMM-CSNet. Another deep unfolding CNN is the ISTA-Net, which is also designed for CS imaging. Similar to the ADMM-Net, the parameters in the ISTA-Net are all learned \cite{8578294}. The ISTA-Net contains several phases, each of which represents an iteration of the ISTA (Fig.~\ref{fig:subfig:istanet}). Each phase of the ISTA-Net includes a forward transform and a symmetric backward transform, where the forward transform is used to replace the hand-crafted sparse transform of the original image in the ISTA, and the backward transform is designed to exhibit a structure symmetric to that of the forward transform. The AMP algorithm can also be used for image denoising, which leads to the denoising AMP (D-AMP) algorithm \cite{7457256}. By unfolding the D-AMP algorithms, Metzler et al. design their learned D-AMP (LDAP) \cite{metzler2017learned}, which can be used to recovery image from different measurement matrices. In LDAP, DnCNN is embedded into the network as a denoiser. Following the deep unfolding principle, Solomon et al. unfold the low-rank plus sparse ISTA to solve the RPCA problem \cite{8836615} more efficiently. Instead of using a fully-connected layer for matrix multiplications, they use the convolutional layers to reduce the number of parameters. The proposed convolutional robust principal component analysis (CORONA) is further used in SR ultrasound to remove the clutter signal.

\par
\begin{figure}
\centering
\vspace{-0.8cm}
\setlength{\abovecaptionskip}{-0.05cm}
\setlength{\belowcaptionskip}{-0.5cm}
\subfigure[A sparse coding based CNN.]{
\label{fig:subfig:sccnn}
\includegraphics[width=0.6\textwidth]{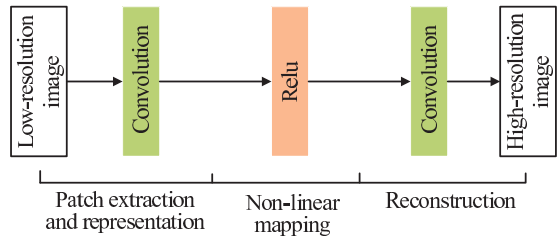}}
\hspace{1in}
\subfigure[The structure of the SCN.]{
\label{fig:subfig:scn}
\includegraphics[width=0.6\textwidth]{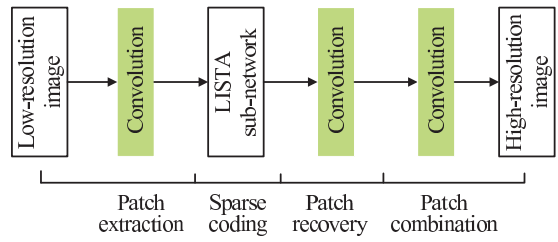}}
\hspace{1in}
\subfigure[The structure of the CSCN.]{
\label{fig:subfig:cscn}
\includegraphics[width=0.6\textwidth]{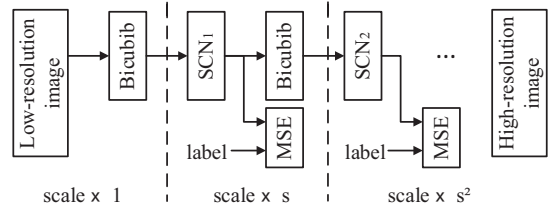}}
\caption{Structured CNNs for image SR.}
\label{fig:subfig2} 
\end{figure}

Another example is the application of image SR. Most related work derives the network with the consideration of sparse coding methods \cite{4587647, 5466111}. Dong et al. use linear transforms to project image patches onto a dictionary and replace the sparse coding solver with a nonlinear transform (Fig.~\ref{fig:subfig:sccnn}) \cite{dong2014learning}. Liu et al. propose the sparse coding based network (SCN) (Fig.~\ref{fig:subfig:scn}), which consists of a patch extraction layer, a LISTA sub-network for sparse coding, an HR patch recovery layer, and a patch combination layer \cite{7466062}. In the SCN, the LISTA sub-network is employed to enforce the sparsity of the representation. In addition, the authors propose a cascade of SCNs (CSCNs) (Fig.~\ref{fig:subfig:cscn}) so that the network can be extended to deal with different scaling factors. In the practical scene where the LR images suffer from various types of corruption, Liu et al. fine-tune the learned SCN with a small amount of training data to adapt the model to the new scenario \cite{7466062}.

\par
Structured CNNs are also proposed for image denoising \cite{8578436} and image restoration \cite{7527621}. For example, to exploit the native non-local self-similarity property of natural images, Lefkimmiatis proposes a CNN based network that uses an extra regularization term in the loss function \cite{8578436}. The key idea is unfolding the proximal gradient method to construct a network graph, where each layer represents one proximal gradient iteration. In \cite{7527621}, Chen and Pock construct the trainable nonlinear reaction diffusion (TNRD) network based on the nonlinear reaction diffusion models for image restoration, which can be thought as a forward convolutional network. Besides, they add a reaction term to adapt to various image processing problems.

\par
Multimodal DL \cite{ngiam2011multimodal} is another promising technique in solving image SR problems and drives plenty of structured CNNs. In multimodal DL for image SR, the input of the network is generally including a LR image and a HR image in a different modality. For example, Marivani et al. use LR near-infrared images and HR RGB images to super-resolve the HR near-infrared images \cite{8803313,8903106}. In \cite{8803313}, they design their learned multimodal convolutional sparse coding (LMCSC) model by unfolding the proximal method that used for solving the convolutional sparse coding with side information. In \cite{8903106}, they turn to solve the appropriate $\ell_{1}-\ell_{1}$ minimization problem for multimodal image SR and design their deep multimodal sparse coding network (DMSC) based on a deep unfolding FNN named learned side-information-driven iterative soft thresholding algorithm (LeSITA). To capture the cross-modality dependency, Deng and Dragotti design a special joint multi-modal dictionary learning (JMDL) algorithm, and unfolding it into a deep coupled ISTA network \cite{8858035}. In particular, they use a layer-wise optimization algorithm (LOA) to solve the multi-layer dictionary learning problem for initialization. In addition to image SR, multimodal DL can also be used in image reconstruction \cite{falvo2019multimodal,8844082}.

\subsection{Recurrent NNs}

Compared with FNNs and CNNs, RNNs are more appropriate in dealing with sequential inputs, such as the time-varying signal \cite{7979523}. Thus, the RNN can be used to solve a sequence of correlated LIPs. Various RNNs for LIPs are summarized in Table \ref{RNN in LIP}.

\begin{table*}[]
\centering  
\setlength{\abovecaptionskip}{0pt}%
\setlength{\belowcaptionskip}{6pt}%
\caption{Details of training some RNNs for LIPs.}\label{RNN in LIP}
\begin{adjustwidth}{-1cm}{-1cm}
\tiny
\begin{tabular}{|p{0.5cm}|p{1.1cm}|p{1.5cm}|p{1.5cm}|p{2cm}|p{1cm}|p{2cm}|p{1.5cm}|}
\hline
Ref. &Application &Input &Output &Loss Function &Initialization &Learning Rate &Optimizer   \\ \hline\hline

\cite{7457684} &MMV problems &The observed signal &The recovered signal &The quadratic error &Small random numbers &Not given  &Not given \\ \hline

\cite{7905830} &MMV problems &The observed signal &The recovered signal &Cross entropy &Small random numbers &Not given   &Backpropagation through time and ADAM \\ \hline

\cite{lyu2019block} &Block-sparsity recovery &The sequence of residual vectors &One-hot vectors &Cross entropy &Not given &$3\times10^{-4}$  &Nesterov accelerated gradient descent \\ \hline

\cite{8296560} &Video SR  &Multiple LR frames &Multiple HR frames &Mean squared error &Not given &Not given  &Root Mean Square Propagation (RMSProp) \cite{hinton2012neural} \\ \hline

\cite{8282261}&Video SR  &Multiple LR frames &Multiple HR frames &The L1 loss &Not given &$10^{-4}$   &ADAM \\ \hline

\cite{7919264} &Video SR  &Multiple LR frames &Multiple HR frames &Mean squared error &Gaussian distribution  &$10^{-4}$   &SGD \\ \hline

\cite{8501928} &Video SR  &Multiple LR frames &Multiple HR frames &Mean squared error &Gaussian distribution &$10^{-4}$ to $10^{-5}$   &RMSProp \\ \hline

\cite{8953613} &Video SR  &Multiple LR frames &Multiple HR frames &The L1 loss  &Follows \cite{7410480} &$10^{-4}$ (decreased by a factor of 10 for half of total 150 epochs)   &ADAM  \\ \hline

\cite{7780550} &Image SR  &LR image patch &HR image patch &Mean squared error  &Follows \cite{7410480} &$10^{-2}$ to $10^{-6}$ (decreased by a factor of 10 if the validation error does not decrease for 5 epochs)   &SGD  \\ \hline

\cite{8425771} &Image SR  &LR image patch &HR image patch &Mean squared error  &Follows \cite{7410480} &$10^{-1}$ (decreased by a factor of 10 every 10 epochs)  &SGD  \\ \hline

\cite{8579237} &Video SR  &Multiple LR frames &Multiple HR frames &The L1 loss for optical flow network and different loss functions for image-reconstruction network  &Not given &$10^{-3}$ to $10^{-5}$ (Polynomial decay)   &Not given  \\ \hline

\cite{8356655} &Image SR  &LR image patch &HR image patch &Mean squared error  &Initialized to 0.5 &$10^{-3}$ for the weights in the output layer while $10^{-2}$ for other layers  &SGD  \\ \hline

\cite{8986369} &Image Denoising  &Noisy image patch &Clean image patch &Not given   &Not given &$10^{-3}$ (decreased by 0.2 every 30 epochs)  &ADAM  \\ \hline

\cite{8425639}  &MRI image reconstruction  &Undersampled image patch &Reconstructed image patch &Mean squared error  &Follows \cite{7410480} &Not given  &ADAM  \\ \hline

\cite{putzky2017recurrent} &Image restoration &Undersampled image patch &Reconstructed image patch &Weighted sum of the individual mean squared error  &Not given &Not given  &Not given  \\ \hline

\cite{7952977}  &Sparse Coding  &The observed signal &The recovered signal  &Mean squared error  &Randomly &$10^{-4}$  &RMSProp   \\ \hline

\cite{8803281} &Sequential signal reconstruction  &The observed signal &The recovered signal  &Mean squared error  &Uniform distribution &$3\times10^{-4}$ &ADAM  \\ \hline

\cite{zhou2018sc2net} &Sparse Coding  &The observed signal &The recovered signal  &The L1 loss for unsupervised SLTSM and softmax for supervised SLTSM  &Not given &Not given &Adadelta \cite{zeiler2012adadelta}  \\ \hline

\end{tabular}
\end{adjustwidth}
\end{table*}

\par
One of the examples is the sparse LIP, especially the structured sparse LIP. In \cite{xin2016maximal}, Xin et al. use an long short-term memory (LSTM) network as an adaptive variant of IHT to allow a longer flow of information to explore the structure of $\mathbf{A}$ in a general sparse LIP. In MMV where the supports of each column are not totally consistent due to the noise or partly innovative sparse pattern in the source, Palangi et al. design an LSTM to capture the unknown dependency between sparse vectors \cite{7457684}. In \cite{7905830}, they further propose a bidirectional LSTM to solve the problem, which uses multiple adjacent predictions. In addition to the MMV problem, the LSTM is used to solve the block-sparsity recovery with unknown cluster patterns in \cite{lyu2019block}.

Besides, the RNN can be used for video SR, where exists spatio-temporal  information between multiple frames. In \cite{8296560}, Li et al. propose a residual recurrent convolutional network (RRCN) for video SR, which integrates the motion compensation into the bidirectional residual convolutional network. However, Lim and Lee consider that the optical flow and motion compensation influence the overall performance, thus they discard this module and use the LSTM as a replacement \cite{8282261}. Different from vanilla RNNs, Huang et al. design a new fully convolutional RNN that uses the weight-sharing convolutional connections to decrease the parameters and uses the 3D feedforward convolutions to capture the short-term fast-varying motions \cite{7919264}. In \cite{8501928}, Li et al. use a very deep non-simultaneous fully recurrent convolutional layers to deal with the visual artifacts that came from fast-moving objects. They also use a new model ensemble strategy to combine their model and the single-image SR model. While the previous networks send the frames into networks together, the recurrent back-projection network (RBPN) proposed by Haris et al. treats them as separate sources \cite{8953613}.

\par
\begin{figure*}
\centering
\vspace{-0.8cm}
\setlength{\abovecaptionskip}{-0.05cm}
\setlength{\belowcaptionskip}{-0.5cm}
\subfigure[The structure of DRCN, which consists a embedding network, an inference network and a reconstruction network.]{
\label{fig:subfig:drcn1}
\includegraphics[width=\textwidth]{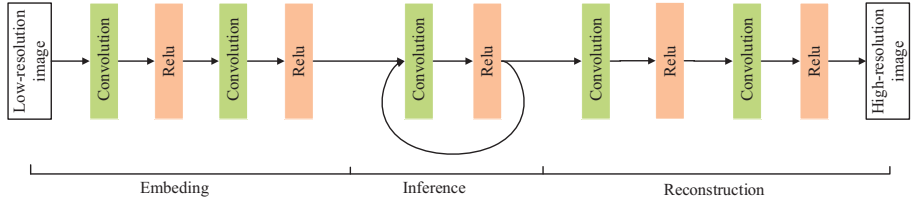}}
\hspace{1in}
\subfigure[The final model of DRCN with recursive-supervision and skip connection. The reconstruction network is shared for recursive predictions. ]{
\label{fig:subfig:drcn2}
\includegraphics[width=\textwidth]{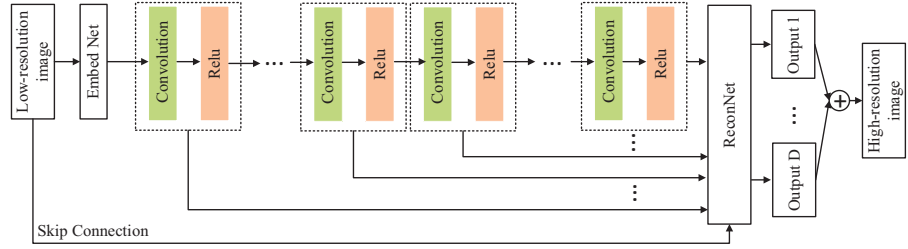}}
\caption{The network structure of deeply-recursive convolutional network.}
\label{fig:subfig3} 
\end{figure*}

The RNN can also be combined with other networks such as CNNs for single image SR and image denoising. For example, Kim et al. incorporate the RNN into a basic CNN for image SR \cite{7780550}. Compared with the SRCNN, the proposed deeply-recursive convolutional network (DRCN) (Fig.~\ref{fig:subfig:drcn1}) has fewer parameters and larger receptive field, which leads to improved quality of image details. However, due to the issue of the exploding/vanishing gradient, this network encounters difficulty in the training with the SGD algorithm. To tackle this problem, they propose the recursive-supervision that uses local outputs of recursions to reconstruct the HR image, together with the skip-connection that transmits the input LR image to the reconstruction layer (Fig.~\ref{fig:subfig:drcn2}). While bicubic interpolation introduces serious visual artifacts under high SR factor, Yang et al. propose a deep recurrent fusion network (DRFN) for single image SR \cite{8425771}. In DRFN, they use the transposed convolution as the upsampling layer and combine different-level features to reconstruct high-quality images. A similar method can also be found in \cite{8579237}, where Wang et al. use convolutional LSTM (ConvLSTM) in the residual block to form their multi-memory CNN (MMCNN) for video SR. In \cite{8356655}, Wang et al. propose a bidirectional recurrent convolutional NN named LFNet for light-field image SR, which uses an implicitly multi-scale fusion to utilize the spatial relations in light-field images. For image denoising, considering that the feature fusion of common CNNs is coarse, Wang et al. use the gated recurrent unit (GRU) to select and combine the features of different layers \cite{8986369}.

\par
For MRI image reconstruction, Qin et al. use a convolutional RNN to explore the dependencies of the temporal sequences \cite{8425639}. In addition, they also combine the network to the traditional optimization algorithms, which form the structured RNN. In \cite{putzky2017recurrent}, Putzky and Welling propose the recurrent inference machines (RIM) for image restoration, which is the unrolling of the inference algorithm. Yang et al. further use the RIM in accelerated photoacoustic tomography (PAT) reconstruction \cite{8856290}, where the forward operator $\mathbf{A}$ is used in the training process.

Structured RNNs are also common when solving sparse LIPs. Similar to structured FNNs, the structured RNNs get inspiration from the traditional iterative algorithms, such as ISTA. Intuitively, the RNN can be used to deal with a sequence of correlated observations in sparse LIPs. For example, in \cite{7952977}, Wisdom et al. solve the sequential sparse LIP with a structured RNN which inspired by the sequential ISTA. Different from the generic stacked RNN, the input of the proposed SISTA-RNN is connected to every iteration layer. In \cite{8803281}, Le et al. design their RNNs for sequential sparse LIP by unfolding the proximal gradient
method that aims to solve the $\ell_{1}-\ell_{1}$ minimization problem. Compared with the stacked RNN, the designed $\ell_{1}-\ell_{1}-$RNN has additional connections between the layers.

\par
In addition, in sparse LIPs, the support of the nonzero elements can be thought as a sequence, and it has been proved that the known part of supports can be used to speed up the convergence. While LISTA uses a fixed learning rate to learn the parameters, Zhou et al. adds an adaptive momentum vector to the network and design their adaptive ISTA \cite{zhou2018sc2net}. They further improve the efficiency of adaptive ISTA by reforming it as an RNN, which can be thought as a variant of the famous LSTM. In addition to the simple, one-step iterative algorithms such as the ISTA, in~\cite{he2017bayesian}, He et al. resemble the complex, multi-loop, majorization-minimization algorithm sparse Bayesian learning (SPL) to an RNN. The proposed network exhibits significantly improved performance in comparison to existing structured FNNs. This method can be applied to many applications including Direction-of-Arrival estimation and 3D photometric stereo recovery.


\subsection{Autoencoders}

AEs are self-supervised feedforward NNs that are usually used for dimension reduction and feature learning~\cite{hinton1994autoencoders,kingma2013auto}. An AE consists of an encoder and a decoder, which learns efficient data coding. The AE aims to learn the useful properties of the data, rather than reproduce the input at the output. Different variants of the basic AE are proposed to force the learning of the useful properties of features, such as the regularized AEs and the sparse AEs. AEs have been used in denoising \cite{8383709,8438540}, modulation classification in communication systems \cite{8038046,8648840} and image classification \cite{8065033,7286736,8421020}. Various AEs for LIPs are summarized in Table \ref{AE in LIP}.

\begin{table*}[]
\centering  
\setlength{\abovecaptionskip}{0pt}%
\setlength{\belowcaptionskip}{6pt}%
\caption{Details of training some AEs for LIPs.}\label{AE in LIP}
\begin{adjustwidth}{-1cm}{-1cm}
\tiny
\begin{tabular}{|p{0.5cm}|p{1.1cm}|p{1.5cm}|p{1.5cm}|p{1.5cm}|p{1.5cm}|p{1.5cm}|p{1.5cm}|}
\hline
Ref. &Application &Input &Output &Loss Function &Initialization &Learning Rate &Optimizer   \\ \hline\hline

\cite{xie2012image} &Image Denoising &Overlapping patches  &Clean patches &The quadratic error with sparsity regularization &Pre-trained stacked denoising auto-encoder  &Not given &Quasi-Newton \cite{le2011optimization} \\ \hline

\cite{agostinelli2013adaptive} &Image Denoising &Overlapping patches  &Clean patches &The quadratic error with sparsity regularization &Pre-trained SSDAs  &Not given &Quasi-Newton\\ \hline

\cite{cho2013simple} &Image Denoising &Overlapping patches  &Clean patches &The quadratic error &Pre-trained single-layer SSDAs  &$10^{-1}$ &SGD \\ \hline

\cite{7279746} &Image Denoising &Overlapping patches  &Clean patches &The quadratic error with KL penalty &Pre-trained multi-layer SDA  &$10^{-1}$ &Not given \\ \hline

\cite{7836672} &Image Denoising &Resized images  &Clean images &Not given &Not given  &Not given &Not given \\ \hline

\cite{7339460} &Image SR  &LR image patch &HR image patch &Mean squared error  &Intrinsic representations &Not given  &Gradient-based methods  \\ \hline

\cite{8758165} &Image SR  &LR image patch &HR image patch &Mean squared error with sparsity constraint  &Not given  &Not given  &Not given   \\ \hline

\cite{8553263} &Image reconstruction &Compressively sampled measurement  &Reconstructed images &Euclidean cost function  &Not given &Not given  &Not given \\ \hline

\cite{8462313} &Sparse coding  &Compressed digits data &Reconstructed digits data &The reconstruction loss and SSIM loss \cite{7797130}  &Same random values &Not given  &ADAM  \\ \hline

\cite{jalali2019using} &Sparse coding  &Compressed digits data &Reconstructed digits data &The L2 loss  &Not given  &Not given  &Not given   \\ \hline

\end{tabular}
\end{adjustwidth}
\end{table*}


The denoising AE (DAE) is the most commonly used AE in solving inverse problems, which is firstly proposed in \cite{vincent2008extracting} to obtain robust features. The DAE tries to reconstruct the signal from its noisy input. In~\cite{xie2012image}, Xie et al. propose the stacked sparse denoising AE (SSDA) for image denoising and blind inpainting, which stacks multiple DAEs and forces parameters to be sparse by employing sparsity regularization. In the training phase, Xie et al. initialize the SSDA with stacked DAs, where each DA is trained one by one, and the input of the successor DA is the output of the predecessor DA rather than the original noisy image. To improve the robustness of the SSDA, Agostinelli et al. propose the adaptive multi-column SSDA (AMC-SSDA), where several SSDAs are learned under different noise levels, and a weight prediction module is learned to combine the results of all SSDAs with different weights~\cite{agostinelli2013adaptive}. While the sparsity regularizer in~\cite{xie2012image} is not computationally efficient for DAEs with multiple hidden layers, Cho improves the performance of the network by forcing the output of the encoder to be sparse \cite{cho2013simple}. The proposed DAE performs well even without sparsity regularization and does not use any prior information about the noise. To enhance the robustness of AE to hybrid noises, Ye et al. add the KL penalty to the loss function, which brings the average activation of the hidden layer close to zero \cite{7279746}. In addition to fully connected AEs, convolutional layers can also be used for AEs. In \cite{7836672}, Gondara uses a DAE constructed using convolutional layers for medical image denoising. However, the previous work in \cite{xie2012image,agostinelli2013adaptive,cho2013simple,7279746,7836672} is inductive. In \cite{8383709}, the AE is further extended for blind image denoising.

\par
The AE can also be used in image SR and reconstruction. In \cite{7339460}, Zeng et al. develop a coupled deep AE (CDA) for single image SR. The CDA contains three parts, two AEs which extract the hidden representations of LR/HR image patches respectively, and a hidden layer which learns the mapping between the two representations. The training process of CDA contains the training of three parts and fine-tuning of the entire network. Considering the problem that the inconsistency between the sparse coefficients of the LR image and HR image influences the SR results, Shao et al. propose coupled sparse AE (CSAE) to learn the mapping between the sparse coefficients of the LR image and HR image \cite{8758165}. The proposed CSAE is used for the spatial resolution of remote sensing images. For image reconstruction, Mehta et al. propose to use AE for CS-based medical image reconstruction to cut off the time for reconstruction \cite{mehta2017rodeo}. Instead of using the Euclidean norm as a cost function, Mehta et al. use a robust $\ell_{1}$ norm. Similar to the work in \cite{mehta2017rodeo}, Gupta and Bhowmick also consider the time-consuming problem in real-time image reconstruction \cite{8553263}. They propose Coupled AE (CAE) to learn the mapping from the measurements to the representation of the target images.

\par
Besides, AEs are also popular in sparse coding. In \cite{barello2018sparse}, Barello et al. design the sparse coding variational AE (SVAE), which is neurally plausible to calculate the neural response of an image patch. To solve the computation problem when using LISTA for convolutional sparse coding, Sreter and Giryes propose the convolutional LISTA, which serves as the sparse encoder in an AE \cite{8462313}. Based on the sparse coding, Jalali and Yuan analyze the performance of AEs for such recovery problems, and proposed a projected gradient descent based algorithm \cite{jalali2019using}.

\par
In addition to the common AEs, AEs can also benefit from the deep unfolding method. In \cite{7010964}, Sprechmann et al. unfold proximal descent algorithms, and then learn the pursuit processes to solve the low-rank models, including the RPCA and non-negative matrix factorization.


%

\subsection{Generative Adversarial Networks}

\begin{table*}[]
\centering  
\setlength{\abovecaptionskip}{0pt}%
\setlength{\belowcaptionskip}{6pt}%
\caption{Details of training some GANs for LIPs.}\label{GAN in LIP}
\begin{adjustwidth}{-1cm}{-1cm}
\tiny
\begin{tabular}{|p{0.5cm}|p{1.2cm}|p{1.2cm}|p{1.2cm}|p{3.2cm}|p{1.5cm}|p{1cm}|p{1cm}|}
\hline
Ref. &Application &Input &Output &Loss Function &Initialization &Learning Rate &Optimizer   \\ \hline\hline

\cite{8578431} &Image Denoising &Noisy image patch &Clean image patch &Mean squared error &Not given &$10^{-3}$ &SGD  \\ \hline

\cite{8546246} &Sparse signal denoising  &Noisy sample  &Denoised sample &Mean squared error and cross-entropy loss  &Not given &Not given &ADAM  \\ \hline

\cite{8340157} &Image Denoising &Noisy image patch &Clean image patch &Wasserstein distance and the perceptual loss &Pre-trained deep CNN   &Not given &SGD  \\ \hline

\cite{8710893}  &Image Denoising &Noisy image patch &Clean image patch &Pixel loss, feature loss, smooth loss and adversarial loss &Not given   &Not given &Not given  \\ \hline

\cite{8941108}  &Image Denoising &Noisy image patch &Clean image patch &Least squares loss, global loss and detail loss  &Not given   &$2\times10^{-4}$ &ADAM \\ \hline

\cite{7934380}  &Image Denoising &Noisy image patch &Clean image patch &Squared error and cross-entropy loss  &Normal distribution   &$2\times10^{-4}$ &ADAM \\ \hline

\cite{8899202} &Image SR &LR image patch &HR image patch &Adversarial loss and pixel-wise loss   &Randomly  &Decreased by 0.5 every 10 epochs &Not given \\ \hline

\cite{8546286} &Joint denoising and SR &Noisy LR subimage &HR subimage &Adversarial loss, mean squared error and VGG based loss   &Randomly  &$10^{-4}$ &Not given \\ \hline

\cite{8736838} &Image SR &LR CT image patch &HR CT image patch &Adversarial loss, cycle consistency loss, identity loss and joint sparsifying transform loss   &Not given  &Not given &Not given \\ \hline

\cite{8099502} &Image SR &LR image patch &HR image patch &Weighted sum of a content loss and an adversarial loss   &Trained MSE-based SRResNet network &$10^{-4}$ to $10^{-5}$ &ADAM \\ \hline

\cite{8553719} &video SR &LR image patch &HR image patch &Perceptual loss (weighted sum of a content loss and an adversarial loss) &Not given &Not given &Not given \\ \hline

\cite{8900228} &Image SR &LR image patch &HR image patch &The sum of a
pixel-wise loss and an adversarial loss&Follows \cite{7410480}  &$2\times10^{-4}$ &ADAM \\ \hline

\end{tabular}
\end{adjustwidth}
\end{table*}

The GAN is originally proposed as a form of the generative model for unsupervised learning, which can also be used for applications involving LIPs. Various GANs for LIPs are summarized in Table \ref{GAN in LIP}.

\par
The main motivation for using GANs for denoising is that GANs can better preserve high-frequency components and image details, while CNNs can easily over-smooth the edges of the image. For image denoising, the generator network is expected to generate the denoised signal, and the discriminator network is used to distinguish the denoised output from the ground truth, which provides provide feedback for the training of the generator network. The application of GANs in denoising could be diverse. For example, Chen et al. proposed a GAN-CNN based blind denoiser, where the generator network is used to estimate the distribution of noisy images and generate paired training data for the training of denoising CNN \cite{8578431}. The network structure of the generator and discriminator can be inspired by various FNNs or CNNs, such as LISTA-GAN \cite{8546246}, VGG-GAN \cite{8340157} and ResNet-GAN \cite{8710893} or special designed \cite{8941108}.

\par
\begin{figure*}[!t]
\centering
\vspace{-0.8cm}
\setlength{\abovecaptionskip}{-0.05cm}
\setlength{\belowcaptionskip}{-0.5cm}
\subfigure[]{
\label{figganep:subfig1}
\includegraphics[width=0.2\textwidth]{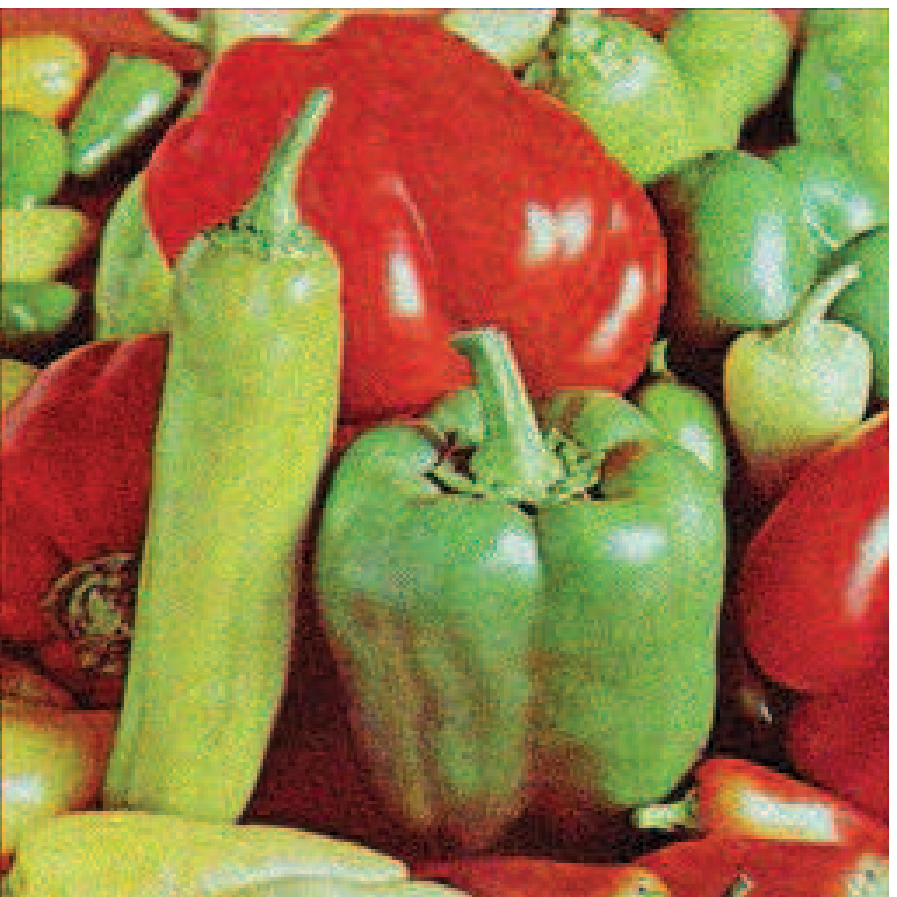}}
\subfigure[]{
\label{figganep:subfig2}
\includegraphics[width=0.2\textwidth]{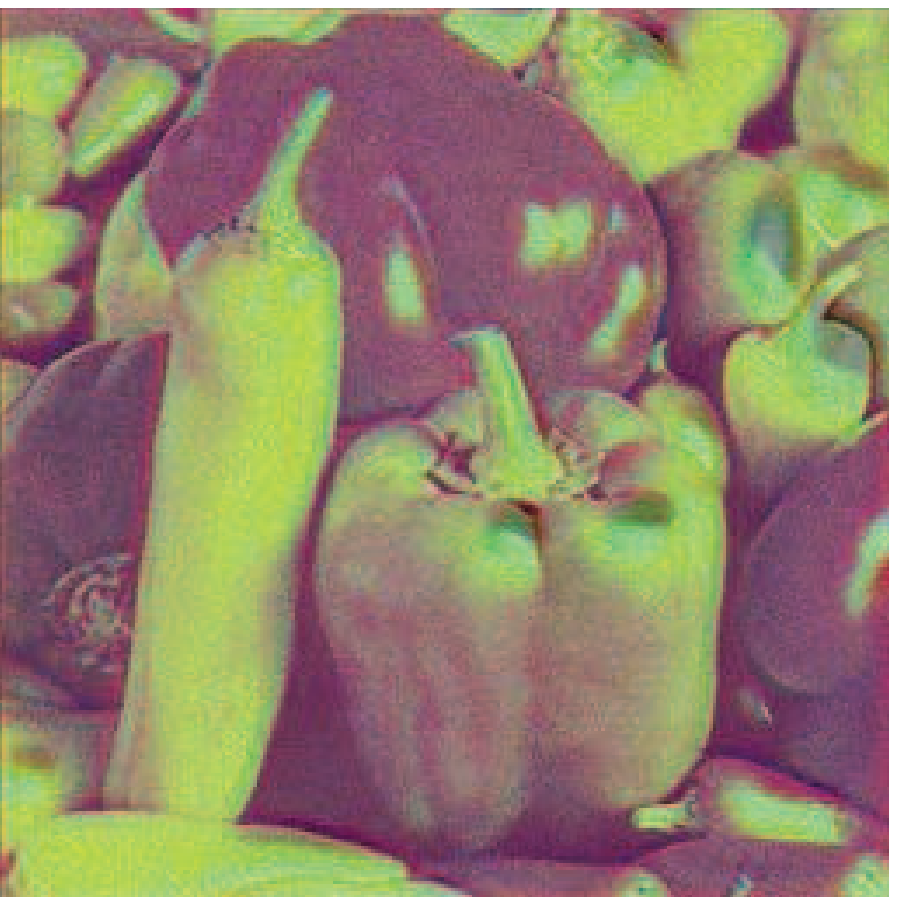}}
\subfigure[]{
\label{figganep:subfig3}
\includegraphics[width=0.2\textwidth]{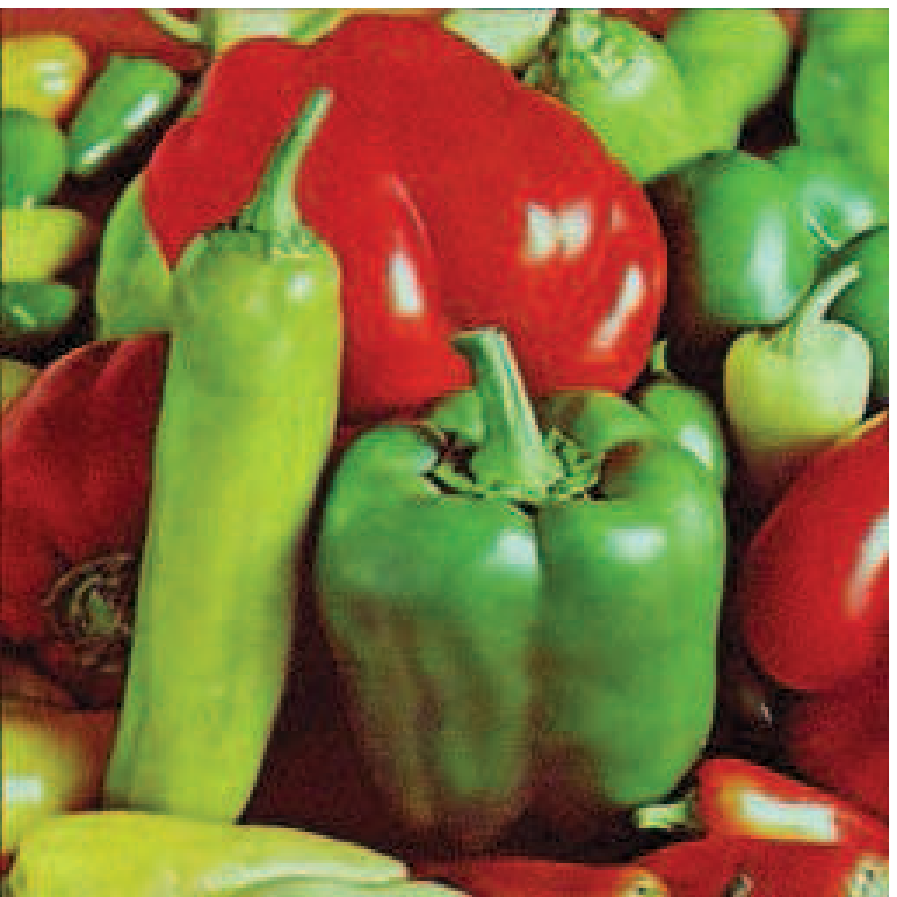}}
\subfigure[]{
\label{figganep:subfig4}
\includegraphics[width=0.2\textwidth]{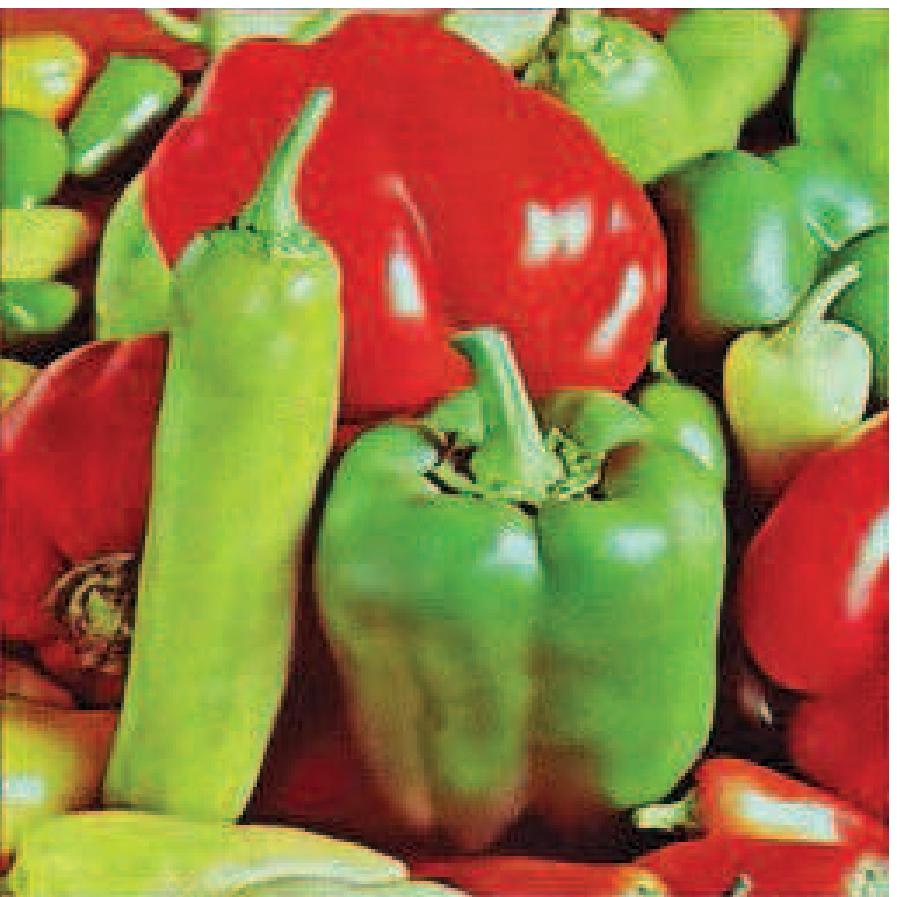}}
\caption{The denoising results with different loss functions. (a)noisy image, (b)denoised image with the adversarial loss (c)denoised image with the adversarial loss and pixel loss, (d)denoised image with the adversarial loss, pixel loss and feature loss.}
\label{fig:ganep} 
\end{figure*}

Another main innovation lies in the design of various loss functions. Wolterink et al. find that the network trained with voxel-wise loss has a higher peak signal-to-noise ratio, while the network trained with adversarial loss better captures image statistics \cite{7934380}. In \cite{8340157}, Yang et al. add the Wasserstein distance and perceptual loss to GANs. The Wasserstein distance, which comes from the optimal transport theory, is used as the discrepancy measure to improve the performance of GANs. The perceptual loss, which calculates the discrepancy between images in an established feature space, is used to suppress noise. Alsaiari et al. use the weighted sum of pixel-to-pixel Euclidean loss, feature loss, smooth loss and adversarial loss \cite{8710893}, while Li and Xiao use the combination of the denoising loss and reconstruction loss. In Fig.~\ref{fig:ganep}, we compare the performance of different loss functions under the same training set and the same network structure. The adversarial loss adapts the binary cross-entropy that comes from the discriminator, and helps to generate images that can deceive the discriminator. It is found that the network that trained with the adversarial loss is hard to convergence and the generated image has higher noise levels. The pixel loss calculates the pixel-to-pixel Euclidean distance between the output and the clean image, and is helpful for correctly filling the noise of the color. However, the network trained with the pixel loss leads to a smooth image. The feature loss, which depends on the features extracted from the convolutional layer, helps to extract features accurately. Thus, the network trained with the adversarial loss, pixel loss and style loss has the best visual quality.

\par
\begin{figure}
\centering
\vspace{-0.8cm}
\setlength{\abovecaptionskip}{-0.05cm}
\setlength{\belowcaptionskip}{-0.5cm}
\subfigure[An hourglass CNN model \cite{8400496}.]{
\label{fig:subfig:gansr1}
\includegraphics[width=0.9\textwidth]{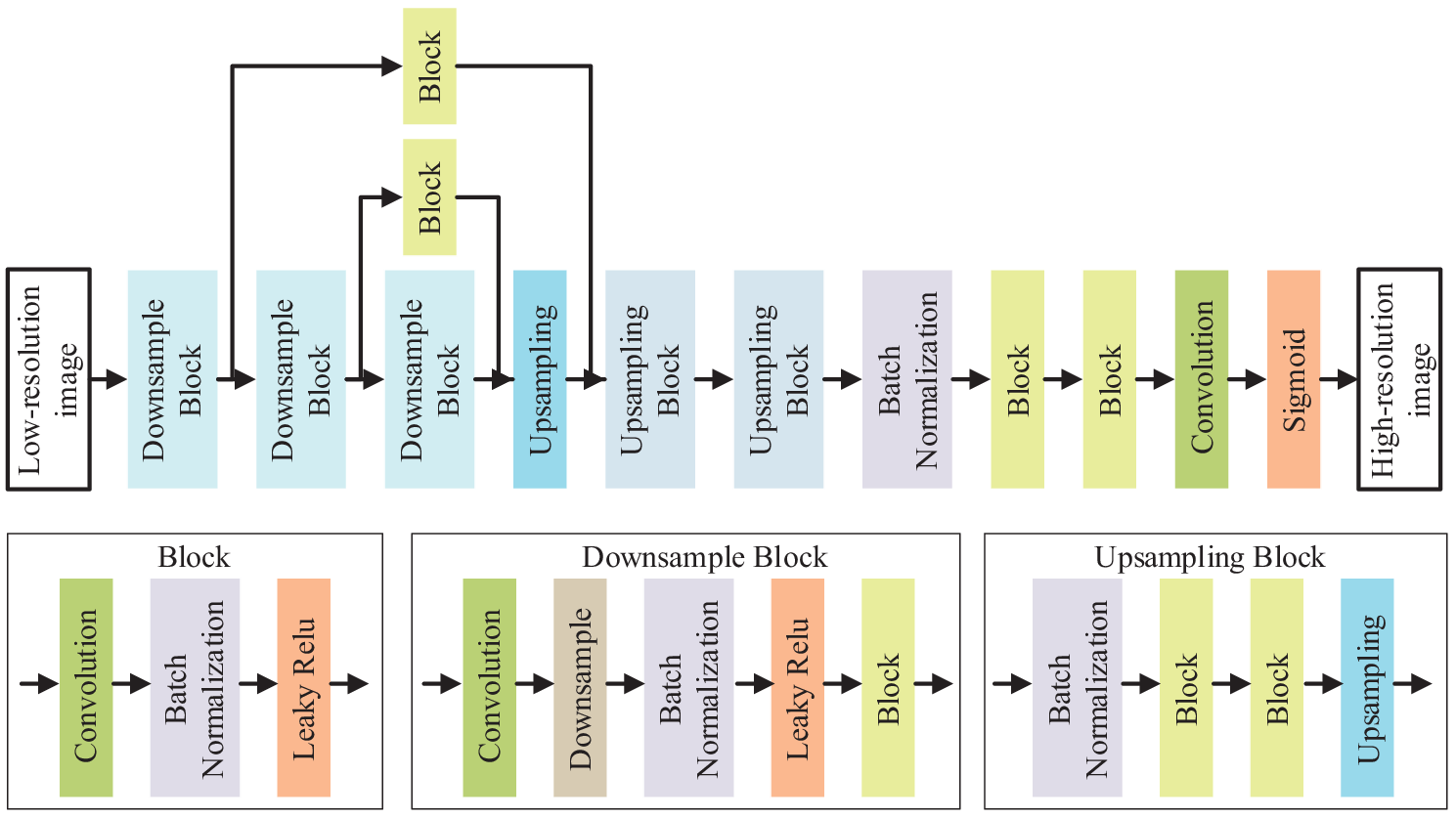}}
\hspace{1in}
\subfigure[A Cycle-in-Cycle network \cite{8575264,8825849}.]{
\label{fig:subfig:gansr2}
\includegraphics[width=0.9\textwidth]{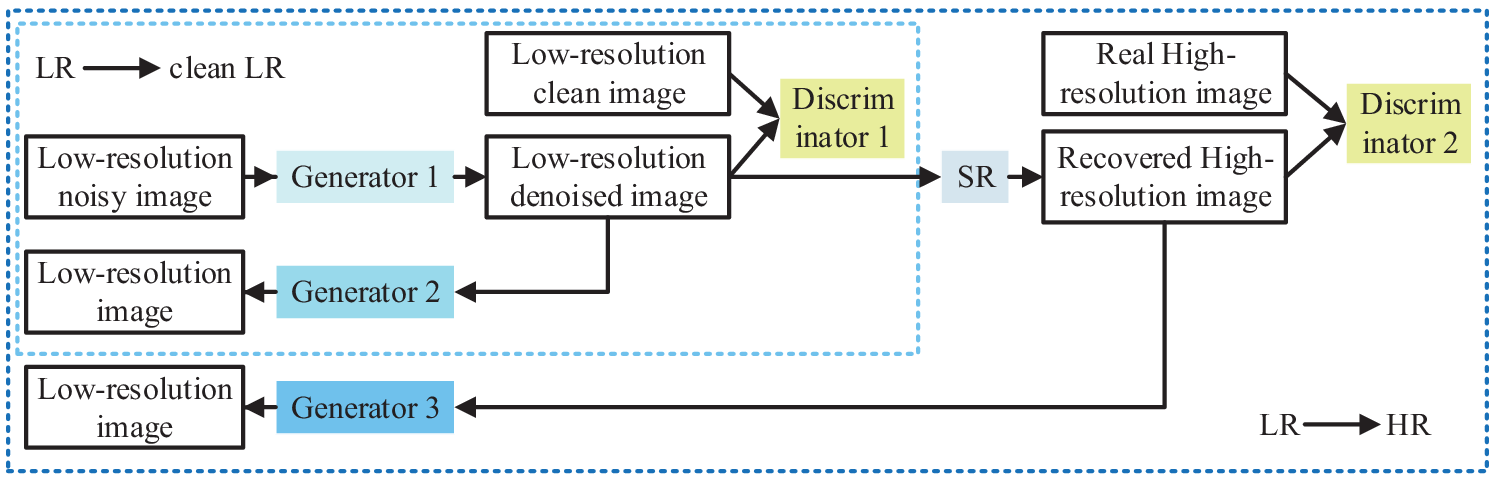}}
\hspace{1in}
\subfigure[A dense block network.]{
\label{fig:subfig:gansr3}
\includegraphics[width=0.9\textwidth]{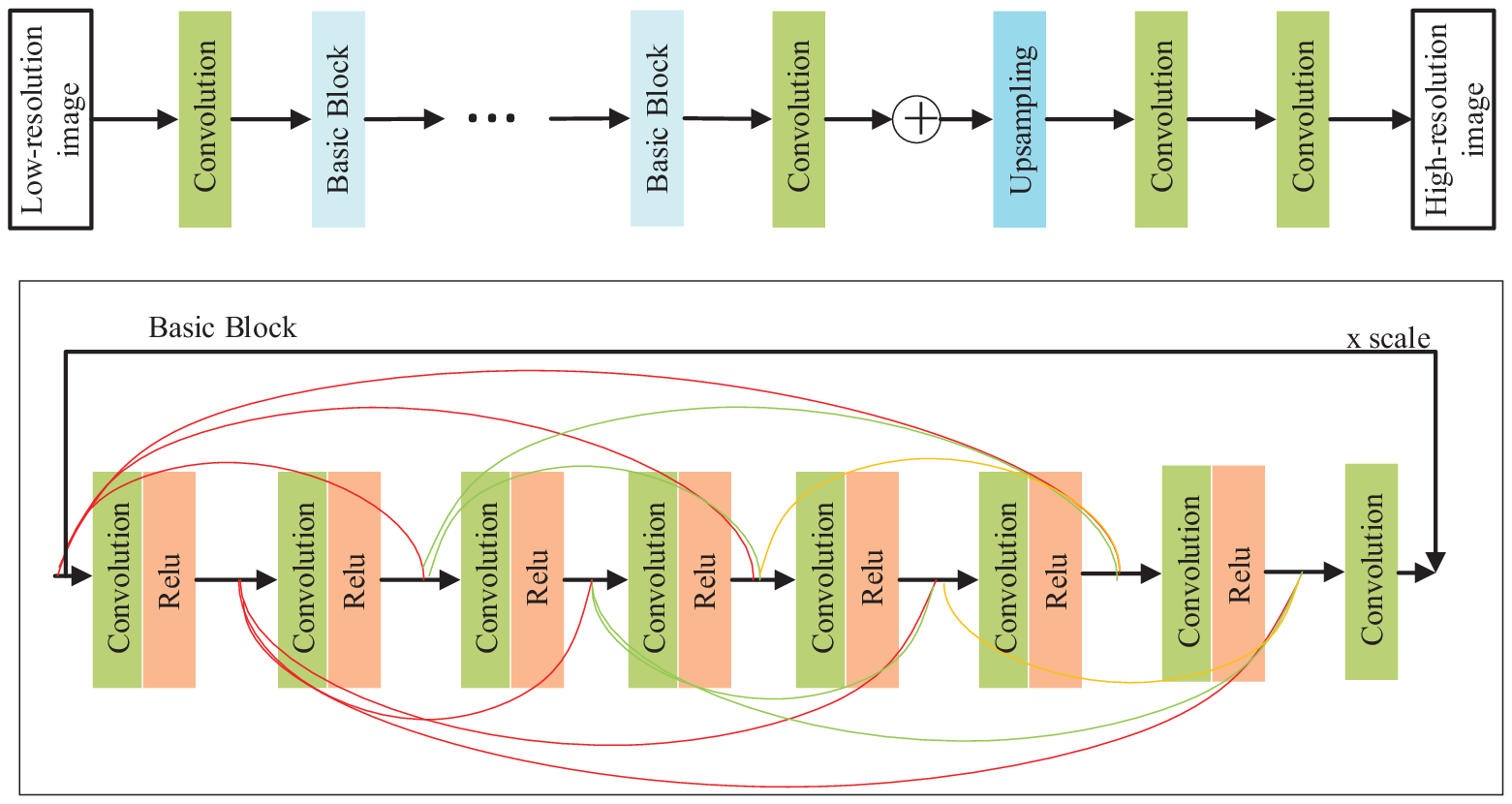}}
\caption{The structure of generator networks of GANs for image SR.}
\label{fig:srgan} 
\end{figure}

GANs have also been employed for image SR, which leads to different innovative designs. A common problem is that the LR images may contain noise, such as the speckle and smudge in synthetic aperture radar images \cite{8899202}. The general method is performing image denoising to LR images firstly, and then reconstructing the HR images. The denoising and DR can be performed with a joint generator network \cite{8899202,8546286} or two generator networks \cite{8736838}. Compared with image denoising, the network structure of generator networks for image SR is more diverse. In Fig.~\ref{fig:srgan}, we show several novel network structures in GANs for image SR, including an hourglass CNN model \cite{8400496}, a Cycle-in-Cycle network \cite{8575264,8825849} and a dense block network \cite{8803711}.

The innovations in loss function also exists in image SR for finer texture details, and most loss function is a weighted sum of several losses. The losses can be classed into adversarial loss, the pixel-based loss and feature map based loss. For example, Ledig et al. use an adversarial loss and a content loss \cite{8099502}, while Chen et al. use an MSE loss, the generative loss and the VGG loss \cite{8546286}. Other loss functions contain the sum of the perceptual loss, MSE-based content loss, and an adversarial loss, which is used \cite{8553719} by Gopan and Kumar, the sum of the pixel-wise loss and adversarial loss used in \cite{8900228} by Jiang et al. and the sum of joint sparsifying transform loss and supervision loss in \cite{8736838} by You et al..

\section{Challenges and Future Research Directions}
In the previous section, we explore several research directions and paradigms on using DL to solve LIPs. It has been observed that DL has brought breakthroughs in many applications. However, there are still many open challenges that require further investigation. In this section, we discuss several potential future research directions in using DL to solve LIPs.

\subsection{Constructing Training Datasets}
In solving LIPs, the performance of DL based methods greatly relies on the data (the input and the label) seen during training, which reflects the functional relationship between model parameters $\mathbf{m}$ and the observed data $\mathbf{d}$. However, just as imperfect mathematical modeling of complex scenarios in traditional methods leads to the model error, imperfect training data in DL methods also leads to the recovery error.

\par
The recovery error caused by the training data may come from the generating process of the training data. In practical scenarios that we do not have access to the real $\mathbf{m}$, a popular method is to artificially generate the training data. However, the artificially generated data may have a distribution that differs from the distribution of the real $\mathbf{m}$. For example, in the sparse LIP, the sparse data $\mathbf{m}$ may contain different sparsity and sparse patterns. In cases that we cannot get the general data $\mathbf{m}$, we may resort to the traditional algorithm, e.g., in the LISTA, the sparse data $\mathbf{m}$ is generated from the traditional CoD algorithm~\cite{gregor2010learning}. However, the traditional algorithms may converge to the non-optimal solution, thus results in errors in the training data. Therefore, a potential research direction is to study the errors contained in the training data and the methods to reduce or even eliminate the recovery error caused by the training data.

\par
The recovery error may also result from the mismatch between the training data and the test data. For example, in image denoising, the mismatch of the noise distributions between the training data and the testing data leads to the performance degeneration~\cite{6247952}. In~\cite{agostinelli2013adaptive}, Agostinelli et al. solve this problem by connecting the networks that are trained under different noise distributions in parallel according to the learned weight. However, such methods increase the model complex and lead to heavy computation. A more straightforward solution is to increase the diversity of training data. In \cite{7839189}, Zhang et al. construct their training data set with different noise distributions and train a single NN to deal with multiple noise distributions. However, this method is not suitable for training time-limited scenarios, since it is impossible to include all the possible $\mathbf{m}$ in a limited training data.

\subsection{Incorporating the Prior Knowledge for Structured Network Designs}
The process of using the DL based method to solve the LIP can be seen as choosing an optimal function from a class of functions defined by a NN for the mapping relationship between model parameters $\mathbf{m}$ and the observed data $\mathbf{d}$. By carefully designing the network architecture, we are designing a class of functions that are closer to the mapping relationship, which helps for faster convergence and optimal solution. However, the design of the network architecture still lacks theoretical support and thus is intractable. Thus, more theoretical explorations are needed.

\par
In LIPs, there usually exists prior knowledge about model parameters, e.g., their spatial distribution or mutual dependence. We expect to gain further improvement in convergence speed and performance by incorporating the prior knowledge into network designs. The structured networks also have profits in other aspects. For example, by limiting the weight sharing between network layers, the LISTA has fewer parameters than the common NN, thus the LISTA is less likely to be over-fitting~\cite{gregor2010learning}. A popular method is to design networks based on the unfolding of iterative algorithms~\cite{gregor2010learning,7905837}. Since the traditional iterative algorithms have calculated an estimation of the LIP, the time-unfolded network can directly obtain a sub-optimal solution without training. As such a structured network can obtain a better solution than the iterative algorithm after training, and needs less training data and time to obtain the same performance compared with the common network. However, unfolding based networks may also converge to a local optimal under the misleading of the iterative algorithm. Therefore, a potential research direction is to investigate the theoretical bound that a structured network can achieve for a specific inverse problem, such as the maximum convergence speed and the highest accuracy. Further research also needs to be done on the design of structured networks that could achieve performance closed to the theoretical bound, in addition to the unfolding based methods. Another potential research direction is the tradeoff between the convergence speed and accuracy. For example, in~\cite{xin2016maximal}, Xin et al. demonstrate that an FNN with independent weights has better estimation accuracy along with the decrease in convergence speed in cases where the linear operator $\mathbf{A}$ has coherent columns.

\subsection{Dealing With High Dimensional Data}
Modern inverse problems increasingly involve high dimensional data such as tensors \cite{8237620,8462592,7460232,8417937}, which usually refer to inter-dimension correlations \cite{8682281}. However, at present, most of the DL based methods for solving LIPs are performed on low-dimensional data, e.g., vectors and matrixes.
A method that using existing models to process high dimensional data is to decrease the dimension of the input data firstly. For example, flattening the three-dimensional tensor into a two-dimensional matrix. However, the dimensionality reduction process is usually accompanied by the loss of the inter-dimension correlations information. A potential solution is to design
specialized networks for high dimensional data processing. For example, The 3-D convolution can be used to explore the spatial and spectral characters of hyperspectral image \cite{8676115,8454887}. Another popular method for high dimensional tensor processing is deep tensor factorization (DTF), which considers the temporal or spatial information. The DTF can extract hierarchical and meaningful features of multi-channel images such as hyperspectral images, thus is popular in image classification and pattern classification \cite{8517386,7902201,8902979}. DTF can also be used in recommender systems \cite{9005677}, scene decomposition \cite{8937263}, and fault diagnosis \cite{9108248}.

\par
Another problem is that processing the high dimensional data needs a larger and deeper network, which means the rapid increase in the number of network parameters and the surge in the demand of hardware with high computational capability. However, DL heavily relies on high-parallel computation of GPUs for training, while GPUs have limited memory, which makes DL based methods encounter computational difficulties when processing high dimensional data. A possible solution is the distributed DL, such as the model parallelism and the data parallelism. In the model parallelism, the whole network is partitioned into small components and then trained in different machines. In the data parallelism, different machines have a complete copy of the entire model and limited training data, then the complete model is calculated by some methods. The model parallelism and the data parallelism can be combined to achieve training acceleration \cite{dean2012large}. Besides, there are several methods to train the distributed NNs, and each method exists many variants \cite{8354695,zhang2015staleness,strom2015scalable,ho2013more}. One of the potential research directions is the maximum accuracy that the distributed DL can get with the specific training algorithm under given conditions such as limited training time or limited training data. Besides, we could also consider the tradeoff between model accuracy and runtime \cite{7837841}.

\subsection{Designing Light and Efficient Architectures}
In general, the DL based methods with more complex networks have better accuracy. However, complex models usually involve a great number of parameters, which increases the difficulty in training and limits its usage in computing resource-constrained applications. Therefore, an important research direction is the design of light and efficient network architectures, which helps to effectively apply the DL models to various hardware platforms \cite{alvarez2016learning,howard2017mobilenets,8578389,lin2017towards}.

\par
A carefully designed network architecture can effectively reduce the redundancy and computation of the DL models, thus speed up the solving process without sacrificing reconstruction accuracy. Representative work include the SqueezeNet \cite{8578814} and the MobileNet \cite{howard2017mobilenets}. Another method is compressing an existing network to decrease the number of parameters and the required computation resource, under the guarantee of reconstruction accuracy \cite{8253600,8547604,8171208,7551397,8686678}. For example, the model cutting method compresses the model by cutting unimportant connections of a trained model according to some effective evaluations \cite{han2019deep}. The network quantization method cuts the redundancy of the data by reducing the length of the code and the number of bits, according to the data distribution in the trained model \cite{hubara2017quantized}. Another efficient method is network binarization, where the original floating-point weights are forced to be $+1$ and $-1$. For a specific LIP, it remains a challenge to choose a suitable method to balance the accuracy and computation speed.

Future AI-driven automation will bring about a step-change in their ability to create efficient, resilient, and also user-centric services. However, the very same algorithms may also cause irreversible environmental damage due to their high energy consumption and lead to serious global sustainability issues. To achieve UN sustainable development goals in the context of lightweight and green AI, we need to reduce the computation and energy consumption.

Model compression approaches are for reducing the sizes of DNN target operations and data access overhead in both training and inference of the DNN. This is highly related to the numbers of neurons and the associated weights in it. Due to the lack of theoretical results on the optimal DNN architecture \footnote{Neuroevolution DL does offer a numerical pathway to finding optimal architectures}. Previous studies have revealed that NNs are typically over-parameterized, and there is significant redundancy that can be exploited \cite{sparsity}. Therefore, it is possible to achieve similar function approximation performance by removing redundant network architecture (e.g. pruning the network) and only retaining useful parts with greatly reduced model size.  The second method is architectural innovations, such as replacing fully-connected layers with convolutional layers that are relatively more compact. Another method is weight quantization. Already, some of the aforementioned DNN compression practices have emerged in recent mobile DL applications.

\subsection{Solving LIPs in Practical Applications}
In practical applications, there is contradiction between the limited training data and training time, and infinite real data and various application scenarios. Although the DL method succeeds in specific scenarios, it takes a very high cost in training different DL models for different application scenarios. Thus, the research on the generalization of the DL models is important and essential, which affects the accuracy of models when applied to new data in practical applications~\cite{azulay2018deep,8601309}.

\par
Most existing methods attribute the poor generalization ability of DL models to the memory of the training data by the NN. Therefore, various regularization methods are adopted to increase the generalization of DL models, including explicit regularization on parameters (e.g., L$1$ regularization), empirically-based regularization (e.g., early stopping~\cite{raskutti2014early} and dropout~\cite{srivastava2014dropout}), and implicit regularization (e.g., data augmentation~\cite{inoue2018data}). In~\cite{zhang2016understanding}, Zhang et al. explore the role of regularization in DL models. They demonstrate that the regularization methods are effective but not necessary for the improvement of the generalization ability. Therefore, a potential research direction is other methods that can improve the generalization of DL models. In addition, the interpretation of the generalization ability of the NN is also promising, which can provide theoretical support and guide for the design of robust models~\cite{neyshabur2017exploring}.

\subsection{Solving Nonlinear Inverse Problems with DL}
While this article focuses on the applications of DL in solving LIPs, there also exist several works in using DL to solve various nonlinear inverse problems, especially in the CS problems with quantized measurements \cite{5306135}. For example, in \cite{9053161}, Takabe et al. propose a complex-field trainable ISTA (C-TISTA) based on the concept of deep unfolding, which aims to solve the complex-field nonlinear inverse problems. In C-TISTA, they use a trainable shrinkage function to utilize various prior information such as sparsity. While Mahabadi et al. try to learn the sampling process of the quantized CS \cite{8637803}, Leinonen and Codreanu directly jointly optimize the whole sampling and recovery process with an encoder and decoder via NNs \cite{9083783}. A similar method for joint optimization of measurement and recovery in can quantized CS also be found in \cite{7560597}, where the NN consists a binary measurement matrix, a non-uniform quantizer, and a non-iterative recovery solver. Considering the high computing and expressive power, the use of NNs in nonlinear inverse problems is a promising direction.

\section{Conclusion}
In this paper, we presented a comprehensive survey of the recent achievements in using DL to solve LIPs. We summarize the use of various DL architectures, optimization algorithms, loss functions and techniques in solving LIPs. For LIPs with structured information, we present how it is used in the design of various DL models. Our hope is that this article can provide guidance for designing NNs for solving various LIPs. In addition to the recent progresses, there are still many open challenges and promising future directions including the construction of training datasets, the design of structured networks, the techniques for high dimensional data processing in NNs, the design of light and efficient network architectures, and the problems in practical applications.

\scriptsize
\bibliography{refs}

\begin{thebibliography}{100}
\expandafter\ifx\csname url\endcsname\relax
  \def\url#1{\texttt{#1}}\fi
\expandafter\ifx\csname urlprefix\endcsname\relax\def\urlprefix{URL }\fi
\expandafter\ifx\csname href\endcsname\relax
  \def\href#1#2{#2} \def\path#1{#1}\fi

\bibitem{backus1968resolving}
G.~{Backus}, F.~{Gilbert}, The resolving power of gross earth data, Geophysical
  Journal International 16~(2) (1968) 169--205.

\bibitem{kabanikhin2008definitions}
S.~I. {Kabanikhin}, Definitions and examples of inverse and ill-posed problems,
  Journal of Inverse and Ill-Posed Problems 16~(4) (2008) 317--357.

\bibitem{8770109}
X.~{Li}, J.~{Fang}, H.~{Duan}, Z.~{Chen}, H.~{Li}, Fast beam alignment for
  millimeter wave communications: A sparse encoding and phaseless decoding
  approach, IEEE Transactions on Signal Processing 67~(17) (2019) 4402--4417.

\bibitem{8122055}
X.~{Li}, J.~{Fang}, H.~{Li}, P.~{Wang}, Millimeter wave channel estimation via
  exploiting joint sparse and low-rank structures, IEEE Transactions on
  Wireless Communications 17~(2) (2018) 1123--1133.

\bibitem{7390294}
W.~{Chen}, I.~J. {Wassell}, Cost-aware activity scheduling for compressive
  sleeping wireless sensor networks, IEEE Transactions on Signal Processing
  64~(9) (2016) 2314--2323.

\bibitem{7031966}
W.~{Chen}, I.~J. {Wassell}, Optimized node selection for compressive sleeping
  wireless sensor networks, IEEE Transactions on Vehicular Technology 65~(2)
  (2016) 827--836.

\bibitem{7293676}
W.~{Chen}, I.~J. {Wassell}, A decentralized bayesian algorithm for distributed
  compressive sensing in networked sensing systems, IEEE Transactions on
  Wireless Communications 15~(2) (2016) 1282--1292.

\bibitem{986949}
J.~F. {Murray}, K.~{Kreutz-Delgado}, An improved focuss-based learning
  algorithm for solving sparse linear inverse problems, in: Conference Record
  of Thirty-Fifth Asilomar Conference on Signals, Systems and Computers
  (Cat.No.01CH37256), Vol.~1, 2001, pp. 347--351 vol.1.

\bibitem{8332502}
X.~{Shen}, Y.~{Gu}, Nonconvex sparse logistic regression with weakly convex
  regularization, IEEE Transactions on Signal Processing 66~(12) (2018)
  3199--3211.

\bibitem{lee2007efficient}
H.~{Lee}, A.~{Battle}, R.~{Raina}, A.~Y. {Ng}, Efficient sparse coding
  algorithms, in: Advances in neural information processing systems, 2007, pp.
  801--808.

\bibitem{8123942}
G.~{Li}, Y.~{Gu}, Restricted isometry property of gaussian random projection
  for finite set of subspaces, IEEE Transactions on Signal Processing 66~(7)
  (2018) 1705--1720.

\bibitem{7101819}
W.~U. {Bajwa}, M.~F. {Duarte}, R.~{Calderbank}, Conditioning of random block
  subdictionaries with applications to block-sparse recovery and regression,
  IEEE Transactions on Information Theory 61~(7) (2015) 4060--4079.

\bibitem{6826555}
P.~{Chen}, I.~W. {Selesnick}, Group-sparse signal denoising: Non-convex
  regularization, convex optimization, IEEE Transactions on Signal Processing
  62~(13) (2014) 3464--3478.

\bibitem{5437428}
R.~G. {Baraniuk}, V.~{Cevher}, M.~F. {Duarte}, C.~{Hegde}, Model-based
  compressive sensing, IEEE Transactions on Information Theory 56~(4) (2010)
  1982--2001.

\bibitem{6967808}
J.~{Fang}, Y.~{Shen}, H.~{Li}, P.~{Wang}, Pattern-coupled sparse bayesian
  learning for recovery of block-sparse signals, IEEE Transactions on Signal
  Processing 63~(2) (2015) 360--372.

\bibitem{7478129}
J.~{Fang}, F.~{Wang}, Y.~{Shen}, H.~{Li}, R.~S. {Blum}, Super-resolution
  compressed sensing for line spectral estimation: An iterative reweighted
  approach, IEEE Transactions on Signal Processing 64~(18) (2016) 4649--4662.

\bibitem{8039502}
W.~{Chen}, Simultaneous sparse bayesian learning with partially shared
  supports, IEEE Signal Processing Letters 24~(11) (2017) 1641--1645.

\bibitem{7558157}
W.~{Chen}, D.~{Wipf}, Y.~{Wang}, Y.~{Liu}, I.~J. {Wassell}, Simultaneous
  bayesian sparse approximation with structured sparse models, IEEE
  Transactions on Signal Processing 64~(23) (2016) 6145--6159.

\bibitem{6800127}
C.~{Chen}, Y.~{Li}, J.~{Huang}, Forest sparsity for multi-channel compressive
  sensing, IEEE Transactions on Signal Processing 62~(11) (2014) 2803--2813.

\bibitem{zou2013segmentation}
W.~{Zou}, K.~{Kpalma}, Z.~{Liu}, J.~{Ronsin}, Segmentation driven low-rank
  matrix recovery for saliency detection, in: 24th British machine vision
  conference (BMVC), 2013, pp. 1--13.

\bibitem{1177153}
R.~{Basri}, D.~W. {Jacobs}, Lambertian reflectance and linear subspaces, IEEE
  Transactions on Pattern Analysis and Machine Intelligence 25~(2) (2003)
  218--233.

\bibitem{koren2009matrix}
Y.~{Koren}, R.~{Bell}, C.~{Volinsky}, Matrix factorization techniques for
  recommender systems, Computer~(8) (2009) 30--37.

\bibitem{7914672}
Z.~{Zhou}, J.~{Fang}, L.~{Yang}, H.~{Li}, Z.~{Chen}, R.~S. {Blum}, Low-rank
  tensor decomposition-aided channel estimation for millimeter wave mimo-ofdm
  systems, IEEE Journal on Selected Areas in Communications 35~(7) (2017)
  1524--1538.

\bibitem{8318613}
L.~{Yang}, J.~{Fang}, H.~{Duan}, H.~{Li}, B.~{Zeng}, Fast low-rank bayesian
  matrix completion with hierarchical gaussian prior models, IEEE Transactions
  on Signal Processing 66~(11) (2018) 2804--2817.

\bibitem{8453881}
W.~{Chen}, Simultaneously sparse and low-rank matrix reconstruction via
  nonconvex and nonseparable regularization, IEEE Transactions on Signal
  Processing 66~(20) (2018) 5313--5323.

\bibitem{candes2011robust}
E.~J. {Cand{\`e}s}, X.~{Li}, Y.~{Ma}, J.~{Wright}, Robust principal component
  analysis?, Journal of the ACM (JACM) 58~(3) (2011) 11.

\bibitem{6138863}
J.~Liu, P.~Musialski, P.~Wonka, J.~Ye, Tensor completion for estimating missing
  values in visual data, IEEE Transactions on Pattern Analysis and Machine
  Intelligence 35~(1) (2013) 208--220.

\bibitem{8861380}
W.~{Chen}, X.~{Gong}, N.~{Song}, Nonconvex robust low-rank tensor
  reconstruction via an empirical bayes method, IEEE Transactions on Signal
  Processing 67~(22) (2019) 5785--5797.

\bibitem{saxena2014noises}
C.~{Saxena}, D.~{Kourav}, Noises and image denoising techniques: a brief
  survey, International journal of Emerging Technology and advanced Engineering
  4~(3) (2014) 878--885.

\bibitem{8705542}
M.~{Chen}, H.~{Zhang}, G.~{Lin}, An adaptive directional non-local means
  algorithm with size-adaptive search window for image denoising, in: 2018 3rd
  International Conference on Smart City and Systems Engineering (ICSCSE),
  2018, pp. 834--839.

\bibitem{7740078}
T.~{Qiao}, J.~{Ren}, Z.~{Wang}, J.~{Zabalza}, M.~{Sun}, H.~{Zhao}, S.~{Li},
  J.~A. {Benediktsson}, Q.~{Dai}, S.~{Marshall}, Effective denoising and
  classification of hyperspectral images using curvelet transform and singular
  spectrum analysis, IEEE Transactions on Geoscience and Remote Sensing 55~(1)
  (2017) 119--133.

\bibitem{4517848}
{Hyung Il Koo}, {Nam Ik Cho}, Image denoising based on a statistical model for
  wavelet coefficients, in: 2008 IEEE International Conference on Acoustics,
  Speech and Signal Processing, 2008, pp. 1269--1272.

\bibitem{4271520}
K.~{Dabov}, A.~{Foi}, V.~{Katkovnik}, K.~{Egiazarian}, Image denoising by
  sparse 3-d transform-domain collaborative filtering, IEEE Transactions on
  Image Processing 16~(8) (2007) 2080--2095.

\bibitem{6392274}
W.~{Dong}, L.~{Zhang}, G.~{Shi}, X.~{Li}, Nonlocally centralized sparse
  representation for image restoration, IEEE Transactions on Image Processing
  22~(4) (2013) 1620--1630.

\bibitem{6909762}
S.~{Gu}, L.~{Zhang}, W.~{Zuo}, X.~{Feng}, Weighted nuclear norm minimization
  with application to image denoising, in: 2014 IEEE Conference on Computer
  Vision and Pattern Recognition, 2014, pp. 2862--2869.

\bibitem{7194811}
X.~{Zeng}, W.~{Bian}, W.~{Liu}, J.~{Shen}, D.~{Tao}, Dictionary pair learning
  on grassmann manifolds for image denoising, IEEE Transactions on Image
  Processing 24~(11) (2015) 4556--4569.

\bibitem{7004012}
S.~K. {Sahoo}, A.~{Makur}, Enhancing image denoising by controlling noise
  incursion in learned dictionaries, IEEE Signal Processing Letters 22~(8)
  (2015) 1123--1126.

\bibitem{6566099}
S.~{Ravishankar}, Y.~{Bresler}, Learning doubly sparse transforms for images,
  IEEE Transactions on Image Processing 22~(12) (2013) 4598--4612.

\bibitem{8438535}
B.~{Wen}, S.~{Ravishankar}, Y.~{Bresler}, Vidosat: High-dimensional sparsifying
  transform learning for online video denoising, IEEE Transactions on Image
  Processing 28~(4) (2019) 1691--1704.

\bibitem{1203207}
{Sung Cheol Park}, {Min Kyu Park}, {Moon Gi Kang}, Super-resolution image
  reconstruction: a technical overview, IEEE Signal Processing Magazine 20~(3)
  (2003) 21--36.

\bibitem{8759425}
W.~{Wang}, J.~{Dong}, S.~{Niu}, Y.~{Chen}, Edge-guided semi-coupled dictionary
  learning super resolution for retina image, in: 2019 IEEE 16th International
  Symposium on Biomedical Imaging (ISBI 2019), 2019, pp. 1631--1634.

\bibitem{7816724}
X.~{Tian}, J.~{Chen}, A fast algorithm for single image super-resolution
  reconstruction via revised statistical prediction model, in: 2016
  International Conference on Information System and Artificial Intelligence
  (ISAI), 2016, pp. 305--309.

\bibitem{8532893}
Z.~{Hu}, T.~{Li}, Y.~{Yang}, X.~{Liu}, H.~{Zheng}, D.~{Liang}, Super-resolution
  pet image reconstruction with sparse representation, in: 2017 IEEE Nuclear
  Science Symposium and Medical Imaging Conference (NSS/MIC), 2017, pp. 1--3.

\bibitem{7351320}
J.~{Choi}, S.~{Bae}, M.~{Kim}, Single image super-resolution based on
  self-examples using context-dependent subpatches, in: 2015 IEEE International
  Conference on Image Processing (ICIP), 2015, pp. 2835--2839.

\bibitem{6588338}
A.~Jalali, P.~Ravikumar, S.~Sanghavi, A dirty model for multiple sparse
  regression, IEEE Transactions on Information Theory 59~(12) (2013)
  7947--7968.

\bibitem{7530147}
A.~H. {Shahana}, V.~{Preeja}, Survey on feature subset selection for high
  dimensional data, in: 2016 International Conference on Circuit, Power and
  Computing Technologies (ICCPCT), 2016, pp. 1--4.

\bibitem{7924323}
X.~{Wang}, Y.~{Gu}, Cross-label suppression: A discriminative and fast
  dictionary learning with group regularization, IEEE Transactions on Image
  Processing 26~(8) (2017) 3859--3873.

\bibitem{qi2018learning}
J.~{Qi}, W.~{Chen}, Learning a discriminative dictionary for classification
  with outliers, Signal Processing 152 (2018) 255--264.

\bibitem{6880772}
W.~{Chen}, I.~J. {Wassell}, M.~R.~D. {Rodrigues}, Dictionary design for
  distributed compressive sensing, IEEE Signal Processing Letters 22~(1) (2015)
  95--99.

\bibitem{5714407}
I.~{To\v{s}i\'{c}}, P.~{Frossard}, Dictionary learning, IEEE Signal Processing
  Magazine 28~(2) (2011) 27--38.

\bibitem{7779008}
S.~{Tariyal}, A.~{Majumdar}, R.~{Singh}, M.~{Vatsa}, Deep dictionary learning,
  IEEE Access 4 (2016) 10096--10109.

\bibitem{8982090}
X.~{Gong}, W.~{Chen}, J.~{Chen}, A low-rank tensor dictionary learning method
  for hyperspectral image denoising, IEEE Transactions on Signal Processing 68
  (2020) 1168--1180.

\bibitem{7914777}
X.~{Ding}, W.~{Chen}, I.~J. {Wassell}, Joint sensing matrix and sparsifying
  dictionary optimization for tensor compressive sensing, IEEE Transactions on
  Signal Processing 65~(14) (2017) 3632--3646.

\bibitem{candes2008restricted}
{E. J. Candes}, The restricted isometry property and its implications for
  compressed sensing, Comptes rendus mathematique 346~(9-10) (2008) 589--592.

\bibitem{blumensath2009iterative}
T.~{Blumensath}, M.~E. {Davies}, Iterative hard thresholding for compressed
  sensing, Applied and computational harmonic analysis 27~(3) (2009) 265--274.

\bibitem{4385788}
J.~A. {Tropp}, A.~C. {Gilbert}, Signal recovery from random measurements via
  orthogonal matching pursuit, IEEE Transactions on Information Theory 53~(12)
  (2007) 4655--4666.

\bibitem{donoho2009message}
D.~L. {Donoho}, A.~{Maleki}, A.~{Montanari}, Message-passing algorithms for
  compressed sensing, Proceedings of the National Academy of Sciences 106~(45)
  (2009) 18914--18919.

\bibitem{8074806}
M.~{Al-Shoukairi}, P.~{Schniter}, B.~D. {Rao}, A gamp-based low complexity
  sparse bayesian learning algorithm, IEEE Transactions on Signal Processing
  66~(2) (2018) 294--308.

\bibitem{gregor2010learning}
K.~{Gregor}, Y.~{LeCun}, Learning fast approximations of sparse coding, in:
  Proceedings of the 27th International Conference on International Conference
  on Machine Learning, 2010, pp. 399--406.

\bibitem{6247952}
H.~C. {Burger}, C.~J. {Schuler}, S.~{Harmeling}, {Image denoising: Can plain
  neural networks compete with {BM3D}?}, in: 2012 IEEE Conference on Computer
  Vision and Pattern Recognition, 2012, pp. 2392--2399.

\bibitem{6781616}
Y.~{Wang}, J.~{Morel}, Can a single image denoising neural network handle all
  levels of gaussian noise?, IEEE Signal Processing Letters 21~(9) (2014)
  1150--1153.

\bibitem{xin2016maximal}
B.~{Xin}, Y.~{Wang}, W.~{Gao}, D.~{Wipf}, B.~{Wang}, Maximal sparsity with deep
  networks?, in: Advances in Neural Information Processing Systems, 2016, pp.
  4340--4348.

\bibitem{wang2016learning}
Z.~{Wang}, Q.~{Ling}, T.~{Huang}, Learning deep $\ell_{0}$ encoders, in: AAAI
  Conference on Artificial Intelligence, 2016, pp. 2194--2200.

\bibitem{7905837}
M.~{Borgerding}, P.~{Schniter}, Onsager-corrected deep learning for sparse
  linear inverse problems, in: 2016 IEEE Global Conference on Signal and
  Information Processing (GlobalSIP), 2016, pp. 227--231.

\bibitem{7934066}
M.~{Borgerding}, P.~{Schniter}, S.~{Rangan}, Amp-inspired deep networks for
  sparse linear inverse problems, IEEE Transactions on Signal Processing
  65~(16) (2017) 4293--4308.

\bibitem{4378954}
K.~{Dabov}, A.~{Foi}, V.~{Katkovnik}, K.~{Egiazarian}, Color image denoising
  via sparse 3d collaborative filtering with grouping constraint in
  luminance-chrominance space, in: 2007 IEEE International Conference on Image
  Processing, Vol.~1, 2007, pp. I -- 313--I -- 316.

\bibitem{1710377}
M.~{Aharon}, M.~{Elad}, A.~{Bruckstein}, K-svd: An algorithm for designing
  overcomplete dictionaries for sparse representation, IEEE Transactions on
  Signal Processing 54~(11) (2006) 4311--4322.

\bibitem{8237387}
J.~{Xu}, L.~{Zhang}, D.~{Zhang}, X.~{Feng}, Multi-channel weighted nuclear norm
  minimization for real color image denoising, in: 2017 IEEE International
  Conference on Computer Vision (ICCV), 2017, pp. 1105--1113.

\bibitem{xu2018trilateral}
J.~{Xu}, L.~{Zhang}, D.~{Zhang}, A trilateral weighted sparse coding scheme for
  real-world image denoising, in: Proceedings of the European Conference on
  Computer Vision (ECCV), 2018, pp. 20--36.

\bibitem{guo2019toward}
S.~{Guo}, Z.~{Yan}, K.~{Zhang}, W.~{Zuo}, L.~{Zhang}, Toward convolutional
  blind denoising of real photographs, in: Proceedings of the IEEE Conference
  on Computer Vision and Pattern Recognition, 2019.

\bibitem{8099777}
T.~{Pl{\"o}tz}, S.~{Roth}, Benchmarking denoising algorithms with real
  photographs, in: 2017 IEEE Conference on Computer Vision and Pattern
  Recognition (CVPR), 2017, pp. 2750--2759.

\bibitem{xu2018real}
J.~{Xu}, H.~{Li}, Z.~{Liang}, D.~{Zhang}, L.~{Zhang}, Real-world noisy image
  denoising: A new benchmark, Real-world noisy image denoising: A new
  benchmark.

\bibitem{8103129}
M.~T. {McCann}, K.~H. {Jin}, M.~{Unser}, Convolutional neural networks for
  inverse problems in imaging: A review, IEEE Signal Processing Magazine 34~(6)
  (2017) 85--95.

\bibitem{8253590}
A.~{Lucas}, M.~{Iliadis}, R.~{Molina}, A.~K. {Katsaggelos}, Using deep neural
  networks for inverse problems in imaging: Beyond analytical methods, IEEE
  Signal Processing Magazine 35~(1) (2018) 20--36.

\bibitem{8723565}
W.~{Yang}, X.~{Zhang}, Y.~{Tian}, W.~{Wang}, J.~{Xue}, Q.~{Liao}, Deep learning
  for single image super-resolution: A brief review, IEEE Transactions on
  Multimedia 21~(12) (2019) 3106--3121.

\bibitem{8962949}
D.~{Liang}, J.~{Cheng}, Z.~{Ke}, L.~{Ying}, Deep magnetic resonance image
  reconstruction: Inverse problems meet neural networks, IEEE Signal Processing
  Magazine 37~(1) (2020) 141--151.

\bibitem{9084378}
G.~{Ongie}, A.~{Jalal}, C.~A. {Metzler}, R.~G. {Baraniuk}, A.~G. {Dimakis},
  R.~{Willett}, Deep learning techniques for inverse problems in imaging, IEEE
  Journal on Selected Areas in Information Theory 1~(1) (2020) 39--56.

\bibitem{arridge2019solving}
S.~{Arridge}, P.~{Maass}, O.~{{\"O}ktem}, C.~{Sch{\"o}nlieb}, Solving inverse
  problems using data-driven models, Acta Numerica 28 (2019) 1--174.

\bibitem{hershey2014deep}
J.~R. Hershey, J.~L. Roux, F.~Weninger, Deep unfolding: Model-based inspiration
  of novel deep architectures, arXiv preprint arXiv:1409.2574.

\bibitem{abadi2016tensorflow}
M.~{Abadi}, P.~{Barham}, J.~{Chen}, Z.~{Chen}, A.~{Davis}, J.~{Dean},
  M.~{Devin}, S.~{Ghemawat}, G.~{Irving}, M.~{Isard}, et~al., Tensorflow: a
  system for large-scale machine learning, in: OSDI, Vol.~16, 2016, pp.
  265--283.

\bibitem{paszke2019pytorch}
A.~{Paszke}, S.~{Gross}, F.~{Massa}, A.~{Lerer}, J.~{Bradbury}, G.~{Chanan},
  T.~{Killeen}, Z.~{Lin}, N.~{Gimelshein}, L.~{Antiga}, et~al., Pytorch: An
  imperative style, high-performance deep learning library, in: Advances in
  Neural Information Processing Systems, 2019, pp. 8024--8035.

\bibitem{daubechies2004iterative}
I.~{Daubechies}, M.~{Defrise}, C.~{De Mol}, An iterative thresholding algorithm
  for linear inverse problems with a sparsity constraint, Communications on
  Pure and Applied Mathematics: A Journal Issued by the Courant Institute of
  Mathematical Sciences 57~(11) (2004) 1413--1457.

\bibitem{7410480}
K.~{He}, X.~{Zhang}, S.~{Ren}, J.~{Sun}, Delving deep into rectifiers:
  Surpassing human-level performance on imagenet classification, in: 2015 IEEE
  International Conference on Computer Vision (ICCV), 2015, pp. 1026--1034.

\bibitem{8545232}
H.~{Zhang}, H.~{Shi}, W.~{Wang}, Cascade deep networks for sparse linear
  inverse problems, in: 2018 24th International Conference on Pattern
  Recognition (ICPR), 2018, pp. 812--817.

\bibitem{chen2018theoretical}
X.~{Chen}, J.~{Liu}, Z.~{Wang}, W.~{Yin}, Theoretical linear convergence of
  unfolded ista and its practical weights and thresholds, in: Advances in
  Neural Information Processing Systems, 2018, pp. 9061--9071.

\bibitem{aberdam2020ada}
A.~{Aberdam}, A.~{Golts}, M.~{Elad}, Ada-lista: Learned solvers adaptive to
  varying models, arXiv preprint arXiv:2001.08456.

\bibitem{liu2018alista}
J.~{Liu}, X.~{Chen}, Z.~{Wang}, W.~{Yin}, Alista: Analytic weights are as good
  as learned weights in lista.

\bibitem{ablin2019learning}
P.~{Ablin}, T.~{Moreau}, M.~{Massias}, A.~{Gramfort}, Learning step sizes for
  unfolded sparse coding, in: Advances in Neural Information Processing
  Systems, 2019, pp. 13100--13110.

\bibitem{scetbon2019deep}
M.~{Scetbon}, M.~{Elad}, P.~{Milanfar}, Deep k-svd denoising, arXiv preprint
  arXiv:1909.13164.

\bibitem{8695874}
D.~{Ito}, S.~{Takabe}, T.~{Wadayama}, Trainable ista for sparse signal
  recovery, IEEE Transactions on Signal Processing 67~(12) (2019) 3113--3125.

\bibitem{8683182}
M.~{Yao}, J.~{Dang}, Z.~{Zhang}, L.~{Wu}, Sure-tista: A signal recovery network
  for compressed sensing, in: ICASSP 2019 - 2019 IEEE International Conference
  on Acoustics, Speech and Signal Processing (ICASSP), 2019, pp. 3832--3836.

\bibitem{fan2018matrix}
J.~{Fan}, J.~{Cheng}, Matrix completion by deep matrix factorization, Neural
  Networks 98 (2018) 34--41.

\bibitem{arora2019implicit}
S.~{Arora}, N.~{Cohen}, W.~{Hu}, Y.~{Luo}, Implicit regularization in deep
  matrix factorization, in: Advances in Neural Information Processing Systems,
  2019, pp. 7413--7424.

\bibitem{8006797}
S.~{Rangan}, P.~{Schniter}, A.~K. {Fletcher}, Vector approximate message
  passing, in: 2017 IEEE International Symposium on Information Theory (ISIT),
  2017, pp. 1588--1592.

\bibitem{9054280}
J.~{Pu}, Y.~{Panagakis}, M.~{Pantic}, Learning differentiable sparse and low
  rank networks for audio-visual object localization, in: ICASSP 2020 - 2020
  IEEE International Conference on Acoustics, Speech and Signal Processing
  (ICASSP), 2020, pp. 8668--8672.

\bibitem{8554065}
J.~{Lewis D.}, V.~{Singhal}, A.~{Majumdar}, Solving inverse problems in imaging
  via deep dictionary learning, IEEE Access 7 (2019) 37039--37049.

\bibitem{singhal2020reconstructing}
V.~{Singhal}, A.~{Majumdar}, Reconstructing multi-echo magnetic resonance
  images via structured deep dictionary learning, Neurocomputing.

\bibitem{singhal2019a}
V.~{Singhal}, A.~{Majumdar}, A domain adaptation approach to solve inverse
  problems in imaging via coupled deep dictionary learning, Pattern Recognition
  (2019) 107163.

\bibitem{8461651}
J.~{Huang}, P.~L. {Dragotti}, A deep dictionary model for image
  super-resolution, in: 2018 IEEE International Conference on Acoustics, Speech
  and Signal Processing (ICASSP), 2018, pp. 6777--6781.

\bibitem{7341021}
X.~{Wang}, Q.~{Tao}, L.~{Wang}, D.~{Li}, M.~{Zhang}, Deep convolutional
  architecture for natural image denoising, in: 2015 International Conference
  on Wireless Communications Signal Processing (WCSP), 2015, pp. 1--4.

\bibitem{8365806}
K.~{Zhang}, W.~{Zuo}, L.~{Zhang}, {FFDNet}: Toward a fast and flexible solution
  for cnn-based image denoising, IEEE Transactions on Image Processing 27~(9)
  (2018) 4608--4622.

\bibitem{7472127}
X.~{Zhang}, R.~{Wu}, Fast depth image denoising and enhancement using a deep
  convolutional network, in: 2016 IEEE International Conference on Acoustics,
  Speech and Signal Processing (ICASSP), 2016, pp. 2499--2503.

\bibitem{8435923}
Y.~{Chang}, L.~{Yan}, H.~{Fang}, S.~{Zhong}, W.~{Liao}, Hsi-denet:
  Hyperspectral image restoration via convolutional neural network, IEEE
  Transactions on Geoscience and Remote Sensing 57~(2) (2019) 667--682.

\bibitem{7780459}
K.~{He}, X.~{Zhang}, S.~{Ren}, J.~{Sun}, Deep residual learning for image
  recognition, in: 2016 IEEE Conference on Computer Vision and Pattern
  Recognition (CVPR), 2016, pp. 770--778.

\bibitem{8454887}
Q.~{Yuan}, Q.~{Zhang}, J.~{Li}, H.~{Shen}, L.~{Zhang}, Hyperspectral image
  denoising employing a spatial¨cspectral deep residual convolutional neural
  network, IEEE Transactions on Geoscience and Remote Sensing 57~(2) (2019)
  1205--1218.

\bibitem{8944535}
A.~{Panda}, R.~{Naskar}, S.~{Rajbans}, S.~{Pal}, A 3d wide residual network
  with perceptual loss for brain mri image denoising, in: 2019 10th
  International Conference on Computing, Communication and Networking
  Technologies (ICCCNT), 2019, pp. 1--7.

\bibitem{8923780}
W.~{Xiaowei}, J.~{Xiong}, R.~{Wang}, W.~J. {Sori}, J.~{Wang}, L.~{Shaohui},
  F.~{Jiang}, Fs-net: Medical image denoising via local receptive field
  smoothing network, in: 2019 IEEE Fourth International Conference on Data
  Science in Cyberspace (DSC), 2019, pp. 70--76.

\bibitem{7839189}
K.~{Zhang}, W.~{Zuo}, Y.~{Chen}, D.~{Meng}, L.~{Zhang}, Beyond a gaussian
  denoiser: Residual learning of deep cnn for image denoising, IEEE
  Transactions on Image Processing 26~(7) (2017) 3142--3155.

\bibitem{8372095}
T.~{Wang}, M.~{Sun}, K.~{Hu}, Dilated deep residual network for image
  denoising, in: 2017 IEEE 29th International Conference on Tools with
  Artificial Intelligence (ICTAI), 2017, pp. 1272--1279.

\bibitem{8741329}
Y.~{Su}, Q.~{Lian}, X.~{Zhang}, B.~{Shi}, X.~{Fan}, Multi-scale cross-path
  concatenation residual network for poisson denoising, IET Image Processing
  13~(8) (2019) 1295--1303.

\bibitem{huber1992robust}
P.~J. {Huber}, Robust estimation of a location parameter, in: Breakthroughs in
  statistics, 1992, pp. 492--518.

\bibitem{7115171}
C.~{Dong}, C.~C. {Loy}, K.~{He}, X.~{Tang}, Image super-resolution using deep
  convolutional networks, IEEE Transactions on Pattern Analysis and Machine
  Intelligence 38~(2) (2016) 295--307.

\bibitem{7780576}
W.~{Shi}, J.~{Caballero}, F.~{Husz¨¢r}, J.~{Totz}, A.~P. {Aitken}, R.~{Bishop},
  D.~{Rueckert}, Z.~{Wang}, Real-time single image and video super-resolution
  using an efficient sub-pixel convolutional neural network, in: 2016 IEEE
  Conference on Computer Vision and Pattern Recognition (CVPR), 2016, pp.
  1874--1883.

\bibitem{dong2016accelerating}
C.~{Dong}, L.~C. {Chen}, X.~{Tang}, Accelerating the super-resolution
  convolutional neural network, in: European conference on computer vision,
  2016, pp. 391--407.

\bibitem{8100101}
W.~{Lai}, J.~{Huang}, N.~{Ahuja}, M.~{Yang}, Deep laplacian pyramid networks
  for fast and accurate super-resolution, in: 2017 IEEE Conference on Computer
  Vision and Pattern Recognition (CVPR), 2017, pp. 5835--5843.

\bibitem{8578277}
M.~{Haris}, G.~{Shakhnarovich}, N.~{Ukita}, Deep back-projection networks for
  super-resolution, in: 2018 IEEE/CVF Conference on Computer Vision and Pattern
  Recognition, 2018, pp. 1664--1673.

\bibitem{7780551}
J.~{Kim}, J.~K. {Lee}, K.~M. {Lee}, Accurate image super-resolution using very
  deep convolutional networks, in: 2016 IEEE Conference on Computer Vision and
  Pattern Recognition (CVPR), 2016, pp. 1646--1654.

\bibitem{8014885}
B.~{Lim}, S.~{Son}, H.~{Kim}, S.~{Nah}, K.~M. {Lee}, Enhanced deep residual
  networks for single image super-resolution, in: 2017 IEEE Conference on
  Computer Vision and Pattern Recognition Workshops (CVPRW), 2017, pp.
  1132--1140.

\bibitem{8099781}
Y.~{Tai}, J.~{Yang}, X.~{Liu}, Image super-resolution via deep recursive
  residual network, in: 2017 IEEE Conference on Computer Vision and Pattern
  Recognition (CVPR), 2017, pp. 2790--2798.

\bibitem{8578442}
K.~{Zhang}, W.~{Zuo}, L.~{Zhang}, Learning a single convolutional
  super-resolution network for multiple degradations, in: 2018 IEEE/CVF
  Conference on Computer Vision and Pattern Recognition, 2018, pp. 3262--3271.

\bibitem{8578427}
A.~{Shocher}, N.~{Cohen}, M.~{Irani}, Zero-shot super-resolution using deep
  internal learning, in: 2018 IEEE/CVF Conference on Computer Vision and
  Pattern Recognition, 2018, pp. 3118--3126.

\bibitem{sun2016deep}
Y.~{Yang}, J.~{Sun}, H.~{Li}, Z.~{Xu}, et~al., Deep {ADMM-Net} for compressive
  sensing {MRI}, in: Advances in neural information processing systems, 2016,
  pp. 10--18.

\bibitem{8578294}
J.~{Zhang}, B.~{Ghanem}, {ISTA-Net}: Interpretable optimization-inspired deep
  network for image compressive sensing, in: 2018 IEEE/CVF Conference on
  Computer Vision and Pattern Recognition, 2018, pp. 1828--1837.

\bibitem{dong2014learning}
C.~{Dong}, C.~C. {Loy}, K.~{He}, X.~{Tang}, Learning a deep convolutional
  network for image super-resolution, in: European conference on computer
  vision, 2014, pp. 184--199.

\bibitem{7410407}
Z.~{Wang}, D.~{Liu}, J.~{Yang}, W.~{Han}, T.~{Huang}, Deep networks for image
  super-resolution with sparse prior, in: 2015 IEEE International Conference on
  Computer Vision (ICCV), 2015, pp. 370--378.

\bibitem{8100106}
S.~{Lefkimmiatis}, Non-local color image denoising with convolutional neural
  networks, in: 2017 IEEE Conference on Computer Vision and Pattern Recognition
  (CVPR), 2017, pp. 5882--5891.

\bibitem{yu2015multi}
F.~{Yu}, V.~{Koltun}, Multi-scale context aggregation by dilated convolutions,
  arXiv preprint arXiv:1511.07122.

\bibitem{bevilacqua2012low}
R.~M. M.~{Bevilacqua}, A.~{Roumy}, C.~{Guillemot}, M.~L. {Alberi-Morel},
  Low-complexity single-image super-resolution based on nonnegative neighbor
  embedding, IEEE Signal Processing Magazine.

\bibitem{zeyde2010single}
R.~{Zeyde}, M.~{Elad}, M.~{Protter}, On single image scale-up using
  sparse-representations, in: International conference on curves and surfaces,
  2010, pp. 711--730.

\bibitem{8578360}
Y.~{Zhang}, Y.~{Tian}, Y.~{Kong}, B.~{Zhong}, Y.~{Fu}, Residual dense network
  for image super-resolution, in: 2018 IEEE/CVF Conference on Computer Vision
  and Pattern Recognition, 2018, pp. 2472--2481.

\bibitem{8579082}
V.~{Lempitsky}, A.~{Vedaldi}, D.~{Ulyanov}, Deep image prior, in: 2018 IEEE/CVF
  Conference on Computer Vision and Pattern Recognition, 2018, pp. 9446--9454.

\bibitem{9022040}
O.~{Sidorov}, J.~Y. {Hardeberg}, Deep hyperspectral prior: Single-image
  denoising, inpainting, super-resolution, in: 2019 IEEE/CVF International
  Conference on Computer Vision Workshop (ICCVW), 2019, pp. 3844--3851.

\bibitem{8581448}
K.~{Gong}, C.~{Catana}, J.~{Qi}, Q.~{Li}, Pet image reconstruction using deep
  image prior, IEEE Transactions on Medical Imaging 38~(7) (2019) 1655--1665.

\bibitem{9060001}
K.~{Gong}, K.~{Kim}, D.~{Wu}, M.~K. {Kalra}, Q.~{Li}, Low-dose dual energy ct
  image reconstruction using non-local deep image prior, in: 2019 IEEE Nuclear
  Science Symposium and Medical Imaging Conference (NSS/MIC), 2019, pp. 1--2.

\bibitem{van2018compressed}
D.~{Van Veen}, A.~{Jalal}, M.~{Soltanolkotabi}, E.~{Price}, S.~{Vishwanath},
  A.~G. {Dimakis}, Compressed sensing with deep image prior and learned
  regularization, arXiv preprint arXiv:1806.06438.

\bibitem{8968714}
J.~{Ren}, J.~{Liang}, Y.~{Zhao}, Soil ph measurement based on compressive
  sensing and deep image prior, IEEE Transactions on Emerging Topics in
  Computational Intelligence 4~(1) (2020) 74--82.

\bibitem{8682856}
J.~{Liu}, Y.~{Sun}, X.~{Xu}, U.~S. {Kamilov}, Image restoration using total
  variation regularized deep image prior, in: ICASSP 2019 - 2019 IEEE
  International Conference on Acoustics, Speech and Signal Processing (ICASSP),
  2019, pp. 7715--7719.

\bibitem{9054218}
G.~{Jagatap}, C.~{Hegde}, High dynamic range imaging using deep image priors,
  in: ICASSP 2020 - 2020 IEEE International Conference on Acoustics, Speech and
  Signal Processing (ICASSP), 2020, pp. 9289--9293.

\bibitem{jagatap2019algorithmic}
G.~{Jagatap}, C.~{Hegde}, Algorithmic guarantees for inverse imaging with
  untrained network priors, in: Advances in Neural Information Processing
  Systems, 2019, pp. 14832--14842.

\bibitem{dittmer2019regularization}
S.~{Dittmer}, T.~{Kluth}, P.~{Maass}, D.~O. {Baguer}, Regularization by
  architecture: A deep prior approach for inverse problems, Journal of
  Mathematical Imaging and Vision (2019) 1--15.

\bibitem{heckel2020compressive}
R.~{Heckel}, M.~{Soltanolkotabi}, Compressive sensing with un-trained neural
  networks: Gradient descent finds the smoothest approximation, arXiv preprint
  arXiv:2005.03991.

\bibitem{8550778}
Y.~{Yang}, J.~{Sun}, H.~{Li}, Z.~{Xu}, Admm-csnet: A deep learning approach for
  image compressive sensing, IEEE Transactions on Pattern Analysis and Machine
  Intelligence 42~(3) (2020) 521--538.

\bibitem{7457256}
C.~A. {Metzler}, A.~{Maleki}, R.~G. {Baraniuk}, From denoising to compressed
  sensing, IEEE Transactions on Information Theory 62~(9) (2016) 5117--5144.

\bibitem{metzler2017learned}
C.~{Metzler}, A.~{Mousavi}, R.~{Baraniuk}, Learned d-amp: Principled neural
  network based compressive image recovery, in: Advances in Neural Information
  Processing Systems, 2017, pp. 1772--1783.

\bibitem{8836615}
O.~{Solomon}, R.~{Cohen}, Y.~{Zhang}, Y.~{Yang}, Q.~{He}, J.~{Luo}, R.~J.~G.
  {van Sloun}, Y.~C. {Eldar}, Deep unfolded robust pca with application to
  clutter suppression in ultrasound, IEEE Transactions on Medical Imaging
  39~(4) (2020) 1051--1063.

\bibitem{4587647}
{Jianchao Yang}, J.~{Wright}, T.~{Huang}, {Yi Ma}, Image super-resolution as
  sparse representation of raw image patches, in: 2008 IEEE Conference on
  Computer Vision and Pattern Recognition, 2008, pp. 1--8.

\bibitem{5466111}
J.~{Yang}, J.~{Wright}, T.~S. {Huang}, Y.~{Ma}, Image super-resolution via
  sparse representation, IEEE Transactions on Image Processing 19~(11) (2010)
  2861--2873.

\bibitem{7466062}
D.~{Liu}, Z.~{Wang}, B.~{Wen}, J.~{Yang}, W.~{Han}, T.~S. {Huang}, Robust
  single image super-resolution via deep networks with sparse prior, IEEE
  Transactions on Image Processing 25~(7) (2016) 3194--3207.

\bibitem{8578436}
S.~{Lefkimmiatis}, Universal denoising networks: A novel {CNN} architecture for
  image denoising, in: 2018 IEEE/CVF Conference on Computer Vision and Pattern
  Recognition, 2018, pp. 3204--3213.

\bibitem{7527621}
Y.~{Chen}, T.~{Pock}, Trainable nonlinear reaction diffusion: A flexible
  framework for fast and effective image restoration, IEEE Transactions on
  Pattern Analysis and Machine Intelligence 39~(6) (2017) 1256--1272.

\bibitem{ngiam2011multimodal}
J.~{Ngiam}, A.~{Khosla}, M.~{Kim}, J.~{Nam}, H.~{Lee}, A.~Y. {Ng}, Multimodal
  deep learning, in: ICML, 2011.

\bibitem{8803313}
I.~{Marivani}, E.~{Tsiligianni}, B.~{Cornelis}, N.~{Deligiannis}, Learned
  multimodal convolutional sparse coding for guided image super-resolution, in:
  2019 IEEE International Conference on Image Processing (ICIP), 2019, pp.
  2891--2895.

\bibitem{8903106}
I.~{Marivani}, E.~{Tsiligianni}, B.~{Cornelis}, N.~{Deligiannis}, Multimodal
  image super-resolution via deep unfolding with side information, in: 2019
  27th European Signal Processing Conference (EUSIPCO), 2019, pp. 1--5.

\bibitem{8858035}
X.~{Deng}, P.~L. {Dragotti}, Deep coupled ista network for multi-modal image
  super-resolution, IEEE Transactions on Image Processing 29 (2020) 1683--1698.

\bibitem{falvo2019multimodal}
A.~{Falvo}, D.~{Comminiello}, S.~{Scardapane}, G.~{Finesi}, M.~{Scarpiniti},
  A.~{Uncini}, A multimodal deep network for the reconstruction of t2w mr
  images, arXiv preprint arXiv:1908.03009.

\bibitem{8844082}
E.~{Tsiligianni}, N.~{Deligiannis}, Deep coupled-representation learning for
  sparse linear inverse problems with side information, IEEE Signal Processing
  Letters 26~(12) (2019) 1768--1772.

\bibitem{7979523}
K.~{Qiu}, X.~{Mao}, X.~{Shen}, X.~{Wang}, T.~{Li}, Y.~{Gu}, Time-varying graph
  signal reconstruction, IEEE Journal of Selected Topics in Signal Processing
  11~(6) (2017) 870--883.

\bibitem{7457684}
H.~{Palangi}, R.~{Ward}, L.~{Deng}, Distributed compressive sensing: A deep
  learning approach, IEEE Transactions on Signal Processing 64~(17) (2016)
  4504--4518.

\bibitem{7905830}
H.~{Palangi}, R.~{Ward}, L.~{Deng}, Reconstruction of sparse vectors in
  compressive sensing with multiple measurement vectors using bidirectional
  long short-term memory, in: 2016 IEEE Global Conference on Signal and
  Information Processing (GlobalSIP), 2016, pp. 192--196.

\bibitem{lyu2019block}
C.~{Lyu}, Z.~{Liu}, L.~{Yu}, Block-sparsity recovery via recurrent neural
  network, Signal Processing 154 (2019) 129--135.

\bibitem{8296560}
D.~{Li}, Y.~{Liu}, Z.~{Wang}, Video super-resolution using motion compensation
  and residual bidirectional recurrent convolutional network, in: 2017 IEEE
  International Conference on Image Processing (ICIP), 2017, pp. 1642--1646.

\bibitem{hinton2012neural}
G.~{Hinton}, N.~{Srivastava}, K.~{Swersky}, Neural networks for machine
  learning lecture 6a overview of mini-batch gradient descent, Cited on 14~(8).

\bibitem{8282261}
B.~{Lim}, K.~M. {Lee}, Deep recurrent resnet for video super-resolution, in:
  2017 Asia-Pacific Signal and Information Processing Association Annual Summit
  and Conference (APSIPA ASC), 2017, pp. 1452--1455.

\bibitem{7919264}
Y.~{Huang}, W.~{Wang}, L.~{Wang}, Video super-resolution via bidirectional
  recurrent convolutional networks, IEEE Transactions on Pattern Analysis and
  Machine Intelligence 40~(4) (2018) 1015--1028.

\bibitem{8501928}
D.~{Li}, Y.~{Liu}, Z.~{Wang}, Video super-resolution using non-simultaneous
  fully recurrent convolutional network, IEEE Transactions on Image Processing
  28~(3) (2019) 1342--1355.

\bibitem{8953613}
M.~{Haris}, G.~{Shakhnarovich}, N.~{Ukita}, Recurrent back-projection network
  for video super-resolution, in: 2019 IEEE/CVF Conference on Computer Vision
  and Pattern Recognition (CVPR), 2019, pp. 3892--3901.

\bibitem{7780550}
J.~{Kim}, J.~K. {Lee}, K.~M. {Lee}, Deeply-recursive convolutional network for
  image super-resolution, in: 2016 IEEE Conference on Computer Vision and
  Pattern Recognition (CVPR), 2016, pp. 1637--1645.

\bibitem{8425771}
X.~{Yang}, H.~{Mei}, J.~{Zhang}, K.~{Xu}, B.~{Yin}, Q.~{Zhang}, X.~{Wei}, Drfn:
  Deep recurrent fusion network for single-image super-resolution with large
  factors, IEEE Transactions on Multimedia 21~(2) (2019) 328--337.

\bibitem{8579237}
Z.~{Wang}, P.~{Yi}, K.~{Jiang}, J.~{Jiang}, Z.~{Han}, T.~{Lu}, J.~{Ma},
  Multi-memory convolutional neural network for video super-resolution, IEEE
  Transactions on Image Processing 28~(5) (2019) 2530--2544.

\bibitem{8356655}
Y.~{Wang}, F.~{Liu}, K.~{Zhang}, G.~{Hou}, Z.~{Sun}, T.~{Tan}, Lfnet: A novel
  bidirectional recurrent convolutional neural network for light-field image
  super-resolution, IEEE Transactions on Image Processing 27~(9) (2018)
  4274--4286.

\bibitem{8986369}
W.~{Wang}, C.~{Pang}, Z.~{Liu}, R.~{Lan}, X.~{Luo}, Srgnet: A gru based feature
  fusion network for image denoising, in: 2019 International Symposium on
  Intelligent Signal Processing and Communication Systems (ISPACS), 2019, pp.
  1--2.

\bibitem{8425639}
C.~{Qin}, J.~{Schlemper}, J.~{Caballero}, A.~N. {Price}, J.~V. {Hajnal},
  D.~{Rueckert}, Convolutional recurrent neural networks for dynamic mr image
  reconstruction, IEEE Transactions on Medical Imaging 38~(1) (2019) 280--290.

\bibitem{putzky2017recurrent}
P.~{Putzky}, M.~{Welling}, Recurrent inference machines for solving inverse
  problems, arXiv preprint arXiv:1706.04008.

\bibitem{7952977}
S.~{Wisdom}, T.~{Powers}, J.~{Pitton}, L.~{Atlas}, Building recurrent networks
  by unfolding iterative thresholding for sequential sparse recovery, in: 2017
  IEEE International Conference on Acoustics, Speech and Signal Processing
  (ICASSP), 2017, pp. 4346--4350.

\bibitem{8803281}
H.~D. {Le}, H.~{Van Luong}, N.~{Deligiannis}, Designing recurrent neural
  networks by unfolding an l1-l1 minimization algorithm, in: 2019 IEEE
  International Conference on Image Processing (ICIP), 2019, pp. 2329--2333.

\bibitem{zhou2018sc2net}
J.~{Zhou}, K.~{Di}, J.~{Du}, X.~{Peng}, H.~{Yang}, S.~{Pan}, I.~W. {Tsang},
  Y.~{Liu}, Z.~{Qin}, R.~S.~M. {Goh}, Sc2net: Sparse lstms for sparse coding,
  in: Thirty-Second AAAI Conference on Artificial Intelligence, 2018.

\bibitem{zeiler2012adadelta}
D.~M. {Zeiler}, Adadelta: an adaptive learning rate method, arXiv preprint
  arXiv:1212.5701.

\bibitem{8856290}
C.~{Yang}, H.~{Lan}, F.~{Gao}, Accelerated photoacoustic tomography
  reconstruction via recurrent inference machines, in: 2019 41st Annual
  International Conference of the IEEE Engineering in Medicine and Biology
  Society (EMBC), 2019, pp. 6371--6374.

\bibitem{he2017bayesian}
H.~{He}, B.~{Xin}, S.~{Ikehata}, D.~{Wipf}, From bayesian sparsity to gated
  recurrent nets, in: Advances in Neural Information Processing Systems, 2017,
  pp. 5554--5564.

\bibitem{hinton1994autoencoders}
G.~E. {Hinton}, R.~S. {Zemel}, Autoencoders, minimum description length and
  helmholtz free energy, in: Advances in neural information processing systems,
  1994, pp. 3--10.

\bibitem{kingma2013auto}
D.~P. {Kingma}, M.~{Welling}, Auto-encoding variational bayes, arXiv preprint
  arXiv:1312.6114.

\bibitem{8383709}
A.~{Majumdar}, Blind denoising autoencoder, IEEE Transactions on Neural
  Networks and Learning Systems 30~(1) (2019) 312--317.

\bibitem{8438540}
A.~{Creswell}, A.~A. {Bharath}, Denoising adversarial autoencoders, IEEE
  Transactions on Neural Networks and Learning Systems 30~(4) (2019) 968--984.

\bibitem{8038046}
A.~{Ali}, F.~{Yangyu}, Automatic modulation classification using deep learning
  based on sparse autoencoders with nonnegativity constraints, IEEE Signal
  Processing Letters 24~(11) (2017) 1626--1630.

\bibitem{8648840}
M.~H. {Shah}, X.~{Dang}, Robust approach for amc in frequency selective fading
  scenarios using unsupervised sparse-autoencoder-based deep neural network,
  IET Communications 13~(4) (2019) 423--432.

\bibitem{8065033}
X.~{Zhang}, Y.~{Liang}, C.~{Li}, N.~{Huyan}, L.~{Jiao}, H.~{Zhou}, Recursive
  autoencoders-based unsupervised feature learning for hyperspectral image
  classification, IEEE Geoscience and Remote Sensing Letters 14~(11) (2017)
  1928--1932.

\bibitem{7286736}
J.~{Geng}, J.~{Fan}, H.~{Wang}, X.~{Ma}, B.~{Li}, F.~{Chen}, High-resolution
  sar image classification via deep convolutional autoencoders, IEEE Geoscience
  and Remote Sensing Letters 12~(11) (2015) 2351--2355.

\bibitem{8421020}
J.~{Feng}, L.~{Liu}, X.~{Cao}, L.~{Jiao}, T.~{Sun}, X.~{Zhang}, Marginal
  stacked autoencoder with adaptively-spatial regularization for hyperspectral
  image classification, IEEE Journal of Selected Topics in Applied Earth
  Observations and Remote Sensing 11~(9) (2018) 3297--3311.

\bibitem{xie2012image}
J.~{Xie}, L.~{Xu}, E.~{Chen}, Image denoising and inpainting with deep neural
  networks, in: Advances in neural information processing systems, 2012, pp.
  341--349.

\bibitem{le2011optimization}
Q.~V. {Le}, J.~{Ngiam}, A.~{Coates}, A.~{Lahiri}, B.~{Prochnow}, A.~Y. {Ng}, On
  optimization methods for deep learning, in: Proceedings of the 28th
  International Conference on International Conference on Machine Learning,
  2011, pp. 265--272.

\bibitem{agostinelli2013adaptive}
F.~{Agostinelli}, M.~R. {Anderson}, H.~{Lee}, Adaptive multi-column deep neural
  networks with application to robust image denoising, in: Advances in Neural
  Information Processing Systems, 2013, pp. 1493--1501.

\bibitem{cho2013simple}
K.~{Cho}, Simple sparsification improves sparse denoising autoencoders in
  denoising highly corrupted images, in: International Conference on Machine
  Learning, 2013, pp. 432--440.

\bibitem{7279746}
X.~{Ye}, L.~{Wang}, H.~{Xing}, L.~{Huang}, Denoising hybrid noises in image
  with stacked autoencoder, in: 2015 IEEE International Conference on
  Information and Automation, 2015, pp. 2720--2724.

\bibitem{7836672}
L.~{Gondara}, Medical image denoising using convolutional denoising
  autoencoders, in: 2016 IEEE 16th International Conference on Data Mining
  Workshops (ICDMW), 2016, pp. 241--246.

\bibitem{7339460}
K.~{Zeng}, J.~{Yu}, R.~{Wang}, C.~{Li}, D.~{Tao}, Coupled deep autoencoder for
  single image super-resolution, IEEE Transactions on Cybernetics 47~(1) (2017)
  27--37.

\bibitem{8758165}
Z.~{Shao}, L.~{Wang}, Z.~{Wang}, J.~{Deng}, Remote sensing image
  super-resolution using sparse representation and coupled sparse autoencoder,
  IEEE Journal of Selected Topics in Applied Earth Observations and Remote
  Sensing 12~(8) (2019) 2663--2674.

\bibitem{8553263}
K.~{Gupta}, B.~{Bhowmick}, Coupled autoencoder based reconstruction of images
  from compressively sampled measurements, in: 2018 26th European Signal
  Processing Conference (EUSIPCO), 2018, pp. 1067--1071.

\bibitem{8462313}
H.~{Sreter}, R.~{Giryes}, Learned convolutional sparse coding, in: 2018 IEEE
  International Conference on Acoustics, Speech and Signal Processing (ICASSP),
  2018, pp. 2191--2195.

\bibitem{7797130}
H.~{Zhao}, O.~{Gallo}, I.~{Frosio}, J.~{Kautz}, Loss functions for image
  restoration with neural networks, IEEE Transactions on Computational Imaging
  3~(1) (2017) 47--57.

\bibitem{jalali2019using}
S.~{Jalali}, X.~{Yuan}, Using auto-encoders for solving ill-posed linear
  inverse problems, arXiv preprint arXiv:1901.05045.

\bibitem{vincent2008extracting}
P.~{Vincent}, H.~{Larochelle}, Y.~{Bengio}, P.~{Manzagol}, Extracting and
  composing robust features with denoising autoencoders, in: Proceedings of the
  25th international conference on Machine learning, 2008, pp. 1096--1103.

\bibitem{mehta2017rodeo}
J.~{Mehta}, A.~{Majumdar}, Rodeo: robust de-aliasing autoencoder for real-time
  medical image reconstruction, Pattern Recognition 63 (2017) 499--510.

\bibitem{barello2018sparse}
G.~{Barello}, A.~{Charles}, J.~{Pillow}, Sparse-coding variational
  auto-encoders, bioRxiv.

\bibitem{7010964}
P.~{Sprechmann}, A.~M. {Bronstein}, G.~{Sapiro}, Learning efficient sparse and
  low rank models, IEEE Transactions on Pattern Analysis and Machine
  Intelligence 37~(9) (2015) 1821--1833.

\bibitem{8578431}
J.~{Chen}, J.~{Chen}, H.~{Chao}, M.~{Yang}, Image blind denoising with
  generative adversarial network based noise modeling, in: 2018 IEEE/CVF
  Conference on Computer Vision and Pattern Recognition, 2018, pp. 3155--3164.

\bibitem{8546246}
K.~{Wu}, C.~{Zhang}, Deep generative adversarial networks for the sparse signal
  denoising, in: 2018 24th International Conference on Pattern Recognition
  (ICPR), 2018, pp. 1127--1132.

\bibitem{8340157}
Q.~{Yang}, P.~{Yan}, Y.~{Zhang}, H.~{Yu}, Y.~{Shi}, X.~{Mou}, M.~K. {Kalra},
  Y.~{Zhang}, L.~{Sun}, G.~{Wang}, Low-dose ct image denoising using a
  generative adversarial network with wasserstein distance and perceptual loss,
  IEEE Transactions on Medical Imaging 37~(6) (2018) 1348--1357.

\bibitem{8710893}
A.~{Alsaiari}, R.~{Rustagi}, A.~{Alhakamy}, M.~M. {Thomas}, A.~G. {Forbes},
  Image denoising using a generative adversarial network, in: 2019 IEEE 2nd
  International Conference on Information and Computer Technologies (ICICT),
  2019, pp. 126--132.

\bibitem{8941108}
L.~{Yang}, H.~{Shangguan}, X.~{Zhang}, A.~{Wang}, Z.~{Han}, High-frequency
  sensitive generative adversarial network for low-dose ct image denoising,
  IEEE Access 8 (2020) 930--943.

\bibitem{7934380}
J.~M. {Wolterink}, T.~{Leiner}, M.~A. {Viergever}, I.~{I\v{s}gum}, Generative
  adversarial networks for noise reduction in low-dose ct, IEEE Transactions on
  Medical Imaging 36~(12) (2017) 2536--2545.

\bibitem{8899202}
F.~{Gu}, H.~{Zhang}, C.~{Wang}, F.~{Wu}, Sar image super-resolution based on
  noise-free generative adversarial network, in: IGARSS 2019 - 2019 IEEE
  International Geoscience and Remote Sensing Symposium, 2019, pp. 2575--2578.

\bibitem{8546286}
L.~{Chen}, W.~{Dan}, L.~{Cao}, C.~{Wang}, J.~{Li}, Joint denoising and
  super-resolution via generative adversarial training, in: 2018 24th
  International Conference on Pattern Recognition (ICPR), 2018, pp. 2753--2758.

\bibitem{8736838}
C.~{You}, G.~{Li}, Y.~{Zhang}, X.~{Zhang}, H.~{Shan}, M.~{Li}, S.~{Ju},
  Z.~{Zhao}, Z.~{Zhang}, W.~{Cong}, M.~W. {Vannier}, P.~K. {Saha}, E.~A.
  {Hoffman}, G.~{Wang}, Ct super-resolution gan constrained by the identical,
  residual, and cycle learning ensemble (gan-circle), IEEE Transactions on
  Medical Imaging 39~(1) (2020) 188--203.

\bibitem{8099502}
C.~{Ledig}, L.~{Theis}, F.~{Husz¨¢r}, J.~{Caballero}, A.~{Cunningham},
  A.~{Acosta}, A.~{Aitken}, A.~{Tejani}, J.~{Totz}, Z.~{Wang}, W.~{Shi},
  Photo-realistic single image super-resolution using a generative adversarial
  network, in: 2017 IEEE Conference on Computer Vision and Pattern Recognition
  (CVPR), 2017, pp. 105--114.

\bibitem{8553719}
K.~{Gopan}, G.~S. {Kumar}, Video super resolution with generative adversarial
  network, in: 2018 2nd International Conference on Trends in Electronics and
  Informatics (ICOEI), 2018, pp. 1489--1493.

\bibitem{8900228}
R.~{Jiang}, X.~{Li}, A.~{Gao}, L.~{Li}, H.~{Meng}, S.~{Yue}, L.~{Zhang},
  Learning spectral and spatial features based on generative adversarial
  network for hyperspectral image super-resolution, in: IGARSS 2019 - 2019 IEEE
  International Geoscience and Remote Sensing Symposium, 2019, pp. 3161--3164.

\bibitem{8400496}
J.~M. {Haut}, R.~{Fernandez-Beltran}, M.~E. {Paoletti}, J.~{Plaza}, A.~{Plaza},
  F.~{Pla}, A new deep generative network for unsupervised remote sensing
  single-image super-resolution, IEEE Transactions on Geoscience and Remote
  Sensing 56~(11) (2018) 6792--6810.

\bibitem{8575264}
Y.~{Yuan}, S.~{Liu}, J.~{Zhang}, Y.~{Zhang}, C.~{Dong}, L.~{Lin}, Unsupervised
  image super-resolution using cycle-in-cycle generative adversarial networks,
  in: 2018 IEEE/CVF Conference on Computer Vision and Pattern Recognition
  Workshops (CVPRW), 2018, pp. 814--81409.

\bibitem{8825849}
Y.~{Zhang}, S.~{Liu}, C.~{Dong}, X.~{Zhang}, Y.~{Yuan}, Multiple cycle-in-cycle
  generative adversarial networks for unsupervised image super-resolution, IEEE
  Transactions on Image Processing 29 (2020) 1101--1112.

\bibitem{8803711}
B.~{Chen}, T.~{Liu}, K.~{Liu}, H.~{Liu}, S.~{Pei}, Image super-resolution using
  complex dense block on generative adversarial networks, in: 2019 IEEE
  International Conference on Image Processing (ICIP), 2019, pp. 2866--2870.

\bibitem{8237620}
W.~{Chen}, N.~{Song}, Low-rank tensor completion: A pseudo-bayesian learning
  approach, in: 2017 IEEE International Conference on Computer Vision (ICCV),
  2017, pp. 3325--3333.

\bibitem{8462592}
L.~{Yuan}, Q.~{Zhao}, J.~{Cao}, High-order tensor completion for data recovery
  via sparse tensor-train optimization, in: 2018 IEEE International Conference
  on Acoustics, Speech and Signal Processing (ICASSP), 2018, pp. 1258--1262.

\bibitem{7460232}
M.~{Baust}, A.~{Weinmann}, M.~{Wieczorek}, T.~{Lasser}, M.~{Storath},
  N.~{Navab}, Combined tensor fitting and tv regularization in diffusion tensor
  imaging based on a riemannian manifold approach, IEEE Transactions on Medical
  Imaging 35~(8) (2016) 1972--1989.

\bibitem{8417937}
A.~{Zare}, A.~{Ozdemir}, M.~A. {Iwen}, S.~{Aviyente}, Extension of pca to
  higher order data structures: An introduction to tensors, tensor
  decompositions, and tensor pca, Proceedings of the IEEE 106~(8) (2018)
  1341--1358.

\bibitem{8682281}
X.~{Gong}, W.~{Chen}, Multi-spectral image denoising with shared dictionaries
  and low-rank representation, in: ICASSP 2019 - 2019 IEEE International
  Conference on Acoustics, Speech and Signal Processing (ICASSP), 2019, pp.
  1707--1711.

\bibitem{8676115}
W.~{Liu}, J.~{Lee}, A 3-d atrous convolution neural network for hyperspectral
  image denoising, IEEE Transactions on Geoscience and Remote Sensing 57~(8)
  (2019) 5701--5715.

\bibitem{8517386}
J.~{Chen}, W.~{Zhang}, Y.~{Qian}, M.~{Ye}, Deep tensor factorization for
  hyperspectral image classification, in: IGARSS 2018 - 2018 IEEE International
  Geoscience and Remote Sensing Symposium, 2018, pp. 4788--4791.

\bibitem{7902201}
J.~{Chien}, Y.~{Bao}, Tensor-factorized neural networks, IEEE Transactions on
  Neural Networks and Learning Systems 29~(5) (2018) 1998--2011.

\bibitem{8902979}
S.~M. {Mohammadi}, S.~{Kouchaki}, S.~{Sanei}, D.~{Dijk}, A.~{Hilton},
  K.~{Wells}, Tensor factorisation and transfer learning for sleep pose
  detection, in: 2019 27th European Signal Processing Conference (EUSIPCO),
  2019, pp. 1--5.

\bibitem{9005677}
Z.~{Chen}, S.~{Gai}, D.~{Wang}, Deep tensor factorization for multi-criteria
  recommender systems, in: 2019 IEEE International Conference on Big Data (Big
  Data), 2019, pp. 1046--1051.

\bibitem{8937263}
J.~{Casebeer}, M.~{Colomb}, P.~{Smaragdis}, Deep tensor factorization for
  spatially-aware scene decomposition, in: 2019 IEEE Workshop on Applications
  of Signal Processing to Audio and Acoustics (WASPAA), 2019, pp. 180--184.

\bibitem{9108248}
L.~{Luo}, L.~{Xie}, H.~{Su}, Deep learning with tensor factorization layers for
  sequential fault diagnosis and industrial process monitoring, IEEE Access 8
  (2020) 105494--105506.

\bibitem{dean2012large}
J.~{Dean}, G.~{Corrado}, R.~{Monga}, K.~{Chen}, M.~{Devin}, M.~{Mao},
  A.~{Senior}, P.~{Tucker}, K.~{Yang}, Q.~V. {Le}, et~al., Large scale
  distributed deep networks, in: Advances in neural information processing
  systems, 2012, pp. 1223--1231.

\bibitem{8354695}
M.~{Langer}, A.~{Hall}, Z.~{He}, W.~{Rahayu}, Mpca sgd¡ªa method for
  distributed training of deep learning models on spark, IEEE Transactions on
  Parallel and Distributed Systems 29~(11) (2018) 2540--2556.

\bibitem{zhang2015staleness}
W.~{Zhang}, S.~{Gupta}, X.~{Lian}, J.~{Liu}, Staleness-aware async-sgd for
  distributed deep learning, arXiv preprint arXiv:1511.05950.

\bibitem{strom2015scalable}
N.~{Strom}, Scalable distributed dnn training using commodity gpu cloud
  computing, in: Sixteenth Annual Conference of the International Speech
  Communication Association, 2015.

\bibitem{ho2013more}
Q.~{Ho}, J.~{Cipar}, H.~{Cui}, S.~{Lee}, J.~K. {Kim}, P.~B. {Gibbons}, G.~A.
  {Gibson}, G.~{Ganger}, E.~P. {Xing}, More effective distributed ml via a
  stale synchronous parallel parameter server, in: Advances in neural
  information processing systems, 2013, pp. 1223--1231.

\bibitem{7837841}
S.~{Gupta}, W.~{Zhang}, F.~{Wang}, Model accuracy and runtime tradeoff in
  distributed deep learning: A systematic study, in: 2016 IEEE 16th
  International Conference on Data Mining (ICDM), 2016, pp. 171--180.

\bibitem{alvarez2016learning}
J.~M. {Alvarez}, M.~{Salzmann}, Learning the number of neurons in deep
  networks, in: Advances in Neural Information Processing Systems, 2016, pp.
  2270--2278.

\bibitem{howard2017mobilenets}
A.~G. {Howard}, M.~{Zhu}, B.~{Chen}, D.~{Kalenichenko}, W.~{Wang}, T.~{Weyand},
  M.~{Andreetto}, H.~{Adam}, Mobilenets: Efficient convolutional neural
  networks for mobile vision applications, arXiv preprint arXiv:1704.04861.

\bibitem{8578389}
G.~{Huang}, S.~{Liu}, L.~v.~d. {Maaten}, K.~Q. {Weinberger}, Condensenet: An
  efficient densenet using learned group convolutions, in: 2018 IEEE/CVF
  Conference on Computer Vision and Pattern Recognition, 2018, pp. 2752--2761.

\bibitem{lin2017towards}
X.~{Lin}, C.~{Zhao}, W.~{Pan}, Towards accurate binary convolutional neural
  network, in: Advances in Neural Information Processing Systems, 2017, pp.
  345--353.

\bibitem{8578814}
X.~{Zhang}, X.~{Zhou}, M.~{Lin}, J.~{Sun}, Shufflenet: An extremely efficient
  convolutional neural network for mobile devices, in: 2018 IEEE/CVF Conference
  on Computer Vision and Pattern Recognition, 2018, pp. 6848--6856.

\bibitem{8253600}
Y.~{Cheng}, D.~{Wang}, P.~{Zhou}, T.~{Zhang}, Model compression and
  acceleration for deep neural networks: The principles, progress, and
  challenges, IEEE Signal Processing Magazine 35~(1) (2018) 126--136.

\bibitem{8547604}
I.~{Oguntola}, S.~{Olubeko}, C.~{Sweeney}, Slimnets: An exploration of deep
  model compression and acceleration, in: 2018 IEEE High Performance extreme
  Computing Conference (HPEC), 2018, pp. 1--6.

\bibitem{8171208}
J.~{Cheng}, J.~{Wu}, C.~{Leng}, Y.~{Wang}, Q.~{Hu}, Quantized cnn: A unified
  approach to accelerate and compress convolutional networks, IEEE Transactions
  on Neural Networks and Learning Systems 29~(10) (2018) 4730--4743.

\bibitem{7551397}
S.~{Han}, X.~{Liu}, H.~{Mao}, J.~{Pu}, A.~{Pedram}, M.~A. {Horowitz}, W.~J.
  {Dally}, Eie: Efficient inference engine on compressed deep neural network,
  in: 2016 ACM/IEEE 43rd Annual International Symposium on Computer
  Architecture (ISCA), 2016, pp. 243--254.

\bibitem{8686678}
C.~{Ding}, S.~{Liao}, Y.~{Wang}, Z.~{Li}, N.~{Liu}, Y.~{Zhuo}, C.~{Wang},
  X.~{Qian}, Y.~{Bai}, G.~{Yuan}, X.~{Ma}, Y.~{Zhang}, J.~{Tang}, Q.~{Qiu},
  X.~{Lin}, B.~{Yuan}, Circnn: Accelerating and compressing deep neural
  networks using block-circulant weight matrices, in: 2017 50th Annual IEEE/ACM
  International Symposium on Microarchitecture (MICRO), 2017, pp. 395--408.

\bibitem{han2019deep}
S.~{Han}, H.~{Mao}and W. J. {Dally}arXiv~preprint arXiv:1510.00149, Deep
  compression: Compressing deep neural networks with pruning, trained
  quantization and huffman coding, arXiv preprint arXiv:1510.00149.

\bibitem{hubara2017quantized}
I.~{Hubara}, M.~{Courbariaux}, D.~{Soudry}, R.~{El-Yaniv}, Y.~{Bengio},
  Quantized neural networks: Training neural networks with low precision
  weights and activations, The Journal of Machine Learning Research 18~(1)
  (2017) 6869--6898.

\bibitem{sparsity}
W.~{Wen}, C.~{Wu}, Y.~{Wang}, Y.~{Chen}, H.~{Li}, Learning structured sparsity
  in deep neural networks, in: nces in neural information processing systems,
  2016, pp. 2082--2090.

\bibitem{azulay2018deep}
A.~{Azulay}, Y.~{Weiss}, Why do deep convolutional networks generalize so
  poorly to small image transformations?, arXiv preprint arXiv:1805.12177.

\bibitem{8601309}
J.~{Su}, D.~V. {Vargas}, K.~{Sakurai}, One pixel attack for fooling deep neural
  networks, IEEE Transactions on Evolutionary Computation (2019) 1--1.

\bibitem{raskutti2014early}
G.~{Raskutti}, M.~J. {Wainwright}and B.~{Yu}, Early stopping and non-parametric
  regression: an optimal data-dependent stopping rule, The Journal of Machine
  Learning Research 15~(1) (2014) 335--366.

\bibitem{srivastava2014dropout}
N.~{Srivastava}, G.~{Hinton}, A.~{Krizhevsky}, I.~{Sutskever},
  R.~{Salakhutdinov}, Dropout: a simple way to prevent neural networks from
  overfitting, The Journal of Machine Learning Research 15~(1) (2014)
  1929--1958.

\bibitem{inoue2018data}
H.~{Inoue}, Data augmentation by pairing samples for images classification,
  arXiv preprint arXiv:1801.02929.

\bibitem{zhang2016understanding}
C.~{Zhang}, S.~{Bengio}, M.~{Hardt}, B.~{Recht}, O.~{Vinyals}, Understanding
  deep learning requires rethinking generalization, arXiv preprint
  arXiv:1611.03530.

\bibitem{neyshabur2017exploring}
B.~{Neyshabur}, S.~{Bhojanapalli}, D.~{McAllester}, N.~{Srebro}, Exploring
  generalization in deep learning, in: Advances in Neural Information
  Processing Systems, 2017, pp. 5947--5956.

\bibitem{5306135}
A.~{Zymnis}, S.~{Boyd}, E.~{Candes}, Compressed sensing with quantized
  measurements, IEEE Signal Processing Letters 17~(2) (2010) 149--152.

\bibitem{9053161}
S.~{Takabe}, T.~{Wadayama}, Y.~C. {Eldar}, Complex trainable ista for linear
  and nonlinear inverse problems, in: ICASSP 2020 - 2020 IEEE International
  Conference on Acoustics, Speech and Signal Processing (ICASSP), 2020, pp.
  5020--5024.

\bibitem{8637803}
R.~K. {Mahabadi}, J.~{Lin}, V.~{Cevher}, A learning-based framework for
  quantized compressed sensing, IEEE Signal Processing Letters 26~(6) (2019)
  883--887.

\bibitem{9083783}
M.~{Leinonen}, M.~{Codreanu}, Quantized compressed sensing via deep neural
  networks, in: 2020 2nd 6G Wireless Summit (6G SUMMIT), 2020, pp. 1--5.

\bibitem{7560597}
B.~{Sun}, H.~{Feng}, K.~{Chen}, X.~{Zhu}, A deep learning framework of
  quantized compressed sensing for wireless neural recording, IEEE Access 4
  (2016) 5169--5178.

\end{thebibliography}
\end{document}